\documentclass[a4paper,11pt]{article}
\usepackage{jheppub}

\usepackage{bm}
\usepackage{amssymb}
\usepackage{amsmath}
\usepackage{appendix}
\usepackage{color}
\usepackage{subfigure}

\usepackage{MnSymbol}
\usepackage{multirow}
\usepackage{diagbox}

\usepackage{slashed}
\usepackage{physics}

\begin{document}

\title{Next-to-leading-order QCD corrections to pion gravitational form factors}

\author[a,b]{Jingshu Dai}\emailAdd{daijs@ihep.ac.cn}
\author[c,a]{Feng Feng}\emailAdd{f.feng@outlook.com}
\author[d,e]{Yu Jia}\emailAdd{yjia@m.scnu.edu.cn}
\author[f,g,h]{Xuan-Bo Tong}\emailAdd{xutong@jyu.fi}
\author[a,b]{Zhe-Yu Wang}\emailAdd{wangzheyu@ihep.ac.cn}

\affiliation[a]{Institute of High Energy Physics, Chinese Academy of Sciences, Beijing 100049, China}
\affiliation[b]{School of Physics, University of Chinese Academy of Sciences, Beijing 100049, China}

\affiliation[c]{China University of Mining and Technology, Beijing 100083, China}

\affiliation[d]{State Key Laboratory of Nuclear Physics and Technology, Institute of Quantum Matter, South China Normal University, Guangzhou 510006, China}

\affiliation[e]{Guangdong Basic Research Center of Excellence for Structure and Fundamental Interactions of Matter, Guangdong Provincial Key Laboratory of Nuclear Science, Guangzhou 510006, China}

\affiliation[f]{Center of Excellence in High Energy Physics, Faculty of Science, Chulalongkorn University, 254 Phaya Thai Rd, Wang Mai, Pathum Wan, Bangkok 10330, Thailand}

\affiliation[g]{Department of Physics, University of Jyväskylä, P.O. Box 35, 40014 University of Jyväskylä, Finland}

\affiliation[h]{Helsinki Institute of Physics, P.O. Box 64, 00014 University of Helsinki, Finland}

\abstract{We calculate the next-to-leading-order (NLO) QCD corrections to the pion gravitational form factors (GFFs) and the pion trace-anomaly form factor at large momentum transfer. Within leading-twist collinear factorization, we express these form factors as convolutions of pion light-cone distribution amplitudes with perturbatively calculable hard-scattering kernels. Ultraviolet renormalization and collinear subtraction are performed consistently using dimensional regularization and the $\overline{\mathrm{MS}}$ scheme. Using recent lattice-QCD inputs for the pion distribution amplitude, we provide numerical predictions and assess the impact of NLO corrections on the normalization and scale dependence of the form factors. Our results provide NLO perturbative benchmarks for experimental, lattice-QCD, and phenomenological studies of the pion's mass and mechanical structure.}

\maketitle
\flushbottom

\section{Introduction}

Elucidating the internal structure of hadrons in terms of fundamental quark and gluon degrees of freedom constitutes a primary objective of QCD. This scientific imperative drives experimental programs at facilities, such as {\tt HERA} and {\tt Jefferson Lab}~\cite{Dudek:2012vr,Accardi:2023chb}, and will remain a cornerstone of the physics case for the
future electron--ion colliders {\tt EIC/EicC}~\cite{Boer:2011fh,Accardi:2012qut,AbdulKhalek:2021gbh,Anderle:2021wcy}.

While electromagnetic form factors
have historically provided essential insights into the spatial distributions of electric charge and magnetization, gravitational form factors (GFFs) offer access to equally fundamental mechanical properties of hadrons. Defined through the off-forward matrix elements of the QCD energy--momentum tensor (EMT)~\cite{Kobzarev:1962wt,Pagels:1966zza,Ji:1996ek}, GFFs encode the spatial distributions of mass~\cite{Ji:1994av,Ji:1995sv,Ji:2021mtz,Hatta:2018sqd,Lorce:2017xzd,Metz:2020vxd,Lorce:2021xku,Yang:2018nqn} and spin~\cite{Jaffe:1989jz,Ji:1996ek,Leader:2013jra,Ji:2020ena,Ji:2020hii}. Furthermore, they characterize the internal momentum flow of hadrons, which can be described through pressure and shear distributions~\cite{Polyakov:2002yz,Polyakov:2018zvc,Shanahan:2018nnv,Lorce:2018egm} or momentum-current multipoles~\cite{Ji:2021mfb} (see also Refs.~\cite{Ji:2025qax,Ji:2025gsq,Ji:2026lyj}).

Experimentally, hadronic GFFs can be accessed via hard exclusive processes, with deeply virtual Compton scattering  serving as the most prominent channel~\cite{Ji:1996ek,Ji:1996nm,Pasquini:2014vua,Burkert:2018bqq,Dutrieux:2021nlz,Burkert:2021ith,Burkert:2023atx,Bhattacharya:2022xxw,Bhattacharya:2023wvy}.
Additionally, the threshold production of heavy quarkonium serves as a complementary probe, providing unique sensitivity to the gluonic contributions to the GFFs of the nucleon~\cite{Kharzeev:1995ij,Kharzeev:1998bz,Hatta:2018ina,Hatta:2019lxo,Kharzeev:2021qkd,Boussarie:2020vmu,Hatta:2021can,Guo:2021ibg,Guo:2023qgu,Guo:2023pqw,Sun:2021gmi,Sun:2021pyw,Mamo:2021tzd,Mamo:2022eui,Wang:2021dis,Duran:2022xag}.

Motivated by recent proposals to probe the proton mass structure via near-threshold $J/\psi$ and $\Upsilon$ production~\cite{Kharzeev:1995ij,Kharzeev:1998bz,Hatta:2018ina,Hatta:2019lxo,Kharzeev:2021qkd,Boussarie:2020vmu,Hatta:2021can,Guo:2021ibg,Guo:2023qgu,Guo:2023pqw,Sun:2021gmi,Sun:2021pyw,Mamo:2021tzd,Mamo:2022eui,Wang:2021dis,Duran:2022xag},
GFFs have attracted significant attention in recent years (for a review, see Ref.~\cite{Burkert:2023wzr}). The GFFs of the proton and pion have been investigated using various complementary approaches, including lattice QCD~\cite{Hagler:2003jd,Gockeler:2003jfa,Hagler:2007xi,Deka:2013zha,Shanahan:2018nnv,Shanahan:2018pib,Hackett:2023nkr,Hackett:2023rif}, perturbative QCD (pQCD)~\cite{Tong:2021ctu,Tong:2022zax,Coriano:2026gyl,Coriano:2026ssa}, small-$x$ approaches~\cite{Hagiwara:2024wqz,Uji:2026zvn}, chiral perturbation theory~\cite{Donoghue:1991qv,Belitsky:2002jp, Diehl:2006ya, Alharazin:2020yjv, Gegelia:2021wnj, Alharazin:2023uhr,Tanaka:2025pny}, light-cone sum rules~\cite{Anikin:2019kwi,Aliev:2020aih}, dispersion relations~\cite{Donoghue:1990xh,Cao:2024zlf,Cao:2025dkv} and holographic methods~\cite{Abidin:2008hn,Abidin:2009hr,Mamo:2021krl,Fujita:2022jus,Fujii:2024rqd,Sain:2025kup,Deng:2026ydm}. In particular, while lattice simulations are well-suited for determining GFFs
in the small $Q^2$ region, pQCD provides a systematic framework for analyzing their behavior at large invariant momentum transfer,
$Q^2 \gg \Lambda_{\mathrm{QCD}}^2$, where short-distance dynamics dominate~\cite{Tong:2021ctu,Tong:2022zax}. The large-$Q^2$ behavior not only elucidates the short-distance structure of mass and mechanical distributions but also imposes crucial constraints on phenomenological parametrizations~\cite{Guo:2021ibg,Duran:2022xag,Guo:2025jiz}. Furthermore, this regime connects directly to heavy quarkonium production, as the relevant near-threshold kinematics typically probe the large-$Q^2$ domain~\cite{Sun:2021gmi,Sun:2021pyw}.

In this work, we focus on the perturbative analysis of the pion GFFs at large momentum transfer.
Among hadrons, the pion occupies a unique position as the lightest QCD bound state and the
pseudo Nambu--Goldstone boson associated with spontaneous chiral symmetry breaking. Its spin-zero nature renders it an exceptionally clean system for probing the fundamental properties encoded in GFFs, free from the complications associated with spin degrees of freedom, thereby considerably simplifying the perturbative analysis. For the total QCD EMT, only two independent GFFs exist for the pion, conventionally denoted as $A^\pi(Q^2)$ and $C^\pi(Q^2)$, which describe the momentum distribution and the internal mechanical (stress) structure, respectively~\cite{Polyakov:2018zvc}. When the quark and gluon contributions are considered separately, an additional form factor, $\overline{C}_{q, g}^\pi(Q^2)$, emerges in each sector. This form factor is intimately connected to the trace of the EMT and, consequently, to the QCD trace anomaly~\cite{Freedman:1974gs,Nielsen:1975ph,Nielsen:1977sy,Collins:1976yq,Tarrach:1981bi} via the matrix element of the gluonic operator $F^2$~\cite{Ji:1994av,Ji:1995sv,Hatta:2018sqd,Boussarie:2020vmu,Ji:2021mtz,Ji:2021qgo}. It therefore plays a pivotal role in elucidating the origin of the pion mass. Experimentally, pion GFFs can be constrained through two-photon pion-pair production~\cite{Kumano:2017lhr,Lorce:2022tiq} and the Sullivan process in electron-proton scattering~\cite{Hatta:2025ryj}.

At large momentum transfer, pion GFFs can be systematically described within the collinear factorization framework~\cite{Avdeenko:1981twg,Lepage:1979za,Brodsky:1981kj,Efremov:1979qk,Chernyak:1977as,Chernyak:1980dj} for hard exclusive processes, an approach that has been widely applied to hadronic electromagnetic form factors (see, e.g., Refs.~\cite{Lepage:1980fj,Chernyak:1983ej,Chernyak:1984bm,Belitsky:2002kj,Chen:2024fhj,Huang:2024ugd,Field:1981wx,Chen:2023byr,Ji:2024iak}). At leading power in $1/Q^2$, these form factors are expressed as convolutions of perturbatively calculable hard-scattering kernels with the universal, nonperturbative pion light-cone distribution amplitude (LCDA)~\cite{Tong:2021ctu,Tong:2022zax}.
For example, the total pion $A$ form factor can be expressed as
\begin{align}
A^\pi(Q^2)
=&\int_0^1 dx\int_0^1 dy\,
\phi_\pi^*(y,\mu_F)
\mathcal A\left(x,y,Q^2,\mu_R^2,\mu_F^2\right)
\phi_\pi(x,\mu_F)~,
\end{align}
where $\phi_\pi(x,\mu_F)$ denotes the pion LCDA, $\mathcal A$ is the corresponding hard-scattering kernel, and $\mu_R$ and $\mu_F$ are the renormalization and factorization scales, respectively.

The factorization formulas and the leading-power behavior of the pion GFFs were first derived at LO in $\alpha_s$~\cite{Tong:2021ctu,Tong:2022zax}. However, higher-order perturbative studies of the pion electromagnetic form factor have demonstrated that radiative corrections can be numerically significant~\cite{Field:1981wx,Ji:2024iak,Chen:2023byr}. A complete NLO calculation is therefore essential for assessing the perturbative stability of pion GFFs and improving the reliability of theoretical predictions.

Compared with the LO calculation, the NLO analysis introduces qualitatively
new theoretical features.
First, infrared singularities arise at the one-loop level and must be systematically absorbed into the pion distribution amplitudes, providing a nontrivial test of collinear factorization. Furthermore, the quark and gluon GFFs
exhibit ultraviolet~(UV) divergences at this order, necessitating nontrivial renormalization and operator mixing
within the QCD EMT sector.
In addition, the pion trace anomaly emerges at the loop level as a manifestation of quantum scale-symmetry breaking~\cite{Freedman:1974gs,Nielsen:1975ph,Nielsen:1977sy,Collins:1976yq,Tarrach:1981bi},
and its explicit realization in the pion GFFs requires careful analysis.

In this work, we present the first complete NLO perturbative QCD analysis of pion GFFs
at large momentum transfer, encompassing both quark and gluon sectors.
Working within the collinear factorization framework, we compute the one-loop corrections
to the hard-scattering kernels associated with the QCD EMT form factor.
Ultraviolet renormalization and collinear subtraction are performed consistently using dimensional regularization in the $\overline{\mathrm{MS}}$ scheme. We explicitly verify the factorization of collinear divergences into the pion
DA and analyze the emergence of the trace anomaly and its relation to the pion GFFs.
Using recent lattice QCD determinations of the pion DAs, we provide numerical predictions
and quantify the impact of NLO corrections on the normalization and renormalization-scale dependence of the GFFs.

Our results provide NLO perturbative benchmarks for pion GFFs at large momentum transfer and complement nonperturbative studies at lower momentum transfer. The methods developed here can also be applied to other hadrons.

\section{Gravitational form factors \label{sec:Introduction}}

In this section, we introduce the symmetric QCD EMT and define the pion GFFs.
We first discuss the decomposition of the EMT into quark and gluon sectors, and the parametrization of the
pion EMT matrix elements in terms of Lorentz-invariant form factors.
We then review the renormalization properties of the quark and gluon EMT operators and the corresponding GFFs,
emphasizing their mixing and scale dependence, which are essential for our NLO perturbative analysis at large momentum transfer. Furthermore, we discuss the relations between the pion GFFs and the $F^2$ scalar form factor through the QCD trace anomaly.

Since we focus on the large-momentum-transfer region, we neglect the pion and light-quark masses in our perturbative calculations. Unless otherwise specified, this massless approximation is adopted throughout this section.

\subsection{Definitions and projections}

We begin with the symmetric (Belinfante) QCD EMT, which can be decomposed into the quark and gluon sectors~\cite{Ji:1996ek}:
\begin{align}
T^{\mu\nu} = T^{\mu\nu}_q + T^{\mu\nu}_g~.
\label{eq:EMT_total}
\end{align}
The gauge-invariant quark and gluon components are given by
\begin{subequations}\label{eq:EMT_qg}
\begin{align}
T^{\mu\nu}_q &= \frac{1}{4} \bar{\psi}_q \left( -i \overleftarrow{D}^\mu \gamma^\nu - i \overleftarrow{D}^\nu \gamma^\mu + i \overrightarrow{D}^\mu \gamma^\nu + i \overrightarrow{D}^\nu \gamma^\mu \right) \psi_q~,\label{eq:EMT_qg:a}  \\
T^{\mu\nu}_g &= -F^{a,\mu\lambda} F^{a,\nu}_{\ \ \lambda} + \frac{1}{4} g^{\mu\nu} F^{a,\sigma\rho} F^a_{\sigma\rho}~,\label{eq:EMT_qg:b}
\end{align}
\end{subequations}
where $F^a_{\mu\nu} = \partial_\mu A^a_\nu - \partial_\nu A^a_\mu - g_s f^{abc} A^b_\mu A^c_\nu$ is the gluon field strength tensor for the gauge field $A^{a,\mu}$. The covariant derivatives are defined as $\overrightarrow{D}_\mu = \overrightarrow{\partial}_\mu + i g_s A^a_\mu T^a$ and $\overleftarrow{D}_\mu = \overleftarrow{\partial}_\mu - i g_s A^a_\mu T^a$, with $T^a$ denoting the generators of the SU(3) gauge group in the fundamental representation. The sum over the $n_f$ active quark flavors is implicit in $T_q^{\mu\nu}$. These EMT operators are explicitly symmetric under the exchange of the Lorentz indices $\mu$ and $\nu$. Possible ghost and gauge-fixing terms are omitted in Eq.~(\ref{eq:EMT_qg}), as they do not contribute to the physical matrix elements of the EMT, and consequently to the GFFs, due to BRST symmetry~\cite{Joglekar:1975nu}. Throughout this work, we focus on the Belinfante-improved (symmetric) EMT $T^{\mu\nu}$, which differs from the canonical (non-symmetric) EMT only by a total derivative term~\cite{BELINFANTE1939887,belinfante1940current}.

Hadronic GFFs are the Lorentz-invariant functions that parametrize off-forward matrix elements of the QCD EMT~\cite{Kobzarev:1962wt,Pagels:1966zza,Ji:1996ek}. They characterize the hadron's response to a weak external gravitational field, in analogy with electromagnetic form factors. For a spin-zero hadron such as the pion, the quark and gluon EMT matrix elements can be parameterized in terms of three independent form factors $A_a,C_a$ and $\overline{C}_a$~\cite{Polyakov:2018zvc}:
\begin{align}
\left\langle P^{\prime}\right| T_a^{\mu \nu}|P\rangle=&2 \overline{P}^\mu \overline{P}^\nu A_a(Q^2) +\frac{1}{2}\left(\Delta^\mu \Delta^\nu-g^{\mu \nu} \Delta^2\right) C_a(Q^2)+2 \Lambda^2 g^{\mu \nu} \overline{C}_a(Q^2)~,
\label{eq_GFF1}
\end{align}
where $a=q,g$. $P$ and $P'$ are the initial and final pion momenta, respectively, which satisfy $P^2=P'^2=0$ in the massless limit. We define the average momentum $\bar P$, the
momentum transfer $\Delta$, and the spacelike momentum-transfer
scale $Q^2$ as:
\begin{subequations}\label{eq:kinematics}
\begin{align}
\overline P&=(P+P')/2,\label{eq:kinematics:a}\\
\Delta&=P'-P,\label{eq:kinematics:b}\\
Q^2&=-\Delta^2\ .\label{eq:kinematics:c}
\end{align}
\end{subequations}
The parameter $\Lambda$ is a nonperturbative mass scale
introduced to make $\overline C_a$ dimensionless.

Summing the quark and gluon contributions gives the
matrix element of the total EMT,
\begin{align}  \left\langle P^{\prime}\right| T^{\mu \nu}|P\rangle=&2 \bar{P}^\mu \bar{P}^\nu A(Q^2) +\frac{1}{2}\left(\Delta^\mu \Delta^\nu-g^{\mu \nu} \Delta^2\right) C(Q^2),
\end{align}
where $A(Q^2)=\sum_{a=q,g} A_a(Q^2)$ and $C(Q^2)=\sum_{a=q,g} C_a(Q^2)$ are the sums of the quark and gluon contributions. Unlike $A(Q^2)$ and $C(Q^2)$, the $\overline{C}$ form factor vanishes in the sum,
\begin{align}
\sum_{a=q,g}  \overline{C}_a(Q^2)=0~.
\label{eq:totalcbar}
\end{align}
This is because the total EMT is conserved $\partial_\mu T^{\mu \nu}=0$ and thus imposes a constraint on the matrix element, $  \Delta_\mu  \left\langle P^{\prime}\right| T^{\mu \nu}|P\rangle=0~
$. Therefore, only two independent form factors remain for the total EMT, namely the $A$- and $C$-GFFs.

As the form factors of QCD EMT matrix elements, $A_a(Q^2), C_a(Q^2)$, and $\overline{C}_a(Q^2)$ correspond to different Lorentz structures and hence encode complementary information about the pion's structure~\cite{Polyakov:2018zvc,Burkert:2023wzr}.
The $A_{a}(Q^2)$-form factor is associated with the momentum density carried by quarks and gluons; after renormalization at a specified scale, its forward value gives the fraction of the pion's momentum carried by partons of type $a$ at that scale~\cite{Ji:1996ek,Shanahan:2018pib,Hackett:2023nkr}. The $C_a(Q^2)$ form factors characterize the internal momentum flow of the pion and are conventionally related to pressure and shear distributions~\cite{Polyakov:2002yz,Polyakov:2018zvc}. The forward value of their sum, $C(0)$, is known as the D-term, while the full momentum-transfer dependence of $C(Q^2)$ enters the conventional definition of the mechanical radius~\cite{Polyakov:2018zvc,Lorce:2018egm,Freese:2021czn}.

In addition, the GFFs $\overline C_{a}(Q^2)$ are closely related to the traces of the quark and gluon EMT operators and play an important role in understanding the pion mass structure~\cite{Ji:1994av,Ji:1995sv,Ji:2021mtz,Ji:2021qgo,Hatta:2018sqd,Boussarie:2020vmu}. We introduce the pion trace-anomaly form factor as the matrix element of the trace of the total EMT,
\begin{align}
\label{eq:traceFF}
\Theta^\pi(Q^2)
\equiv
g_{\mu\nu}\langle P'|T^{\mu\nu}|P\rangle~.
\end{align}
In the chiral limit, this form factor is governed by the QCD trace anomaly, whose relation to the gluonic scalar operator $F^2$ and the pion GFFs will be discussed in Sec.~\ref{sec:renormalization}.

\subsection{Renormalization of the GFFs and the \texorpdfstring{$F^2$}{F2} scalar form factor}
\label{sec:renormalization}
We now briefly review the renormalization properties of the EMT and their implications for the pion GFFs and the $F^2$ gluonic scalar form factor.
The total EMT is conserved as a consequence of spacetime translation
invariance,
\begin{align}
\partial_\mu T^{\mu\nu}=0~.
\end{align}
As the generator of translations, its normalization is fixed by the corresponding Ward identities. Consequently, the total EMT is UV finite and renormalization-scale independent~\cite{Freedman:1974gs,Nielsen:1975ph,Nielsen:1977sy,Collins:1976yq}. In contrast, the individual quark and gluon components are not separately conserved~\cite{Hatta:2018sqd,Tanaka:2018nae}. Their divergences are related to twist-four quark-gluon operators~\cite{Tanaka:2018wea}:
\begin{align}
& \partial_\nu T_q^{\mu \nu}=\bar{\psi}_f g F^{\mu \nu} \gamma_\nu \psi_f~, \\
& \partial_\nu T_g^{\mu \nu}=F_\nu{ }^\mu D_\alpha F^{\alpha \nu}~.
\end{align}
The quark and gluon EMT operators therefore require UV renormalization and mix with each other. Throughout this work, we use dimensional regularization with $d=4-2\epsilon$ and the $\overline{\mathrm{MS}}$ renormalization scheme.

The renormalization of the quark and gluon EMT operators has been studied in detail~\cite{Ji:1995sv,Hatta:2018sqd,Tanaka:2018nae,Tanaka:2022wzy}. At one-loop order, the renormalized operators are
related to their bare counterparts through:
\begin{subequations}\label{eq:reEMT}
\begin{align}
T_{q,R}^{\mu\nu}(\mu)
=&\,T_q^{\mu\nu}
+\frac{\alpha_s(\mu)}{4\pi\epsilon_{\rm UV}}
\left(
\frac{8C_F}{3}T_q^{\mu\nu}
-\frac{2n_f}{3}T_g^{\mu\nu}
\right)+\mathcal O(\alpha_s^2)~,\label{eq:reEMT:a}
\\
T_{g,R}^{\mu\nu}(\mu)
=&\,T_g^{\mu\nu}
+\frac{\alpha_s(\mu)}{4\pi\epsilon_{\rm UV}}
\left(
-\frac{8C_F}{3}T_q^{\mu\nu}
+\frac{2n_f}{3}T_g^{\mu\nu}
\right)+\mathcal O(\alpha_s^2)~.\label{eq:reEMT:b}\end{align}
\end{subequations}
Here, $\alpha_s(\mu)$ is the renormalized strong
coupling at the scale $\mu$. The color factors are
$C_F=(N_c^2-1)/(2N_c)$ and $C_A=N_c$, with $N_c=3$.
The mixing terms cancel
upon summing the two operators, giving
$T_{q,R}^{\mu\nu}+T_{g,R}^{\mu\nu}=T^{\mu\nu}$,
consistent with the non-renormalization of the
total EMT.

The renormalized pion GFFs are defined through
the corresponding EMT matrix elements:
\begin{align}
\langle P'|T_{a,R}^{\mu\nu}(\mu)|P\rangle
=&\,2\bar P^\mu\bar P^\nu A_a^R(Q^2,\mu)
+\frac12
\left(
\Delta^\mu\Delta^\nu-g^{\mu\nu}\Delta^2
\right)C_a^R(Q^2,\mu)
\notag\\
&+2\Lambda^2g^{\mu\nu}
\overline C_a^R(Q^2,\mu)~,
\qquad a=q,g.
\label{eq_GFF2}
\end{align}
This decomposition has the same Lorentz structures
as the bare parametrization in Eq.~\eqref{eq_GFF1}.
Hereafter, the superscript or subscript $R$ denotes
renormalized quantities, while quantities without
this designation refer to bare operators unless
otherwise specified.

Since the renormalization in Eq.~\eqref{eq:reEMT}
acts on the complete Lorentz tensor, the three
pion GFFs obey the same quark--gluon mixing relation.
Denoting by $X_a$ any of the form factors
$A_a$, $C_a$, or $\overline C_a$, we obtain
\begin{align}
\begin{pmatrix}
X_q^R(Q^2,\mu)\\
X_g^R(Q^2,\mu)
\end{pmatrix}
=&
\left[
\begin{pmatrix}
1&0\\
0&1
\end{pmatrix}
+
\frac{\alpha_s(\mu)}{4\pi\epsilon_{\rm UV}}
\begin{pmatrix}
\displaystyle\frac{8C_F}{3}
&\displaystyle-\frac{2n_f}{3}\\[6pt]
\displaystyle-\frac{8C_F}{3}
&\displaystyle\frac{2n_f}{3}
\end{pmatrix}
\right]
\begin{pmatrix}
X_q(Q^2)\\
X_g(Q^2)
\end{pmatrix}~,
\qquad X=A,C,\overline C~.
\label{eq:renor_A}
\end{align}
The mixing contributions cancel in the sum
of the quark and gluon GFFs, consistent with
the conservation of the total EMT.

For the perturbative calculation below, the GFFs $A_a$ and
$C_a$ can be extracted from the EMT matrix elements using
the following traceless projection tensors:
\begin{subequations}\label{eq_tensor}
\begin{align}
A_a(Q^2)
=&
\frac{
8(d-1)\Gamma_{1,\mu\nu}
-2d\Gamma_{2,\mu\nu}
}{
(d-2)d(\Delta^2)^2
}
\langle P'|T_a^{\mu\nu}|P\rangle~,\label{eq_tensor:a}
\\
C_a(Q^2)
=&
\frac{
2(d-1)d\Gamma_{2,\mu\nu}
-8\Gamma_{1,\mu\nu}
}{
(d-2)d(\Delta^2)^2
}
\langle P'|T_a^{\mu\nu}|P\rangle~.\label{eq_tensor:b}
\end{align}
\end{subequations}
Here,
$\Gamma_1^{\mu\nu}
=d\bar P^\mu\bar P^\nu-\bar P^2g^{\mu\nu}$
and
$\Gamma_2^{\mu\nu}
=\Delta^\mu\Delta^\nu-\Delta^2g^{\mu\nu}/d$,
which satisfy $g_{\mu\nu}\Gamma_{1,2}^{\mu\nu}=0$.
Consequently, the contributions proportional to
$g^{\mu\nu}\overline C_a$ do not enter these projections.
The same projectors apply to both bare and renormalized
EMT matrix elements.

The remaining form factor $\overline C_a$
can be related to the trace of the EMT
matrix element. Taking the trace of the
bare decomposition in $d$
dimensions gives
\begin{align}
2\Lambda^2\overline C_a(Q^2)
=&
\frac{1}{d}
\langle P'|g_{\mu\nu}T_a^{\mu\nu}|P\rangle
-\frac{Q^2}{2d}
\left[
A_a(Q^2)+(d-1)C_a(Q^2)
\right]~.
\label{eq:Cbar_bare}
\end{align}
An analogous relation holds for the renormalized
EMT matrix elements.

At the classical level, the EMT is traceless in four dimensions in the massless limit. Quantum corrections, however, break scale invariance and generate the well-known trace anomaly~\cite{Freedman:1974gs,Nielsen:1975ph,Nielsen:1977sy,Collins:1976yq,Tarrach:1981bi}. In dimensional regularization with $d=4-2\epsilon$, the bare EMT operators satisfy~\cite{Hatta:2018sqd}
\begin{subequations}\label{eq:bareTrace}
\begin{align}
g_{\mu\nu}T_g^{\mu\nu}
=&-\frac{\epsilon}{2}F^2~,\label{eq:bareTrace:a}
\\
g_{\mu\nu}T_q^{\mu\nu}=&0~,\label{eq:bareTrace:b}
\end{align}
\end{subequations}
where $F^2=F^a_{\rho\sigma}F^{a,\rho\sigma}$
and the quark equation of motion has been used. Although the bare gluon trace is proportional to $\epsilon$, the UV divergences of the composite operator $F^2$ generate a finite anomalous contribution after renormalization.

In massless QCD, the gluonic scalar operator $F^2$ is multiplicatively renormalizable in the physical operator sector. In the $\overline{\mathrm{MS}}$ scheme, its renormalization relation through one-loop order reads~\cite{Tarrach:1981bi}
\begin{equation}
(F^2)_R
=
\left(
1-
\frac{\alpha_s(\mu)}{4\pi}
\frac{\beta_0}{\epsilon_{\rm UV}}
+\mathcal O(\alpha_s^2)
\right)F^2~,
\label{eq:FFR}
\end{equation}
where $\beta_0=11C_A/3-2n_f/3$, with $n_f$ denoting the number of active quark flavors. The UV pole in this renormalization relation, combined with the explicit factor of $\epsilon$ in Eq.~\eqref{eq:bareTrace}, gives rise to the finite QCD trace anomaly. The trace of the total  EMT therefore satisfies~\cite{Nielsen:1977sy,Collins:1976yq}
\begin{align}
g_{\mu\nu}T^{\mu\nu}
=
\frac{\beta(g_R)}{2g_R}(F^2)_R~,
\label{eq:TraceR}
\end{align}
where $g_R$ is the renormalized strong coupling and $\beta(g_R)$ is the QCD beta function, whose leading contribution reads
\begin{align}
\frac{\beta(g_R)}{2g_R}
=
-\frac{\beta_0}{2}
\frac{\alpha_s(\mu)}{4\pi}
+\mathcal O(\alpha_s^2)~.
\end{align}

The trace anomaly is also distributed between
the renormalized quark and gluon EMT operators.
Indeed, taking the $d$-dimensional trace of
Eq.~\eqref{eq:reEMT}, using
Eq.~\eqref{eq:bareTrace}, and expressing the
result in terms of $(F^2)_R$, we obtain~\cite{Hatta:2018sqd,Tanaka:2018nae}
\begin{subequations}\label{eq:qgtraceEMT}
\begin{align}
g_{\mu\nu}T_{g,R}^{\mu\nu}
=&
-\frac{\alpha_s(\mu)}{4\pi}
\frac{11C_A}{6}(F^2)_R
+\mathcal O(\alpha_s^2)~,\label{eq:qgtraceEMT:a}
\\
g_{\mu\nu}T_{q,R}^{\mu\nu}
=&
\frac{\alpha_s(\mu)}{4\pi}
\frac{n_f}{3}(F^2)_R
+\mathcal O(\alpha_s^2)~.\label{eq:qgtraceEMT:b}
\end{align}
\end{subequations}
Their sum reproduces the total trace anomaly in Eq.~\eqref{eq:TraceR}, which is renormalization-group invariant, although its decomposition into the individual quark and gluon contributions depends on the renormalization scheme and scale.

Taking the trace of the renormalized GFF
decomposition in Eq.~\eqref{eq_GFF2} and using
Eq.~\eqref{eq:qgtraceEMT}, we obtain
\begin{subequations}\label{eq:Cbarrelation}
\begin{align}
2\Lambda^2\overline C_g^R(Q^2,\mu)
=&
-\frac{\alpha_s(\mu)}{16\pi}
\frac{11C_A}{6}
\langle P'|(F^2)_R|P\rangle
-\frac{Q^2}{8}
\left[
A_g^R(Q^2,\mu)+3C_g^R(Q^2,\mu)
\right]~,\label{eq:Cbarrelation:a}
\\
2\Lambda^2\overline C_q^R(Q^2,\mu)
=&
\frac{\alpha_s(\mu)}{16\pi}
\frac{n_f}{3}
\langle P'|(F^2)_R|P\rangle
-\frac{Q^2}{8}
\left[
A_q^R(Q^2,\mu)+3C_q^R(Q^2,\mu)
\right]~.\label{eq:Cbarrelation:b}
\end{align}
\end{subequations}
Here, the anomaly coefficients are retained
through one-loop order, as required for the NLO
pion GFF calculation. These relations connect
the renormalized $\overline C_{q,g}$ form factors
to the other GFFs and the gluonic scalar
matrix element. They provide an independent
constraint on the form factors obtained through
the direct EMT renormalization in
Eq.~\eqref{eq:renor_A}.

Finally, using EMT conservation, which requires $\overline C_q^R(Q^2,\mu)+\overline C_g^R(Q^2,\mu)=0$, we obtain the relation between the total pion GFFs and the trace anomaly:
  \begin{align}
\Theta^\pi(Q^2)=\frac{Q^2}{2}\big[A(Q^2)+3C(Q^2)\big]
=\frac{\beta(g_{R})}{2 g_{R}}\langle P'|  (F^2)_R|P\rangle~.
\label{eq:F2fromGFFs}
 \end{align}
This identity provides a direct connection between the pion GFFs and the QCD trace anomaly and serves as an important consistency check of our NLO perturbative calculation.

  \section{Perturbative analysis at large momentum transfer}
Having introduced the pion GFFs and their renormalization properties,
we now calculate their leading-power contributions at large spacelike
momentum transfer, $Q^2\gg\Lambda_{\mathrm{QCD}}^2$.
Within leading-twist collinear factorization, the pion is described
by its minimal quark--antiquark Fock state, with the large momentum
transfer shared between its constituents through hard-gluon exchange~\cite{Lepage:1979za,Brodsky:1981kj,Efremov:1979qk,Chernyak:1977as,Chernyak:1980dj,Chernyak:1983ej,Belitsky:2002kj}.
Long-distance collinear dynamics are encoded in the pion LCDAs,
while the short-distance interaction with the EMT is described by
perturbatively calculable hard-scattering kernels.
We determine these kernels by matching partonic EMT matrix elements
to their factorized expressions, extending the LO results of
Refs.~\cite{Tong:2021ctu,Tong:2022zax} to NLO in both the quark
and gluon sectors.

At NLO, collinear singularities must be absorbed into the pion LCDAs, whereas UV divergences require quark--gluon EMT operator mixing. After introducing necessary conventions, we first calculate the amplitudes with bare EMT insertions and perform collinear subtraction, followed by the operator renormalization discussed in Sec.~\ref{sec:renormalization}. The $d$-dimensional trace of the bare gluon EMT also provides access to the $F^2$ scalar form factor, allowing us to determine the pion trace anomaly through NLO. Finally, we express the finite hard-scattering kernels in the Gegenbauer basis to obtain master formulas for numerical applications.

   \subsection{Factorization formula and conventions}
We work with massless external pion momenta, $P^2=P'^2=0$. At leading power in $1/Q^2$, the pion matrix elements of the renormalized quark and gluon EMT operators can be factorized as
\begin{align}
\langle P'|T_{a,R}^{\mu\nu}(\mu_R)|P\rangle
=&\int_0^1 dx\int_0^1 dy\,
\phi_\pi^*(y,\mu_F)
H_{a,R}^{\mu\nu}(x,y,Q,\mu_F,\mu_R)
\phi_\pi(x,\mu_F)~,
\qquad a=q,g.
\label{GFFconv}
\end{align}
Here, $\mu_F$ and $\mu_R$ denote the factorization and renormalization scales, respectively. The kernel $H_{a,R}^{\mu\nu}$ corresponds to the renormalized EMT matrix element. We also introduce $H_a^{\mu\nu}$  to denote the kernel associated with the bare EMT operator, which may still contain operator UV divergences. Throughout this work, we use the same renormalization scale $\mu_R$ for the strong coupling and the quark and gluon EMT operators.

$\phi_\pi\left(x, \mu_F\right)$ is the renormalized twist-2 pion LCDA at the scale $\mu_F$, which represents the probability amplitude for
finding the valence $q \bar q$ pair in the initial pion with the constituents carrying the longitudinal momentum $x P$ and $(1-x)P$. We adopt the normalization convention of Ref.~\cite{Tong:2022zax}. Its bare counterpart, $\phi_\pi(x)$, is defined through the light-cone matrix element~\cite{Radyushkin:1977gp,Efremov:1979qk,Tong:2022zax}:
\begin{align}
\label{eq:pionLCDA}
i\sqrt{6}\phi_\pi(x)
=& \int \frac{d \xi^-}{2\pi} e^{i x P^+ \xi^-}\langle 0|\bar{d}(0)[0, \xi^-] \gamma^+ \gamma_{5} u( \xi^-)| \pi(P)\rangle~,
\end{align}
Here, $[0,\xi^-]$ is the Wilson line along the light-cone direction, ensuring gauge invariance. With this convention,
$\int_0^1 dx\,\phi_\pi(x)=f_\pi/\sqrt{6}$, where $f_\pi\simeq131~\mathrm{MeV}$ is the pion decay constant. Upon UV renormalization, the pion LCDA acquires a dependence on the factorization scale $\mu_F$, governed by the Efremov--Radyushkin--Brodsky--Lepage (ERBL) evolution equation~%
\cite{Efremov:1978rn,Lepage:1979zb}:
\begin{equation}
\label{EQ:evolution}
\frac{\mathrm{d}\phi_\pi\left(x, \mu_F\right)}{\mathrm{d\;ln}\mu_F^2}=\int^1_0\mathrm{d}y\;V(x,y)\phi_\pi\left(y, \mu_F\right)~,
\end{equation}
where the ERBL kernel admits the perturbative expansion
\begin{align}
    V=\frac{\alpha_s}{\pi}V^{(0)}+\left(\frac{\alpha_s}{\pi}\right)^2V^{(1)}+\cdots~.
\end{align}
The LO kernel reads
\begin{equation}
    V^{(0)}(x,y)=\frac{C_F}{2}\left[\frac{x}{y}\Big(1+\frac{1}{y-x}\Big)\theta(y-x)+\{x\to\bar{x},y\to\bar{y}\}\right]_+~,
    \label{eq:VLO}
\end{equation}
where $\bar{x}=1-x$ and $\bar{y}=1-y$, and the plus prescription
acts on the first argument.
For the NLO numerical analysis presented later, we also employ the NLO ERBL kernel $V^{(1)}$, whose explicit expression can be found in Refs.~\cite{Dittes:1983dy,Mikhailov:1984ii,Mueller:1994cn}.

The renormalized hard-scattering kernels can be calculated perturbatively and expanded in powers of the renormalized coupling $\alpha_s(\mu_R)$ as
\begin{equation}
H_{a,R}^{\mu\nu}(x,y)
=
\frac{4\pi\alpha_s(\mu_R)C_F}{Q^2}
\left[
H_{a,R}^{\mu\nu,(0)}(x,y)
+\frac{\alpha_s(\mu_R)}{\pi}
H_{a,R}^{\mu\nu,(1)}(x,y)
+\mathcal O(\alpha_s^2)
\right]~,
\label{eq:Hexpan}
\end{equation}
where we have factored out the common prefactor $4\pi\alpha_sC_F/Q^2$ and suppressed the scale arguments for brevity.
The hard-scattering kernels admit the same Lorentz decomposition as the corresponding renormalized pion EMT matrix elements in Eq.~\eqref{eq_GFF2}. We write
\begin{align}
H_{a,R}^{\mu\nu}(x,y)
=&\,2\bar P^\mu\bar P^\nu
\mathcal A_{a,R}(x,y)
+\frac12
\left(\Delta^\mu\Delta^\nu+Q^2g^{\mu\nu}\right)
\mathcal C_{a,R}(x,y)
\notag\\
&+2\Lambda^2g^{\mu\nu}
\overline{\mathcal C}_{a,R}(x,y)~.
\label{eq:Hparame}
\end{align}
The scalar kernels determine the corresponding renormalized
pion GFFs through the factorization formula
\begin{align}
X_a^R(Q^2,\mu_R)
=&\int_0^1 dx\int_0^1 dy\,
\phi_\pi^*(y,\mu_F)
\mathcal X_{a,R}
\left(x,y,Q,\mu_R,\mu_F\right)
\phi_\pi(x,\mu_F)~,
\end{align}
where $X=A,C,\overline C$ and
$\mathcal X_{a,R}$ denotes
$\mathcal A_{a,R}$, $\mathcal C_{a,R}$, or
$\overline{\mathcal C}_{a,R}$, respectively.
Its perturbative coefficients
$\mathcal X_{a,R}^{(k)}$ are defined by decomposing
$H_{a,R}^{\mu\nu,(k)}$ into the same Lorentz structures
as in Eq.~\eqref{eq:Hparame}.
The same decomposition applies to kernels with
bare EMT insertions, for which the subscript $R$
is omitted.

   \subsection{NLO calculation of hard coefficients}
\label{sec:hardcoefficients}
\begin{figure}
    \centering
    \includegraphics[width=0.3\linewidth]{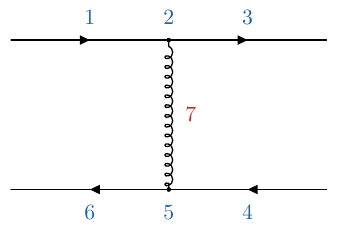}
   \caption{LO diagrams for the EMT amplitude
$\mathcal{T}^{\mu\nu}_a(u,v)$ in Eq.~\eqref{eq:Tqqb}
for a collinear quark--antiquark state.
The numbered positions indicate insertion sites of the
EMT operator.
Blue and red numbers correspond to insertions of its quark
and gluon components, $T_q^{\mu\nu}$ and $T_g^{\mu\nu}$,
respectively.}
    \label{fig:LOFeyDiag}
\end{figure}

To determine the renormalized hard-scattering kernels introduced above, we first calculate the partonic amplitudes with bare EMT insertions and extract the corresponding kernels through collinear subtraction. The remaining UV divergences are subsequently removed by quark--gluon EMT operator mixing. We regulate UV and IR singularities in
$d=4-2\epsilon$ dimensions and adopt the
$\overline{\mathrm{MS}}$ scheme for renormalization.

\paragraph{Partonic matching and LO coefficients.}

The hard-scattering kernels are extracted by replacing the pion states with on-shell, massless, collinear quark--antiquark states in the color-singlet pseudoscalar channel. We assign momenta $uP$ and $\bar uP$ to the incoming quark and antiquark, and $vP'$ and $\bar vP'$ to their outgoing counterparts, with $\bar u=1-u$ and $\bar v=1-v$. The corresponding partonic amplitude for the bare EMT is
\begin{align}
         \mathcal{T}^{\mu\nu}_a(u,v)\equiv	\left\langle q(vP')\overline{q}((1-v)P')\right| T_a^{\mu \nu}|q(uP)\overline{q}((1-u)P)\rangle~.
         \label{eq:Tqqb}
\end{align}
Its perturbative expansion in the renormalized coupling $\alpha_s(\mu_R)$ reads
\begin{align}
\mathcal{T}^{\mu\nu}_a(u,v)=&\frac{4\pi\alpha_s(\mu_R)C_F}{Q^2}\left[\mathcal{T}^{\mu\nu,(0)}_a(u,v)+\frac{\alpha_s(\mu_R)}{\pi}\mathcal{T}^{\mu\nu,(1)}_a(u,v)+\cdots\right]~.
\end{align}
  Similarly, the pion LCDA appearing in the factorization formula is replaced by the renormalized partonic LCDA.
 It is perturbatively calculable:
\begin{equation}
    \Phi_{q \bar q }(x|u;\mu_F)=\Phi^{(0)}(x|u)+\frac{\alpha_s(\mu_R)}{\pi}\Phi^{(1)}(x|u;\mu_F,\mu_R)+\cdots.
\end{equation}
At LO in $\alpha_s$, $\Phi^{(0)}(x|u)=\delta(x-u)$, so that the matching gives
\begin{align}
H^{\mu\nu,(0)}_a(x,y)=&\mathcal{T}^{\mu\nu,(0)}_a(x,y)~.
\end{align}
The relevant diagrams are shown in Fig.~\ref{fig:LOFeyDiag}. The four-dimensional LO kernels were obtained in Ref.~\cite{Tong:2021ctu}. For the NLO calculation, we retain their dimensional dependence and obtain
\begin{subequations}\label{eq:H(0)}
\begin{align}
    {\cal A}^{(0)}_g(x,y)=& \frac{1}{x\bar{x}}+\frac{1}{y\bar{y}}+\left(\frac{1}{2x\bar{x}y\bar{y}}-\frac{1}{x\bar{x}}-\frac{1}{y\bar{y}}\right)\epsilon\label{eq:H(0):a}\\
     {\cal C}^{(0)}_g(x,y)=& \frac{1}{x\bar{x}}+\frac{1}{y\bar{y}}-\left(\frac{1}{2x\bar{x}y\bar{y}}+\frac{1}{x\bar{x}}+\frac{1}{y\bar{y}}\right)\epsilon\label{eq:H(0):b}\\
    {\cal A}^{(0)}_q(x,y)=&\frac{3-2x\bar{x}-2y\bar{y}}{2x\bar{x}y\bar{y}}+\left(-\frac{1}{x\bar{x}y\bar{y}}+\frac{1}{x\bar{x}}+\frac{1}{y\bar{y}}\right)\epsilon\label{eq:H(0):c}\\    {\cal C}^{(0)}_q(x,y)=&-\frac{1+2x\bar{x}+2y\bar{y}}{2x\bar{x}y\bar{y}}+\left(\frac{1}{x\bar{x}}+\frac{1}{y\bar{y}}\right)\epsilon\label{eq:H(0):d}\\
     \overline{\cal C}^{(0)}_g(x,y)=&-\overline{\cal C}^{(0)}_q(x,y)=-\frac{Q^2}{4\Lambda^2}\left(\frac{1}{x\bar{x}}+\frac{1}{y\bar{y}}\right)(1-\epsilon)~.\label{eq:H(0):e}
\end{align}
\end{subequations}
Here, we have used the symmetry of the pion LCDA under $x\leftrightarrow\bar x$ for simplification.
At momentum fractions $u,v$, their dependence on $\epsilon$ can be organized as
\begin{align}
\mathcal A_a^{(0)}(\epsilon)&=(1-\epsilon)\mathcal A_a^{(0)}(0)+\frac{\epsilon}{2u\bar u\,v\bar v},\notag\\
\mathcal C_a^{(0)}(\epsilon)&=(1-\epsilon)\mathcal C_a^{(0)}(0)-\frac{\epsilon}{2u\bar u\,v\bar v},\qquad
\overline{\mathcal C}_a^{(0)}(\epsilon)=(1-\epsilon)\overline{\mathcal C}_a^{(0)}(0).
\label{eq:Born_epsilon}
\end{align}
The $\mathcal O(\epsilon)$ terms contribute to the finite NLO coefficients when multiplied by UV or collinear poles. The four-dimensional limit is therefore taken only after the corresponding subtractions have been performed.

Summing the quark and gluon contributions, the LO
hard-scattering coefficients for the total EMT simplify to
\begin{subequations}\label{eq:LO_total_kernels}
\begin{align}
\mathcal A^{(0)}(x,y,\epsilon)
&=
\mathcal A_q^{(0)}(x,y,\epsilon)
+\mathcal A_g^{(0)}(x,y,\epsilon)
=\frac{3-\epsilon}{2x\bar x y\bar y}~,\label{eq:LO_total_kernels:a}
\\
\mathcal C^{(0)}(x,y,\epsilon)
&=
\mathcal C_q^{(0)}(x,y,\epsilon)
+\mathcal C_g^{(0)}(x,y,\epsilon)
=-\frac{1+\epsilon}{2x\bar x y\bar y}~,\label{eq:LO_total_kernels:b}
\end{align}
\end{subequations}
while
$\overline{\mathcal C}^{(0)}
=\overline{\mathcal C}_q^{(0)}
 +\overline{\mathcal C}_g^{(0)}=0$.

\begin{figure}
    \centering
    \includegraphics[width=0.9\linewidth]{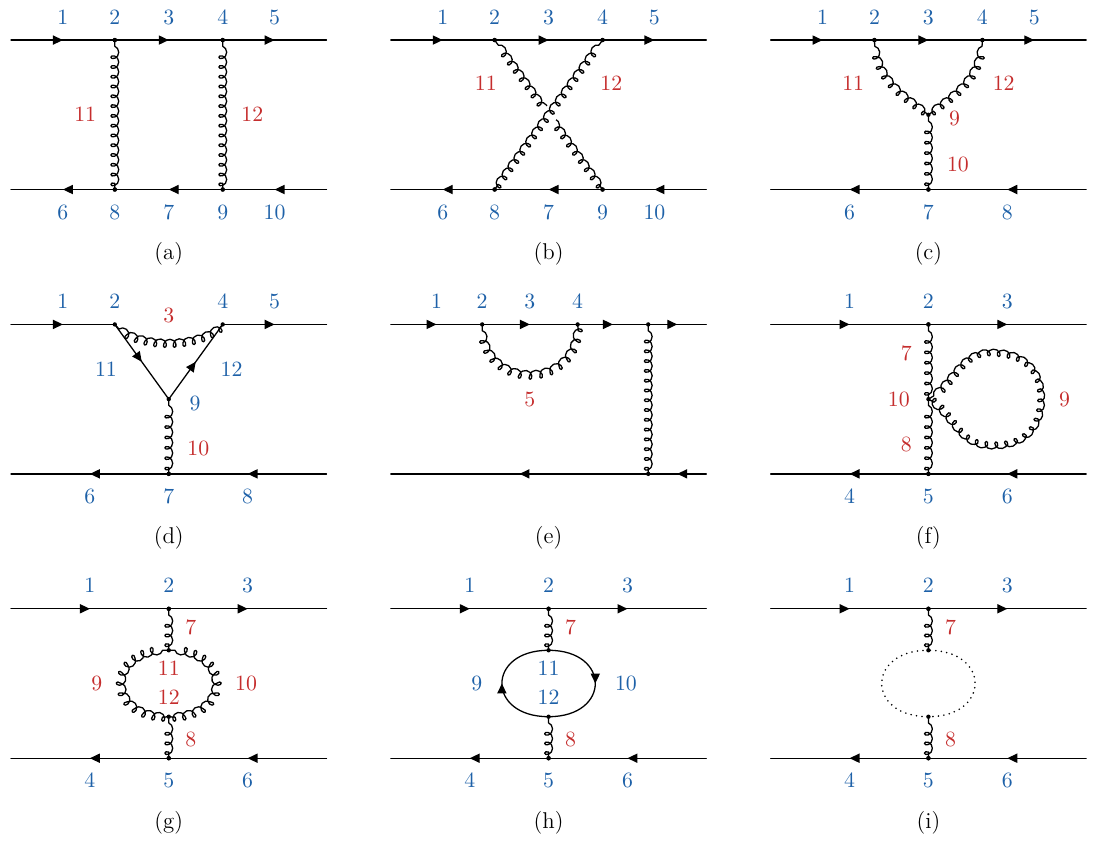}
   \caption{NLO diagrams for the EMT amplitude
$\mathcal{T}^{\mu\nu}_a(u,v)$ in Eq.~\eqref{eq:Tqqb}
for a collinear quark--antiquark state.
The numbered positions indicate insertion sites of the
EMT operator.
Blue and red numbers correspond to insertions of its quark
and gluon components, $T_q^{\mu\nu}$ and $T_g^{\mu\nu}$,
respectively. The left--right mirror image of diagram (e) and the diagrams
obtained from (c) and (d) by interchanging the upper and lower
fermion lines are implied but not shown.}
    \label{fig:NLOFeyDiag}
\end{figure}
\paragraph{One-loop amplitudes and collinear subtraction.}

Expanding the partonic factorization formula through NLO yields
\begin{align}
H^{\mu\nu,(1)}_a(u,v) &=\mathcal{T}^{\mu\nu,(1)}_a(u,v)-\int_0^1 d y H^{\mu\nu,(0)}_a(u,y)\Phi^{(1)}(y|v)\notag\\ &\quad-\int_0^1 d x\Phi^{(1)}(x|u)H^{\mu\nu,(0)}_a(x,v)~.
    \label{eq:nlomatching}
\end{align}
The NLO contribution to the renormalized partonic LCDA contains the collinear divergence
\begin{equation}
\Phi^{(1)}(x|u)
=-\left(\frac{1}{\epsilon_{\rm IR}}+\ln \frac{\mu_R^2}{\mu_F^2}\right)V^{(0)}(x,u)~,
\end{equation}
where $V^{(0)}$ is the LO ERBL kernel in Eq.~\eqref{eq:VLO}. These contributions cancel the collinear poles of the partonic
EMT amplitude in Eq.~\eqref{eq:nlomatching}, leaving kernels
that are free of collinear divergences.

 The evaluation of the NLO partonic EMT amplitude $\mathcal{T}^{\mu\nu(1)}_a(u,v)$ involves a total of 157 Feynman diagrams. Representative diagram topologies are shown in Fig.~\ref{fig:NLOFeyDiag}, where the numbered positions indicate all possible insertions of the EMT operator. In total, 120 diagrams contribute to the quark EMT sector, while 37 diagrams contribute to the gluon EMT sector. Here, we employ HepLib~\cite{Feng:2021kha} and FeynArts~\cite{Hahn:2000kx} to generate Feynman diagrams and
the corresponding amplitudes for the reaction $q\bar q+\mathcal{G}^*\to q\bar q$.%

After summing the contributions from the one-loop diagrams,
the NLO partonic amplitude $\mathcal{T}^{\mu\nu,(1)}_a$ can be decomposed into the three Lorentz structures:
\begin{align}
\mathcal{T}^{\mu\nu,(1)}_a
={}&2\bar{P}^\mu\bar{P}^\nu
\left[\left(\frac{1}{\epsilon_{\rm UV}}+L_R\right)A_{{\rm UV},a}
+\left(\frac{1}{\epsilon_{\rm IR}}+L_R\right)A_{{\rm IR},a}+f_{A,a}\right]
\notag\\
&+\frac{1}{2}(\Delta^\mu\Delta^\nu+Q^2g^{\mu\nu})
\left[\left(\frac{1}{\epsilon_{\rm UV}}+L_R\right)C_{{\rm UV},a}
+\left(\frac{1}{\epsilon_{\rm IR}}+L_R\right)C_{{\rm IR},a}+f_{C,a}\right]
\notag\\
&+2\Lambda^2g^{\mu\nu}
\left[\left(\frac{1}{\epsilon_{\rm UV}}+L_R\right)\overline C_{{\rm UV},a}
+\left(\frac{1}{\epsilon_{\rm IR}}+L_R\right)\overline C_{{\rm IR},a}+f_{\overline C,a}\right]
\label{eq:Tpole}
\end{align}
 Here, the momentum-fraction arguments are suppressed, $L_R=\ln(\mu_R^2/Q^2)$, and $1/\epsilon_{\rm UV}$ and $1/\epsilon_{\rm IR}$ denote the UV and IR poles, respectively. The functions $f_{A,a}$, $f_{C,a}$, and $f_{\overline C,a}$ denote the finite parts.%

For the gluon EMT, the UV poles and the IR poles are
\begin{subequations}\label{eq:gluonpole}
\begin{align}
    A_{\text{UV},g}(u,v)=&\frac{2C_F}{3}\frac{3-2u\bar{u}-2v\bar{v}}{2u\bar{u}v\bar{v}}-\frac{n_f}{6}\left(\frac{1}{u\bar{u}}+\frac{1}{v\bar{v}}\right)~,\label{eq:gluonpole:a}\\
    C_{\text{UV},g}(u,v)=&-\frac{2C_F}{3}\frac{1+2u\bar{u}+2v\bar{v}}{2u\bar{u}v\bar{v}}-\frac{n_f}{6}\left(\frac{1}{u\bar{u}}+\frac{1}{v\bar{v}}\right)~,\label{eq:gluonpole:b}\\
     \overline{C}_{\text{UV},g}(u,v)=&\frac{Q^2}{4\Lambda^2}\left(\frac{n_f}{6}+\frac{2C_F}{3}\right)\left(\frac{1}{u\bar{u}}+\frac{1}{v\bar{v}}\right)~,\label{eq:gluonpole:c}\\
     A_{\text{IR},g}(u,v)=&-\frac{C_F}{4}\left[3\left(\frac{1}{u\bar{u}}+\frac{1}{v\bar{v}}\right)+2\left(\frac{\log u}{u}+\frac{\log v}{v}+\frac{\log \bar{u}}{\bar{u}}+\frac{\log \bar{v}}{\bar{v}}\right)\right]~,\label{eq:gluonpole:d}\\
     C_{\text{IR},g}(u,v)=& A_{\text{IR},g}(u,v)~,\label{eq:gluonpole:e}\\
     \overline{C}_{\text{IR},g}(u,v)=&-\frac{Q^2}{4\Lambda^2}A_{\text{IR},g}(u,v)~.\label{eq:gluonpole:f}
\end{align}
\end{subequations}
For the quark EMT, they are given by
\begingroup
\footnotesize
\medmuskip=3mu\relax
\begin{subequations}\label{eq:quarkpole}
\begin{align}
    A_{\text{UV},q}(u,v)=&-\frac{2C_F}{3}\frac{3-2u\bar{u}-2v\bar{v}}{2u\bar{u}v\bar{v}}+\frac{1}{6}n_f\left(\frac{1}{u\bar{u}}+\frac{1}{v\bar{v}}\right)~,\label{eq:quarkpole:a}\\
    C_{\text{UV},q}(u,v)=&\frac{2C_F}{3}\frac{1+2u\bar{u}+2v\bar{v}}{2u\bar{u}v\bar{v}}+\frac{1}{6}n_f\left(\frac{1}{u\bar{u}}+\frac{1}{v\bar{v}}\right)~,\label{eq:quarkpole:b}\\
     \overline{C}_{\text{UV},q}(u,v)=&-\frac{Q^2}{4\Lambda^2}\left(\frac{n_f}{6}+\frac{2C_F}{3}\right)\left(\frac{1}{u\bar{u}}+\frac{1}{v\bar{v}}\right)~,\label{eq:quarkpole:c}\\
     A_{\text{IR},q}(u,v)=&-\frac{C_F}{4}\left[3\left(\frac{3-u\bar{u}-v\bar{v}}{u\bar{u}v\bar{v}}\right)+\frac{3-2u\bar{u}}{u\bar{u}}\left(\frac{\log v}{v}+\frac{\log \bar{v}}{\bar{v}}\right)+\frac{3-2v\bar{v}}{v\bar{v}}\left(\frac{\log u}{u}+\frac{\log \bar{u}}{\bar{u}}\right)\right]~,\label{eq:quarkpole:d}\\
     C_{\text{IR},q}(u,v)=&\frac{C_F}{4}\left[3\left(\frac{1+u\bar{u}+v\bar{v}}{u\bar{u}v\bar{v}}\right)+\frac{1+2u\bar{u}}{u\bar{u}}\left(\frac{\log v}{v}+\frac{\log \bar{v}}{\bar{v}}\right)+\frac{1+2v\bar{v}}{v\bar{v}}\left(\frac{\log u}{u}+\frac{\log \bar{u}}{\bar{u}}\right)\right]~,\label{eq:quarkpole:e}\\
     \overline{C}_{\text{IR},q}(u,v)=&-\frac{Q^2}{4\Lambda^2}\frac{C_F}{4}\left[3\left(\frac{1}{u\bar{u}}+\frac{1}{v\bar{v}}\right)+2\left(\frac{\log v}{v}+\frac{\log \bar{v}}{\bar{v}}+\frac{\log u}{u}+\frac{\log \bar{u}}{\bar{u}}\right)\right]~.\label{eq:quarkpole:f}
\end{align}
\end{subequations}
\endgroup

The IR poles arise from collinear divergences and can be absorbed into
the pion LCDAs through the matching formula in
Eq.~\eqref{eq:nlomatching}. For each Lorentz projection, the
corresponding pole residue can be expressed as convolutions of
the LO hard-scattering coefficient with the LO ERBL kernel:
\begin{align}
X_{{\rm IR},a}(u,v)
=&-\int_0^1dy\,
\mathcal X_a^{(0)}(u,y)V^{(0)}(y,v)
-\int_0^1dx\,
V^{(0)}(x,u)\mathcal X_a^{(0)}(x,v)~,
\label{eq:IR_convolutions}
\end{align}
where $X=A,C,\overline C$, and $\mathcal X_a^{(0)}$ denotes
$\mathcal A_a^{(0)}$, $\mathcal C_a^{(0)}$, and
$\overline{\mathcal C}_a^{(0)}$, respectively.
All LO coefficients in these convolutions are evaluated at
$\epsilon=0$.
These convolutions reproduce the IR-pole residues in
Eqs.~\eqref{eq:gluonpole} and \eqref{eq:quarkpole},
confirming their cancellation against the partonic LCDA
contributions in Eq.~\eqref{eq:nlomatching}.
The resulting hard-scattering kernels are therefore free of
collinear divergences, providing a nontrivial consistency check
of collinear factorization at NLO.

\paragraph{EMT operator renormalization and NLO coefficients.}

The UV divergences of the NLO amplitude $\mathcal{T}^{\mu\nu,(1)}_a$ shown in Eq.~\eqref{eq:Tpole} arise from the composite EMT insertions. Their residues satisfy
\begin{equation}
X_{{\rm UV},q}+X_{{\rm UV},g}=0,\qquad X=A,C,\overline C~,
\label{eq:total_EMT_UV}
\end{equation}
as required by the conservation of the total EMT. The individual quark and gluon amplitudes are renormalized through the operator mixing in Eq.~\eqref{eq:reEMT}:
\begin{align}
\begin{pmatrix}
\mathcal T_{q,R}^{\mu\nu,(1)}\\
\mathcal T_{g,R}^{\mu\nu,(1)}
\end{pmatrix}
=&
\begin{pmatrix}
\mathcal T_q^{\mu\nu,(1)}\\
\mathcal T_g^{\mu\nu,(1)}
\end{pmatrix}
+\frac{1}{4\epsilon_{\rm UV}}
\begin{pmatrix}
\displaystyle\frac{8C_F}{3}
&\displaystyle-\frac{2n_f}{3}\\[6pt]
\displaystyle-\frac{8C_F}{3}
&\displaystyle\frac{2n_f}{3}
\end{pmatrix}
\begin{pmatrix}
\mathcal T_q^{\mu\nu,(0)}(\epsilon)\\
\mathcal T_g^{\mu\nu,(0)}(\epsilon)
\end{pmatrix}~.
\label{eq:partonic_UV_renormalization}
\end{align}
The counterterms act on the
$d$-dimensional LO amplitudes, whose
$\mathcal O(\epsilon)$ terms must be retained
to obtain the finite NLO coefficients.

Since Eq.~\eqref{eq:partonic_UV_renormalization}
acts on the complete Lorentz tensor, the same
renormalization applies to all three GFF structures. Indeed, the UV-pole coefficients in
Eqs.~\eqref{eq:gluonpole} and \eqref{eq:quarkpole}
satisfy
\begin{align}
\begin{pmatrix}
X_{{\rm UV},q}\\
X_{{\rm UV},g}
\end{pmatrix}
=
\frac14
\begin{pmatrix}
\displaystyle-\frac{8C_F}{3}
&\displaystyle\frac{2n_f}{3}\\[6pt]
\displaystyle\frac{8C_F}{3}
&\displaystyle-\frac{2n_f}{3}
\end{pmatrix}
\begin{pmatrix}
\mathcal X_q^{(0)}(0)\\
\mathcal X_g^{(0)}(0)
\end{pmatrix},
\qquad X=A,C,\overline C~,
\end{align}
where $\mathcal X_a^{(0)}$ denotes the corresponding
LO coefficient. This relation confirms that the
UV poles of all three Lorentz structures are canceled
by the counterterms in
Eq.~\eqref{eq:partonic_UV_renormalization}.
Together with the collinear subtraction discussed
above, this yields finite renormalized hard-scattering
kernels for the factorization formula in Eq.~\eqref{GFFconv}.

Using the Lorentz decomposition in Eq.~\eqref{eq:Hparame},
the finite NLO coefficients can be expressed in terms of
the one-loop finite remainders defined in
Eq.~\eqref{eq:Tpole}. With $L_R$ defined above and
$L_F=\ln(\mu_F^2/Q^2)$, we obtain
\begin{subequations}\label{eq:NLO_scalar_kernels}
\begin{align}
\mathcal A_{a,R}^{(1)}(u,v)
=&\,
(L_R+1)A_{{\rm UV},a}(u,v)
+(L_F+1)A_{{\rm IR},a}(u,v)
\notag\\
&+f_{A,a}(u,v)+s_a\mathcal E_a(u,v),\label{eq:NLO_scalar_kernels:a}
\\
\mathcal C_{a,R}^{(1)}(u,v)
=&\,
(L_R+1)C_{{\rm UV},a}(u,v)
+(L_F+1)C_{{\rm IR},a}(u,v)
\notag\\
&+f_{C,a}(u,v)-s_a\mathcal E_a(u,v),\label{eq:NLO_scalar_kernels:b}
\\
\overline{\mathcal C}_{a,R}^{(1)}(u,v)
=&\,
(L_R+1)\overline C_{{\rm UV},a}(u,v)
+(L_F+1)\overline C_{{\rm IR},a}(u,v)
\notag\\
&+f_{\overline C,a}(u,v),
\qquad s_q=1,\quad s_g=-1~.\label{eq:NLO_scalar_kernels:c}
\end{align}
\end{subequations}
The functions $\mathcal E_a$ arise from the additional
$\mathcal O(\epsilon)$ terms in the LO $A$ and $C$
coefficients, beyond their common $(1-\epsilon)$ factor
in Eq.~\eqref{eq:Born_epsilon}. These terms generate
finite contributions when multiplied by the poles in
the EMT-operator and collinear subtractions. Explicitly,
\begin{subequations}\label{eq:finite_Born_contributions}
\begin{align}
\mathcal E_q(u,v)
=&\,
\frac{1}{u\bar u\,v\bar v}
\left[
\frac{C_F}{4}
\left(
\bar u\ln u+u\ln\bar u
+\bar v\ln v+v\ln\bar v
\right)
+\frac{13C_F}{12}
-\frac{n_f}{12}
\right],\label{eq:finite_Born_contributions:a}
\\
\mathcal E_g(u,v)
=&\,
-\frac{1}{u\bar u\,v\bar v}
\left[
\frac{C_F}{4}
\left(
\bar u\ln u+u\ln\bar u
+\bar v\ln v+v\ln\bar v
\right)
+\frac{5C_F}{12}
+\frac{n_f}{12}
\right]~.\label{eq:finite_Born_contributions:b}
\end{align}
\end{subequations}
For the $\overline C$ projection, the LO coefficient
is proportional to $(1-\epsilon)$, so its finite
subtraction contributions are equal to the corresponding
UV- and IR-pole residues. These contributions are already
included in the constant terms accompanying $L_R$ and
$L_F$ in Eq.~\eqref{eq:NLO_scalar_kernels:c}, and no
additional term analogous to $\mathcal E_a$ is required.
The complete expressions for the NLO hard-scattering
coefficients are collected in
Appendix~\ref{app:explicit-nlo-kernels}.

Summing the quark and gluon contributions, we obtain
the total NLO coefficients
\begin{subequations}\label{eq:finite_Born_contributions2}
\begin{align}
\mathcal A^{(1)}(u,v)
=&\,
\mathcal A_{q,R}^{(1)}(u,v)
+\mathcal A_{g,R}^{(1)}(u,v)
\notag\\
=&\,
(L_F+1)A_{\rm IR}(u,v)
+f_A(u,v)
+\mathcal E_q(u,v)-\mathcal E_g(u,v)~,\label{eq:finite_Born_contributions2:a}
\\
\mathcal C^{(1)}(u,v)
=&\,
\mathcal C_{q,R}^{(1)}(u,v)
+\mathcal C_{g,R}^{(1)}(u,v)
\notag\\
=&\,
(L_F+1)C_{\rm IR}(u,v)
+f_C(u,v)
-\mathcal E_q(u,v)+\mathcal E_g(u,v)~,\label{eq:finite_Born_contributions2:b}
\\
\overline{\mathcal C}^{(1)}(u,v)
=&\,0~.\label{eq:finite_Born_contributions2:c}
\end{align}
\end{subequations}
Here, $X_{\rm IR}=X_{{\rm IR},q}+X_{{\rm IR},g}$
and $f_X=f_{X,q}+f_{X,g}$ for $X=A,C$.
The quark--gluon operator-mixing contributions
cancel in the total EMT, as required by EMT
conservation. The total EMT is renormalization-scale independent, while the individual quark and gluon GFFs retain their
operator-scale dependence.%

\subsection{Pion trace anomaly and the \texorpdfstring{$F^2$}{F2} scalar form factor at NLO}
The NLO results of the pion GFFs obtained in the previous section also allow us to investigate the pion trace anomaly form factor $\Theta^\pi(Q^2)$.
As discussed in Sec.~\ref{sec:renormalization}, the trace of the total EMT in the massless limit is related to the renormalized gluonic scalar operator $(F^2)_R$ through Eq.~\eqref{eq:F2fromGFFs}.
This relation provides two complementary approaches to calculating the trace anomaly. The first is to take the trace of the EMT matrix element directly using the pion GFFs, while the second is to calculate the renormalized $F^2$ scalar form factor and combine it with the QCD beta function~\cite{Tong:2022zax}.

We first consider the direct trace of the EMT matrix element. The classical EMT is traceless in the massless limit, while quantum corrections give rise to the trace anomaly.
Consequently, the trace starts one order higher in $\alpha_s$ than the pion
GFFs themselves. Although the pion GFFs have been calculated through NLO in
the previous section, their trace therefore determines the anomaly only at
its leading nonvanishing order. Obtaining the NLO correction to the trace
anomaly through this approach would require the pion GFFs to be calculated
through NNLO.

Substituting our NLO results for the total pion GFFs into Eq.~\eqref{eq:F2fromGFFs}, we obtain the leading nonvanishing contribution to the pion trace anomaly form factor:
\begin{align}
\Theta^\pi(Q^2)\Big\vert_{\rm LO}
&=
\frac{Q^2}{2}
\left[
 A(Q^2)
+3 C(Q^2)
\right]\Big\vert_{\rm NLO}
\notag\\
&=
-\frac{\beta_0C_F}{2}\,
\alpha_s^2(\mu_R)
\int_0^1 dx\int_0^1dy\,
\phi_\pi^*(y,\mu_F)
\frac{1}{x\bar{x}y\bar{y}}
\phi_\pi(x,\mu_F)
~.
\end{align}
Here, we have used the LO relation $A_{\rm LO}(Q^2)+3C_{\rm LO}(Q^2)=0$. This result agrees with Ref.~\cite{Tong:2022zax}, where the same LO contribution was obtained from a direct calculation of the $F^2$ scalar form factor. The agreement provides a nontrivial consistency check of our NLO GFF calculation.

To determine the NLO correction to the trace-anomaly form factor $\Theta^\pi(Q^2)$, we turn to
the second approach and calculate the renormalized $F^2$ scalar form factor through NLO.
 At leading power in the large-$Q^2$ expansion, the bare $F^2$ scalar form factor can be factorized as
\begin{equation}
\label{FFconv}
\left\langle P' \left| F^2 \right| P \right\rangle
=
\int dx\,dy\,
\phi_\pi^*(y,\mu_F)
H^{ F^2}
\left(
x,y
\right)
\phi_\pi(x,\mu_F)~,
\end{equation}
where the scale dependence of the hard-scattering kernel is suppressed for brevity. Its perturbative expansion in the renormalized coupling reads
\begin{equation}
H^{F^2}(x,y)
=
4\pi\alpha_s(\mu_R) C_F
\left[
H^{F^2,(0)}(x,y)
+
\frac{\alpha_s(\mu_R)}{\pi}H^{F^2,(1)}(x,y)
+
\mathcal{O}(\alpha_s^2)
\right]~.
\end{equation}
  The LO kernel was obtained in Ref.~\cite{Tong:2022zax}. Here, we calculate its NLO correction and perform the corresponding UV renormalization. Since the QCD beta-function factor starts at $\mathcal O(\alpha_s)$, the NLO $F^2$ scalar form factor, together with the two-loop beta function, allows us to determine the trace anomaly through $\mathcal O(\alpha_s^3)$.

To extract the bare $F^2$ hard-scattering kernel, we use the relation between the bare gluon EMT and the scalar operator $F^2$ given in Eq.~\eqref{eq:bareTrace}. Taking its matrix element in $d=4-2\epsilon$ dimensions, we obtain:
\begin{align}
\label{eq:FF}
\left\langle P' \left| F^2 \right| P \right\rangle
&=
-\frac{2}{\epsilon}
g_{\mu\nu}\left\langle P' \left| T_{g}^{\mu\nu}\right| P \right\rangle
\notag \\
&=
-\frac{Q^2}{\epsilon}\left[A_g\left(Q^2\right)+(3-2 \epsilon) C_g\left(Q^2\right)+\frac{4(4-2 \epsilon) \Lambda^2}{Q^2} \bar{C}_g\left(Q^2\right)\right]~.
\end{align}
The explicit factor $1/\epsilon$ implies that the gluon EMT matrix element must be retained through $\mathcal O(\epsilon)$ to determine the finite contribution to the $F^2$ scalar form factor.

Applying this relation to the hard-scattering kernels, and factoring out the overall perturbative prefactors, we obtain
\begin{align}
H^{{F^2},(k)}(x,y,\epsilon)
&=
-\frac{2}{Q^2\epsilon}
g_{\mu\nu}H_g^{\mu\nu,(k)}(x,y,\epsilon)
\notag\\
&=
-\frac{1}{\epsilon}
\bigg[
{\cal A}_g^{(k)}(x,y,\epsilon)
+(3-2\epsilon){\cal C}_g^{(k)}(x,y,\epsilon)
+\frac{4(4-2\epsilon)\Lambda^2}{Q^2}
\overline{\cal C}_g^{(k)}(x,y,\epsilon)
\bigg]~.
\end{align}
Here, $H_g^{\mu\nu,(k)}$ denotes the perturbative coefficient of the bare gluon EMT hard-scattering kernel, following the same normalization in Eq.~\eqref{eq:Hexpan}, while ${\cal A}_g^{(k)}$, ${\cal C}_g^{(k)}$, and $\overline{\cal C}_g^{(k)}$ denote the corresponding bare scalar coefficients in the Lorentz decomposition following Eq.~\eqref{eq:Hparame}.

At LO, using the $d$-dimensional gluon EMT kernels in Eq.~\eqref{eq:H(0)}, we obtain
\begin{align}
H_R^{{ F^2},(0)}(x,y,\epsilon)
=
H^{{ F^2},(0)}(x,y,\epsilon)
=
\frac{1-\epsilon}{x\bar{x}y\bar{y}}~.
\label{eq:HFF(0)}
\end{align}
Its four-dimensional limit reproduces the LO result of Ref.~\cite{Tong:2022zax}. The LO $F^2$ scalar kernel also provides a consistency
check of the renormalized GFFs obtained in Sec.~\ref{sec:hardcoefficients}.
Using the trace relations in Eq.~\eqref{eq:Cbarrelation},
the NLO hard-scattering coefficients must satisfy
\begin{subequations}\label{eq:Cbar_NLO_check}
\begin{align}
\overline{\mathcal C}_{g,R}^{(1)}(x,y)
=&-\frac{Q^2}{16\Lambda^2}
\left[
\mathcal A_{g,R}^{(1)}(x,y)
+3\mathcal C_{g,R}^{(1)}(x,y)
\right]
-\frac{11C_A}{192}\frac{Q^2}{\Lambda^2}
H_R^{F^2,(0)}(x,y),\label{eq:Cbar_NLO_check:a}
\\
\overline{\mathcal C}_{q,R}^{(1)}(x,y)
=&-\frac{Q^2}{16\Lambda^2}
\left[
\mathcal A_{q,R}^{(1)}(x,y)
+3\mathcal C_{q,R}^{(1)}(x,y)
\right]
+\frac{n_f}{96}\frac{Q^2}{\Lambda^2}
H_R^{F^2,(0)}(x,y)~.\label{eq:Cbar_NLO_check:b}
\end{align}
\end{subequations}
Here, all hard-scattering coefficients are evaluated
in four dimensions, with their scale arguments
suppressed for brevity. These relations provide
a nontrivial consistency check of the NLO
$\overline C_{q,g}$ kernels obtained through
the direct renormalization of the EMT.
Their sum is also consistent with
$\overline{\mathcal C}_{q,R}^{(1)}
+\overline{\mathcal C}_{g,R}^{(1)}=0$
and the total trace-anomaly result derived above.

At NLO, the $d$-dimensional trace of the one-loop bare gluon EMT kernel gives
\begin{align}
H^{{F^2},(1)}(x,y)
&=
-\frac{1}{\epsilon}
\left[
{\cal A}^{(1)}_{g}(x,y)
+
(3-2\epsilon){\cal C}^{(1)}_{g}(x,y)
+
\frac{4\Lambda^2}{Q^2}(4-2\epsilon)
\overline{\cal C}^{(1)}_{g}(x,y)
\right]
\notag\\
&=
\frac{\beta_0}{4\epsilon_{\rm UV}}
H^{{ F^2},(0)}(x,y)\big|_{\epsilon=0}
+
f^{ F^2}
\left(
x,y%
\right)~,
\label{eq:FFNLO}
\end{align}
Here, the function $f^{ F^2}$ denotes the finite remainder. Since the collinear divergences have already been subtracted in obtaining ${\cal A}_g^{(1)}$, ${\cal C}_g^{(1)}$, and $\overline{\cal C}_g^{(1)}$,  the derived $F^2$ hard-scattering kernels are infrared finite.

The remaining UV divergence in Eq.~\eqref{eq:FFNLO}
is removed by the $F^2$-operator renormalization
in Eq.~\eqref{eq:FFR}, as discussed in
Sec.~\ref{sec:renormalization}.
The resulting renormalized NLO hard-scattering
kernel is finite and reads
\begin{align}
H_R^{F^2,(1)}(x,y)
=f^{F^2}(x,y)+\frac{\beta_0}{4}H^{F^2,(0)}(x,y)\big|_{\epsilon=0}~.
\end{align}
The finite term proportional to $\beta_0$ arises
from the $\mathcal O(\epsilon)$ part of the LO kernel
in Eq.~\eqref{eq:HFF(0)} multiplied by the UV pole
of the operator counterterm.
The full analytic expression for $H_R^{F^2,(1)}$
is given in Appendix~\ref{app:explicit-nlo-kernels}.

Having obtained the $F^2$ scalar form factor through NLO, we can determine
the trace anomaly form factor $\Theta^\pi(Q^2)$ one perturbative order beyond that accessible from the
direct trace of the NLO GFFs. The beta-function factor has the expansion
\begin{equation}
\frac{\beta(g_R)}{2g_R}
=
-\frac{\beta_0}{8\pi}\alpha_s(\mu_R)
-\frac{\beta_1}{32\pi^2}\alpha_s^2(\mu_R)
+\mathcal{O}(\alpha_s^3)~,
\end{equation}
where
$\beta_1=\frac{34}{3}C_A^2
-\left(\frac{10}{3}C_A+2C_F\right)n_f$.
Since the $F^2$ form factor starts at $\mathcal{O}(\alpha_s)$,
the trace anomaly starts at $\mathcal{O}(\alpha_s^2)$, and its NLO correction
is of $\mathcal{O}(\alpha_s^3)$. The latter receives two contributions: the NLO $F^2$ hard-scattering kernel multiplied by $\beta_0$, and the LO kernel multiplied by the two-loop coefficient $\beta_1$.

For the canonical
scale choice $\mu_R=\mu_F=Q$, the pion trace anomaly at NLO is therefore
\begin{align}
\Theta^\pi(Q^2)
=&
\int_0^1 dx\int_0^1dy\,
\phi_\pi^*(y,Q)\phi_\pi(x,Q)
\Bigg\{
-\frac{\beta_0 C_F}{2}
\alpha_s^2(Q)
H^{\rm F^2,(0)}_R(x,y)
\notag\\
&\hspace{1.5cm}
-\frac{C_F}{\pi}\alpha_s^3(Q)
\left[
\frac{\beta_0}{2}
H^{F^2,(1)}_R(x,y)
+
\frac{\beta_1}{8}
H^{F^2,(0)}_R(x,y)
\right]
\Bigg\}
+
\mathcal{O}(\alpha_s^4)~,
\end{align}
where the hard-scattering kernels are understood in their four-dimensional limit. Thus, taking the trace of the NLO pion GFFs determines the leading nonvanishing contribution to the trace anomaly, whereas the NLO calculation of the $F^2$ scalar form factor extends the prediction through $\mathcal O(\alpha_s^3)$. Obtaining the same accuracy directly from the pion EMT matrix element would require the GFFs through NNLO.

\subsection{Master formulas of the pion GFFs at NLO}
\label{eq:masterformula}

With the factorization formulas established to NLO, numerical predictions for the pion GFFs and the pion trace anomaly can in principle be obtained for any given parametrization of the pion LCDA. In practice, however, a direct numerical evaluation of the convolution integrals can become unstable because of spurious singularities appearing in the NLO hard-scattering kernels. Although these singularities cancel in the complete convolution and do not correspond to physical divergences, they complicate a straightforward numerical integration. To circumvent this difficulty, we perform the convolution integrals analytically by expanding the pion LCDA in Gegenbauer polynomials.

The leading-twist pion LCDA can be expanded in eigenfunctions of the LO evolution kernel as
\begin{align}
\phi_\pi(x,\mu_F)
&=
\frac{f_\pi}{\sqrt{2N_c}}
\sum_{n=0}^{\infty}
a_{2n}(\mu_F)\psi_{2n}(x),
\label{eq_pionDAGegen}
\\
\psi_n(x)
&=
6x\bar{x}\,C_n^{3/2}(2x-1),
\end{align}
where $C_n^{3/2}(x)$ denotes the Gegenbauer polynomial. $a_0=1$, and the nonperturbative information is encoded in the higher Gegenbauer moments $a_{2n}(\mu_F)$. For $N_c=3$, this expansion satisfies $\int_0^1 dx\,\phi_\pi(x,\mu_F)=f_\pi/\sqrt{6}$ and is therefore consistent with the normalization convention introduced above.

Substituting Eq.~\eqref{eq_pionDAGegen} into the factorization formula \eqref{GFFconv}, the pion EMT matrix element can be written as
\begin{align}
\langle P' | T_{a,R}^{\mu\nu} | P \rangle
&=
\frac{4\pi\alpha_s(\mu_R)C_Ff_\pi^2}{2N_cQ^2}
\sum_{k=0}^{\infty}
\left(
\frac{\alpha_s(\mu_R)}{\pi}
\right)^k
\sum_{m,n}
a_{2m}(\mu_F)a_{2n}(\mu_F)
T_{a,R,mn}^{\mu\nu,(k)},
\end{align}
where the Gegenbauer-projected hard-scattering kernels are defined by
\begin{align}
T_{a,R,mn}^{\mu\nu,(k)}
&\equiv
\psi_{2m}(x)
\otimes
H_{a,R}^{\mu\nu,(k)}(x,y)
\otimes
\psi_{2n}(y).
\end{align}

In the following, we set $\mu_F=\mu_R=\mu$.
The pion GFFs can then be expressed as
\begin{subequations}\label{eq:GFF_moment_expansions}
\begin{align}
A^R_a(Q^2,\mu)
&=
\frac{4\pi\alpha_s(\mu)C_Ff_\pi^2}{2N_cQ^2}
\sum_{k=0}^{\infty}
\left(
\frac{\alpha_s(\mu)}{\pi}
\right)^k
\sum_{m,n}
a_{2m}(\mu)a_{2n}(\mu)
\mathcal{H}^{A_a,(k)}_{mn}(Q,\mu),\label{eq:GFF_moment_expansions:a}
\\
C^R_a(Q^2,\mu)
&=
\frac{4\pi\alpha_s(\mu)C_Ff_\pi^2}{2N_cQ^2}
\sum_{k=0}^{\infty}
\left(
\frac{\alpha_s(\mu)}{\pi}
\right)^k
\sum_{m,n}
a_{2m}(\mu)a_{2n}(\mu)
\mathcal{H}^{C_a,(k)}_{mn}(Q,\mu),\label{eq:GFF_moment_expansions:b}
\\
\overline{C}^R_a(Q^2,\mu)
&=
\frac{4\pi\alpha_s(\mu)C_Ff_\pi^2}{2N_cQ^2}
\sum_{k=0}^{\infty}
\left(
\frac{\alpha_s(\mu)}{\pi}
\right)^k
\sum_{m,n}
a_{2m}(\mu)a_{2n}(\mu)
\mathcal{H}^{\overline{C}_a,(k)}_{mn}(Q,\mu),\label{eq:GFF_moment_expansions:c}
\end{align}
\end{subequations}
where the corresponding projected hard-scattering kernels are defined by
\begin{subequations}\label{eq:GFF_projected_kernels}
\begin{align}
\mathcal{H}^{A_a,(k)}_{mn}
&\equiv
\psi_{2m}(x)
\otimes
{\cal A}^{(k)}_{a,R}(x,y)
\otimes
\psi_{2n}(y),\label{eq:GFF_projected_kernels:a}
\\
\mathcal{H}^{C_a,(k)}_{mn}
&\equiv
\psi_{2m}(x)
\otimes
{\cal C}^{(k)}_{a,R}(x,y)
\otimes
\psi_{2n}(y),\label{eq:GFF_projected_kernels:b}
\\
\mathcal{H}^{\overline{C}_a,(k)}_{mn}
&\equiv
\psi_{2m}(x)
\otimes
\overline{\cal C}^{(k)}_{a,R}(x,y)
\otimes
\psi_{2n}(y).\label{eq:GFF_projected_kernels:c}
\end{align}
\end{subequations}
Similarly, the $F^2$ scalar form factor can be written as
\begin{align}
\left\langle P' \left| (F^2)_R \right| P \right\rangle
&=
\frac{4\pi\alpha_s(\mu)C_Ff_\pi^2}{2N_c}
\sum_{k=0}^{\infty}
\left(
\frac{\alpha_s(\mu)}{\pi}
\right)^k
\sum_{m,n}
a_{2m}(\mu)a_{2n}(\mu)
\mathcal{H}^{F^2,(k)}_{mn}(Q,\mu),
\end{align}
with the projected hard-scattering kernel defined as
\begin{align}
\mathcal{H}^{F^2,(k)}_{mn}
&\equiv
\psi_{2m}(x)
\otimes
H_R^{F^2(k)}(x,y)
\otimes
\psi_{2n}(y).
\end{align}

The projected hard-scattering kernels for $\overline C_a$ are not independent. Using the relations between the renormalized $\overline C_a$ GFFs and the $F^2$ form factor given in Eq.~\eqref{eq:Cbarrelation}, one obtains at LO
\begin{subequations}\label{eq:Cbar_moments_LO}
\begin{align}
\mathcal{H}^{\overline{C}_g,(0)}_{mn}
&=
-\frac{Q^2}{16\Lambda^2}
\left[
\mathcal{H}^{A_g,(0)}_{mn}
+
3\mathcal{H}^{C_g,(0)}_{mn}
\right],\label{eq:Cbar_moments_LO:a}
\\
\mathcal{H}^{\overline{C}_q,(0)}_{mn}
&=
-\frac{Q^2}{16\Lambda^2}
\left[
\mathcal{H}^{A_q,(0)}_{mn}
+
3\mathcal{H}^{C_q,(0)}_{mn}
\right]~,\label{eq:Cbar_moments_LO:b}
\end{align}
\end{subequations}
which satisfy
$\mathcal{H}^{\overline{C}_g,(0)}_{mn}
+\mathcal{H}^{\overline{C}_q,(0)}_{mn}=0$.
At NLO, the LO $F^2$ hard-scattering kernel also contributes, giving
\begin{subequations}\label{eq:Cbar_moments_NLO}
\begin{align}
\mathcal{H}^{\overline{C}_g,(1)}_{mn}
&=
-\frac{Q^2}{16\Lambda^2}
\left[
\mathcal{H}^{A_g,(1)}_{mn}
+
3\mathcal{H}^{C_g,(1)}_{mn}
\right]
-
\frac{11C_A}{192}\frac{Q^2}{\Lambda^2}
\mathcal{H}^{F^2,(0)}_{mn},\label{eq:Cbar_moments_NLO:a}
\\
\mathcal{H}^{\overline{C}_q,(1)}_{mn}
&=
-\frac{Q^2}{16\Lambda^2}
\left[
\mathcal{H}^{A_q,(1)}_{mn}
+
3\mathcal{H}^{C_q,(1)}_{mn}
\right]
+
\frac{n_f}{96}\frac{Q^2}{\Lambda^2}
\mathcal{H}^{F^2,(0)}_{mn}~,\label{eq:Cbar_moments_NLO:b}
\end{align}
\end{subequations}
which satisfy
$\mathcal{H}^{\overline{C}_g,(1)}_{mn}
+\mathcal{H}^{\overline{C}_q,(1)}_{mn}=0$.

The convolutions of the Gegenbauer polynomials with the LO and NLO hard-scattering kernels can be evaluated analytically using \textsc{PolyLogTools}~\cite{Duhr:2019tlz}. At LO, the projected kernels take the particularly simple form
\begin{align}
\mathcal{H}^{A_g,(0)}_{00}
&=12,
&
\mathcal{H}^{A_g,(0)}_{i0}
=
\mathcal{H}^{A_g,(0)}_{0i}
&=6
\qquad (i\geq1),
&
\mathcal{H}^{A_g,(0)}_{ij}
&=0
\qquad (i,j\geq1),
\notag\\
\mathcal{H}^{A_q,(0)}_{00}
&=42,
&
\mathcal{H}^{A_q,(0)}_{i0}
=
\mathcal{H}^{A_q,(0)}_{0i}
&=48
\qquad (i\geq1),
&
\mathcal{H}^{A_q,(0)}_{ij}
&=54
\qquad (i,j\geq1),
\notag\\
\mathcal{H}^{C_q,(0)}_{00}
&=-30,
&
\mathcal{H}^{C_q,(0)}_{i0}
=
\mathcal{H}^{C_q,(0)}_{0i}
&=-24
\qquad (i\geq1),
&
\mathcal{H}^{C_q,(0)}_{ij}
&=-18
\qquad (i,j\geq1),
\notag\\
\mathcal{H}^{C_g,(0)}_{ij}
&=
\mathcal{H}^{A_g,(0)}_{ij},
&
\mathcal{H}^{F^2,(0)}_{ij}
&=36.
\label{eq:Gegen_LO}
\end{align}
The corresponding NLO projected kernels are considerably more involved and are collected in Appendix~\ref{sec:GegenbauerHK}. Together with the Gegenbauer moments $a_{2n}(\mu)$, these analytical results provide the master formulas used in our numerical evaluation of the pion GFFs and the pion trace anomaly.

\section{Numerical predictions}

In this section, we present numerical predictions for the
pion GFFs and the trace anomaly using the master formulas
derived in Sec.~\ref{eq:masterformula}. The principal
nonperturbative inputs are the Gegenbauer moments
$a_{2n}(\mu)$ of the pion LCDA. Its shape has been
investigated using various approaches~\cite{Chernyak:1983ej,Mikhailov:2016klg,Bakulev:2001pa,
Ball:2006wn,Stefanis:2020rnd,Cheng:2020vwr,
Stefanis:2014nla,Chang:2013pq,Raya:2015gva,
Lan:2019rba,Choi:2014ifm,Chang:2016ouf,
RQCD:2019osh,LatticeParton:2022zqc}.
In particular, the RQCD Collaboration determined
the second Gegenbauer moment~\cite{RQCD:2019osh},
\begin{align}
a_2(2\,{\rm GeV})
=0.116^{+0.019}_{-0.020}~.
\end{align}
The Lattice Parton Collaboration (LPC) has also
investigated the $x$-dependent pion LCDA using
large-momentum effective theory
(LaMET)~\cite{LatticeParton:2022zqc}.
For our numerical analysis, we use the updated LPC
Gegenbauer moments~\cite{LPCnew}:
\begin{align}
a_2(2\,{\rm GeV})&=0.178\pm0.048~,
&
a_4(2\,{\rm GeV})&=0.073\pm0.026~,
\notag\\
a_6(2\,{\rm GeV})&=0.045\pm0.027~.
\end{align}
The two determinations of $a_2$ are compatible
within their quoted uncertainties, while the
additional LPC moments allow us to examine the
sensitivity to the shape of the LCDA.

We consider both inputs using their central values.
At $\mu_0=2\,{\rm GeV}$, moments beyond $a_2$ are
set to zero for the RQCD input, while moments
beyond $a_6$ are set to zero for the LPC input.
The pion LCDA is evolved to the factorization
scale $\mu_F$ using NLO ERBL evolution~\cite{Meli__1999}, and the running coupling
$\alpha_s(\mu_R)$ is evaluated at two-loop
accuracy using the \texttt{FAPT} package~\cite{Bakulev_2013}.
We take $\mu_R=\mu_F=\mu=Q$ as the central scale
choice and estimate the residual perturbative
uncertainty by varying the common scale within
$Q/\sqrt{2}\leq\mu\leq\sqrt{2}Q$.
The resulting bands reflect only this correlated
scale variation and do not include uncertainties
from the Gegenbauer moments, LCDA truncation,
or power corrections.

\begin{figure}[t]
    \centering
    \includegraphics[width=0.49\textwidth]
    {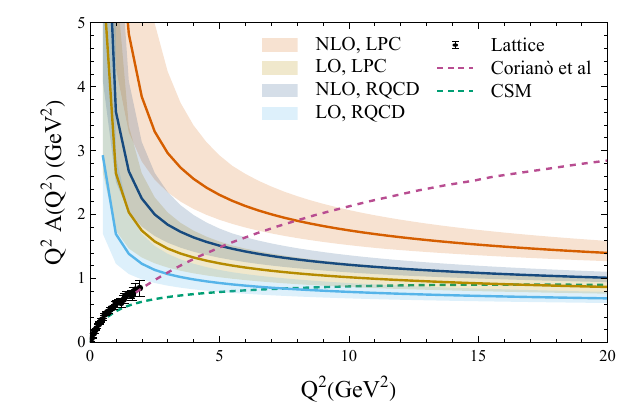}
    \hspace{-0.01\textwidth}
    \includegraphics[width=0.49\textwidth]
    {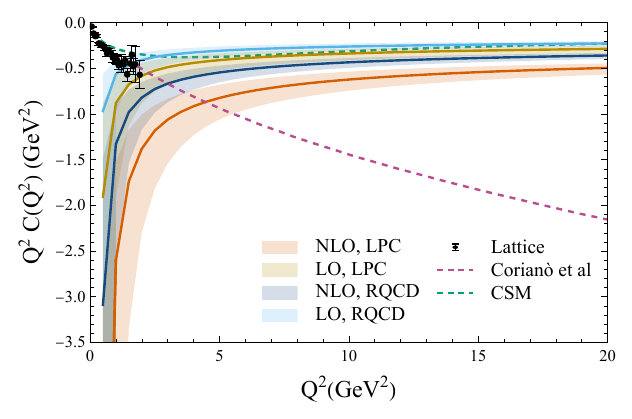}
    \caption{
    The total pion GFFs $A(Q^2)$ (left) and
    $C(Q^2)$ (right), multiplied by $Q^2$.
    The bands correspond to the LO and NLO
    predictions with $\mu_R=\mu_F=\mu$ varied
    within $Q/\sqrt{2}\leq\mu\leq\sqrt{2}Q$,
    around the central choice $\mu=Q$.
    Results from lattice QCD
    ~\cite{Hackett:2023nkr}, the continuum
    Schwinger-function method (CSM)
    ~\cite{Xu:2023izo}, and $k_T$ factorization
    ~\cite{Coriano:2026gyl}
    are shown for comparison.
    }
    \label{fig:1}
\end{figure}

\begin{figure}[t]
    \centering
    \includegraphics[width=1\linewidth]
    {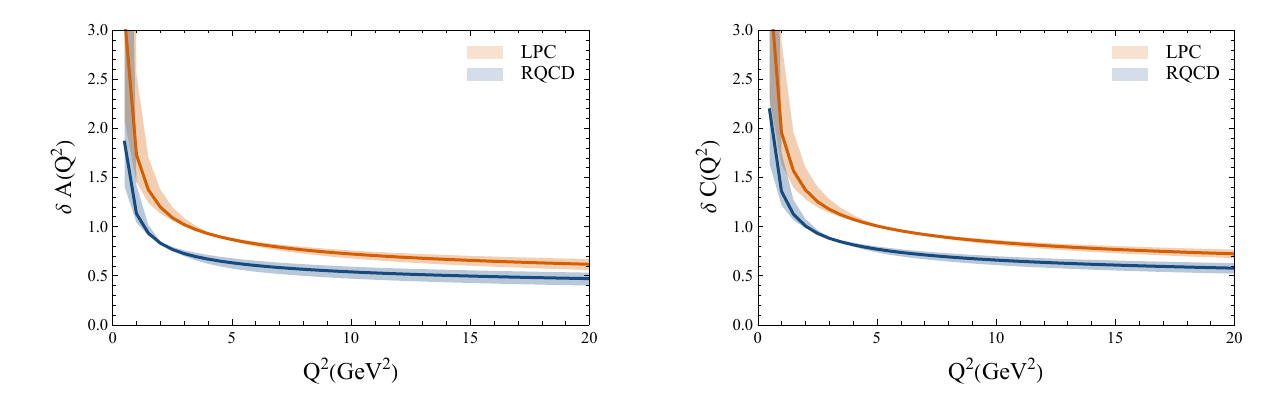}
    \caption{
    Relative NLO corrections to the total pion
    GFFs $A(Q^2)$ and $C(Q^2)$ for the LPC
    and RQCD inputs.
    }
    \label{fig:correction}
\end{figure}

Figure~\ref{fig:1} shows the total pion GFFs
$A(Q^2)$ and $C(Q^2)$,
each multiplied by $Q^2$
to remove the leading $1/Q^2$ suppression and
highlight the residual logarithmic dependence.
The NLO corrections preserve the LO signs,
$A>0$ and $C<0$, while enhancing the magnitudes
of both form factors. The LPC input
generally gives larger values of $|A|$ and
$|C|$ than the RQCD input, illustrating
their sensitivity to the Gegenbauer moments.
The lattice QCD and CSM predictions provide complementary information at low $Q^2$, where our leading-power perturbative calculation is not reliable. The comparison is therefore qualitative rather than quantitative. The $k_T$-factorization results of Corian\`o \textit{et al.}~\cite{Coriano:2026gyl} exhibit a different $Q^2$ dependence, although a direct quantitative comparison is complicated by the differences in theoretical frameworks and nonperturbative inputs.

To assess perturbative convergence, we define
the relative NLO correction for a generic
form factor $X$ as
\begin{align}
\delta X(Q^2)
\equiv
\frac{\Delta X_{\rm NLO}(Q^2)}
{X_{\rm LO}(Q^2)},
\qquad
\Delta X_{\rm NLO}
\equiv X_{\rm NLO}-X_{\rm LO}~.
\end{align}
As shown in figure~\ref{fig:correction},
$\delta A$ and $\delta C$ are positive and
decrease with increasing $Q^2$.
Nevertheless, the corrections remain comparable
to the LO contributions even for
$Q^2\gtrsim10\,{\rm GeV}^2$, indicating
rather slow perturbative convergence.
Although the scale-variation bands narrow
toward larger $Q^2$, this reduction alone
does not imply that uncalculated higher-order
contributions are small.

\begin{figure}[t]
    \centering
    \includegraphics[width=1\linewidth]
    {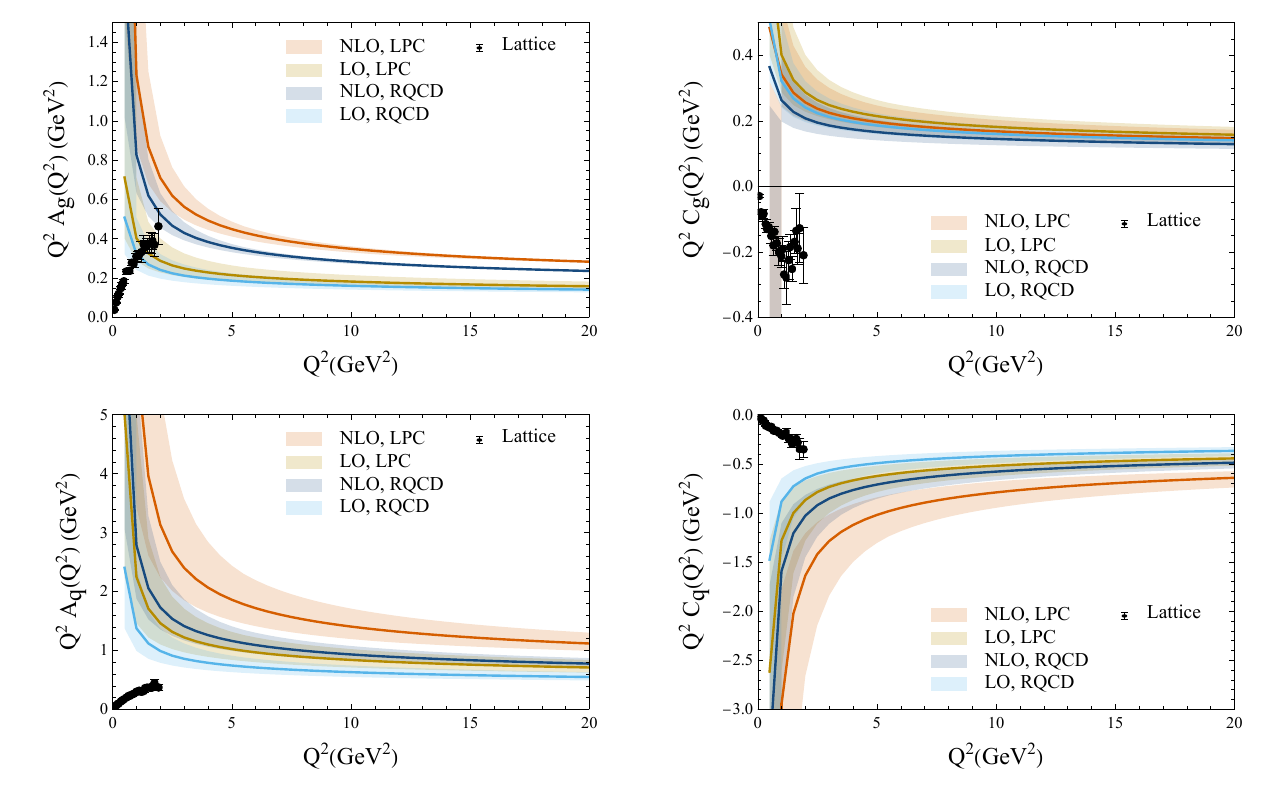}
    \caption{
    The renormalized quark and gluon GFFs
    $A_g^R$, $C_g^R$, $A_q^R$, and $C_q^R$
    at LO and NLO, multiplied by $Q^2$.
    The central scale is $\mu_R=\mu_F=Q$,
    with bands obtained from correlated
    scale variation.
    Lattice QCD results from
    Ref.~\cite{Hackett:2023nkr}, renormalized
    in the $\overline{\rm MS}$ scheme at
    $\mu=2\,{\rm GeV}$, are shown for comparison.
    }
    \label{fig:2}
\end{figure}

\begin{figure}[t]
    \centering
    \includegraphics[width=1\linewidth]
    {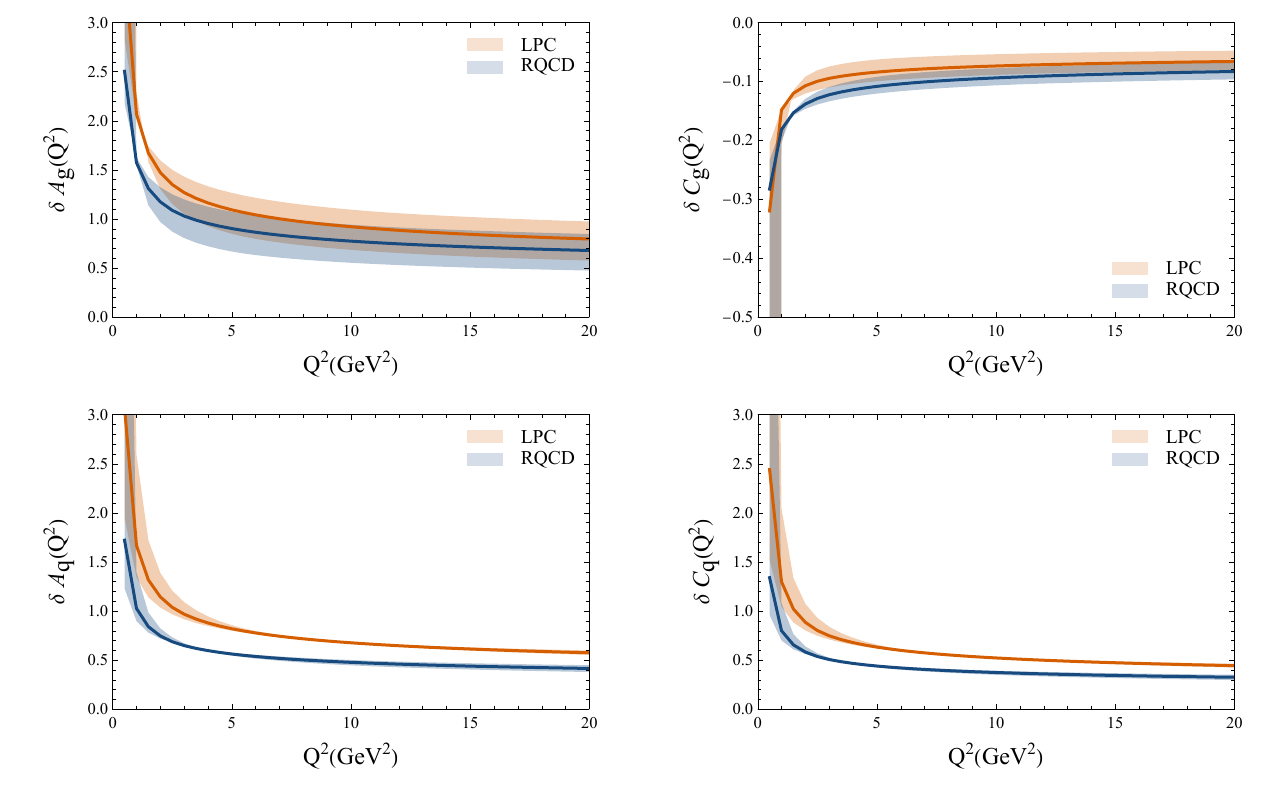}
    \caption{
    Relative NLO corrections to the individual
    quark and gluon GFFs,
    $\delta {A_g}$, $\delta {C_g}$,
    $\delta {A_q}$, and $\delta {C_q}$.
    The upper and lower panels correspond
    to the gluon and quark sectors,
    respectively.
    }
    \label{fig:correctionGQ}
\end{figure}

The separate quark and gluon GFFs are
shown in figure~\ref{fig:2}.
The quark sector dominates both $A$ and the
magnitude of $C$ over most of the displayed
range. In particular, our perturbative result
for $C_g^R$ is positive at sufficiently large
$Q^2$, in contrast to the negative values
found in lattice QCD at low $Q^2$~\cite{Hackett:2023nkr}. This contrast suggests
a possible sign change between the nonperturbative
and perturbative regimes, although a comparison at
a common operator renormalization scale is needed
to assess this possibility. Our central predictions
use $\mu=Q$, whereas the lattice results are quoted
at $\mu=2\,{\rm GeV}$.

The perturbative behavior of $C_g^R$ also
differs from that of the other GFFs.
As illustrated in figure~\ref{fig:correctionGQ},
the NLO contributions to $A_q^R$, $A_g^R$,
and $C_q^R$ have the same signs as their
respective LO terms, whereas the NLO
correction to $C_g^R$ has the opposite sign.
The resulting negative relative correction
partially cancels the LO gluon contribution.
Since $C_g^R$ is numerically small, its
relative correction is sensitive to such
cancellations and should not be interpreted
as evidence of particularly rapid
perturbative convergence.

\begin{figure}[t]
    \centering
    \includegraphics[width=1\linewidth]
    {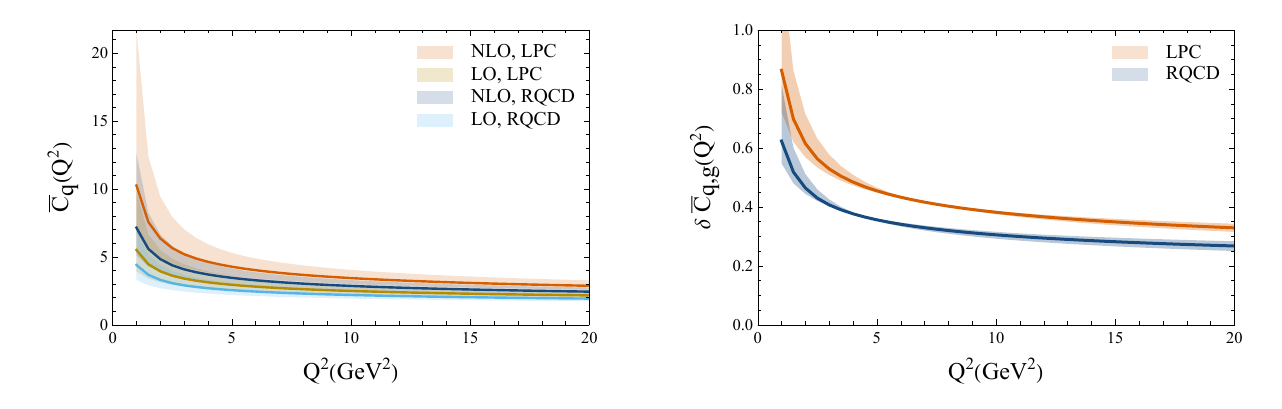}
    \caption{
    The renormalized pion GFF
    $\overline C_q^R(Q^2)$ at LO and NLO
    (left), and its relative NLO correction
    $\delta\overline C_q$ (right).
    The normalization uses $\Lambda=m_\pi$.
    The gluon contribution satisfies
    $\overline C_g^R=-\overline C_q^R$
    by EMT conservation.
    }
    \label{fig:Cbar}
\end{figure}

Figure~\ref{fig:Cbar} shows
$\overline C_q^R(Q^2)$, with the reference
mass scale chosen as $\Lambda=m_\pi$.
This choice fixes the normalization of the
dimensionless form factor, since the corresponding
EMT matrix element involves the combination
$2\Lambda^2\overline C_q^R$.
Unlike $A_a^R$ and $C_a^R$, the
$\overline C_a^R$ form factors are not
suppressed by $1/Q^2$ at leading power.
More importantly, the quark and gluon
contributions are constrained by
$\overline C_q^R+\overline C_g^R=0$,
which is explicitly satisfied by our NLO
results. Thus, only one independent
$\overline C$ form factor is displayed,
and the two sectors have identical
relative NLO corrections.
The moderate corrections at larger $Q^2$
are consistent with an improved perturbative
description of this particular form factor,
although its normalization remains sensitive
to the scale choice at lower $Q^2$.

\begin{figure}[t]
    \centering
    \includegraphics[width=1\linewidth]
    {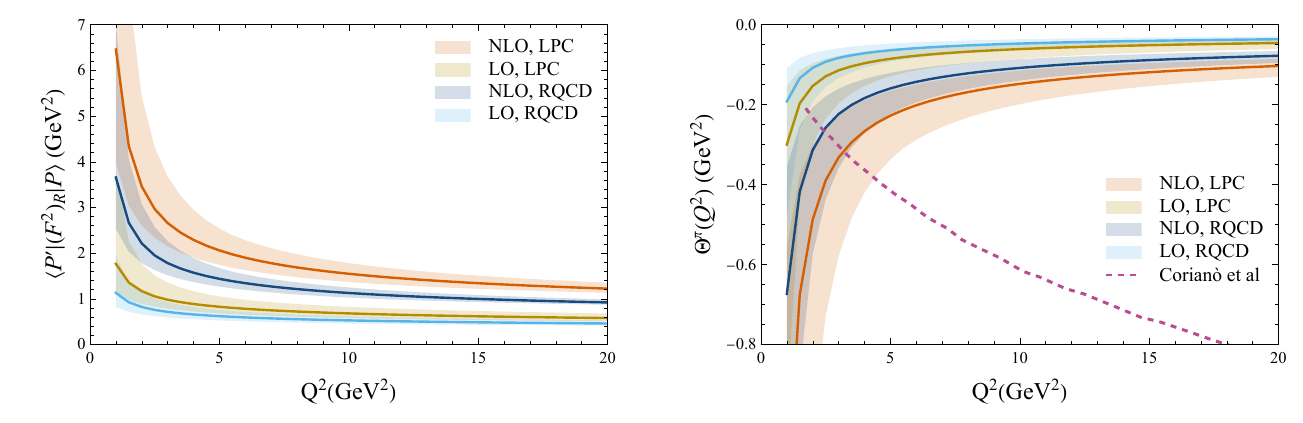}
    \caption{
    The renormalized scalar matrix element
    $\langle P'|(F^2)_R|P\rangle$ (left)
    and the pion trace-anomaly form factor
    $\Theta^\pi(Q^2)$ (right) at LO and NLO.
    The trace-anomaly results of
    Corian\`o \textit{et al.}
    ~\cite{Coriano:2026gyl,Coriano:2026ssa}
    are included for comparison. Both quantities are given in $\mathrm{GeV}^2$.
    }
    \label{fig:FFTr}
\end{figure}

\begin{figure}[t]
    \centering
    \includegraphics[width=1\linewidth]
    {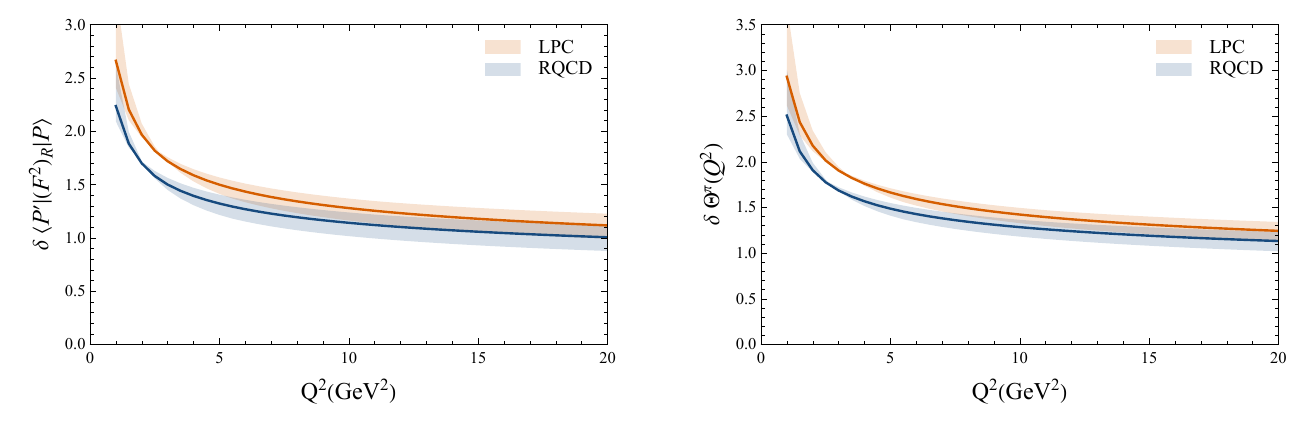}
    \caption{
    Relative NLO corrections to the
    $F^2$ scalar form factor (left)
    and the pion trace-anomaly form factor
    $\Theta^\pi(Q^2)$ (right).
    }
    \label{figFFTrcorrect}
\end{figure}

Figure~\ref{fig:FFTr} presents our LO and NLO predictions for the renormalized gluonic scalar matrix element $\langle P'|(F^2)_R|P\rangle$ and the pion trace-anomaly form factor $\Theta^\pi(Q^2)$. Similar to the $\overline{C}_{q,g}$ GFFs, these quantities are not explicitly suppressed by $1/Q^2$ at leading power. Their perturbative contributions start at $\mathcal O(\alpha_s)$ and $\mathcal O(\alpha_s^2)$, respectively, with the remaining $Q$-dependence governed by the running coupling and LCDA evolution. The scalar matrix element is positive, while $\Theta^\pi(Q^2)$ is negative, reflecting the negative QCD beta function. Interestingly, the lattice-QCD study by the $\chi$QCD Collaboration~\cite{Wang:2024lrm} finds that the pion gluonic trace-anomaly form factor changes sign and becomes negative with increasing $Q^2$. Despite the different pion masses and kinematic regimes, the two approaches agree on the negative sign of the trace-anomaly form factor.

In the right panel of figure~\ref{fig:FFTr}, we compare our trace-anomaly prediction with the results of Corian\`o \textit{et al.}~\cite{Coriano:2026gyl}, which exhibit a markedly different $Q^2$ dependence. Their calculations employ a $k_T$-factorization framework with Sudakov resummation, whereas ours is based on leading-twist collinear factorization with the complete NLO hard-scattering kernels. Given the differences in the theoretical frameworks, perturbative treatments, and nonperturbative inputs, the origin of this discrepancy remains to be clarified.

Figure~\ref{figFFTrcorrect} shows the relative NLO corrections to the $F^2$ scalar form factor and the pion trace anomaly. Their difference can be understood from the trace-anomaly relation in Eq.~\eqref{eq:TraceR}. For $\mu_R=\mu_F=Q$, we obtain
\begin{align}
\delta \Theta^\pi(Q^2)-\delta  \langle P'|(F^2)_R| P\rangle
=
\frac{\beta_1}{4\beta_0}
\frac{\alpha_s(Q)}{\pi}
+\mathcal O(\alpha_s^2)~.
\end{align}
For $n_f=3$, this difference reduces to $16\alpha_s(Q)/(9\pi)$, independent of the pion LCDA at this order. Thus, the larger relative NLO correction to the trace anomaly arises from the additional contribution of the two-loop QCD beta function. The sizable corrections to both quantities, even at the upper end of the displayed $Q^2$ range, indicate relatively slow perturbative convergence and motivate further higher-order studies.

\section{Summary and outlook}

In this work, we have calculated the NLO QCD corrections to the pion GFFs at large $Q^2$ within the leading-twist collinear factorization framework. We have derived the one-loop hard-scattering kernels for both the quark and gluon sectors. The collinear divergences are absorbed into the pion LCDAs, while the UV divergences are removed through  quark--gluon EMT operator mixing, providing nontrivial consistency checks of our calculation.

We have also investigated the pion trace-anomaly form factor, $\Theta^\pi (Q^2)$, whose leading nonvanishing contribution is reproduced by taking the trace of the NLO GFFs. By calculating the $F^2$ scalar form factor through NLO and incorporating the two-loop QCD beta function, we further determine the pion trace anomaly through $\mathcal O(\alpha_s^3)$.

Our results confirm the distinct large-$Q^2$ power behavior of
the pion GFFs. The $A_{q,g}$ and $C_{q,g}$ form factors
exhibit the expected $1/Q^2$ suppression, whereas
$\overline C_{q,g}$, the $F^2$ scalar form factor, and the
trace-anomaly form factor $\Theta^\pi (Q^2)$ exhibit no power-law dependence on
$Q^2$ at leading power, apart from logarithmic corrections.
The quark and gluon contributions to $\overline C$ cancel
exactly, as required by conservation of the total EMT.


Using recent lattice-QCD determinations of the pion LCDA, we have presented numerical predictions for the individual quark and gluon GFFs, their sums, and the pion trace anomaly. The total GFFs satisfy $A(Q^2)>0$ and $C(Q^2)<0$ in the perturbative region, with NLO corrections increasing their magnitudes. The quark contributions dominate both form factors, while the positive gluon contribution to $C_g^R$ at large $Q^2$ contrasts with lattice results at low momentum transfer. The renormalized $F^2$ scalar form factor is positive, whereas the pion trace-anomaly form factor $\Theta^\pi(Q^2)$ is negative, consistent with the QCD beta function.

Although the residual renormalization-scale dependence generally decreases at larger $Q^2$, the NLO corrections remain sizable for pion GFFs, indicating relatively slow perturbative convergence. Further higher-order calculations will be valuable for improving the precision of these predictions. Our results provide perturbative benchmarks for pion GFFs at large $Q^2$ and complementary constraints for future experimental measurements, lattice-QCD studies, and phenomenological descriptions of the pion internal structure.

\begin{acknowledgments}

We thank Jun Hua for providing the latest pion DA moments from the {\tt LPC} Collaboration prior to their arXiv submission.
  The work of J.~D., Y.~J. and Z.-Y.~W. is supported in part by the NNSFC under Grant No.~12475090.
  The work of F.~F. is supported in part by the NNSFC under Grant No.~12275353.
 The work of X-B.~T. is supported in part by the Research Council of Finland, the Centre of Excellence in Quark Matter and projects 338263 and 359902, and by the European Research Council (ERC, grant agreements No. ERC-2023-101123801 GlueSatLight and No. ERC-2018-ADG-835105 YoctoLHC).
\end{acknowledgments}
\appendix

\section{NLO hard-scattering kernels}
\label{appendix_A}
In this appendix, we summarize the full analytic expressions of the NLO hard-scattering kernels and their asymptotic forms in the relevant endpoint limits.
\subsection{Full analytic expressions}
\label{app:explicit-nlo-kernels}
With $\bar x=1-x$ and $L_{\mu_i}=\ln(\mu_i^2/Q^2)$, $i=R,F$, define
\begin{align}
 \mathcal S f(u,v)&=\frac{f(u,v)+f(\bar u,v)}{2},\notag\\*
 \mathcal L_1(x)&=\bar x\ln x+x\ln\bar x,\notag\\*
 \mathcal L_2(x)&=\bar x\ln^2x+x\ln^2\bar x
                   -2L_{\mu_F}\mathcal L_1(x),\notag\\*
 \Phi(u,v)&=\operatorname{Li}_2(u)-\operatorname{Li}_2(\bar u)
            -\operatorname{Li}_2(v)+\operatorname{Li}_2(\bar v)
            -\ln u\ln\bar v+\ln\bar u\ln v.
 \label{eq:five-notation}
\end{align}
The NLO kernels for the $A_{a},C_{a}$ and $\overline{C}_a$ GFFs are
\begin{align}
 \mathcal A_{g,R}^{(1)}(u,v)
 &=\mathcal G(u,v)+3C_A\mathcal J(u,v)+3C_F\mathcal B(u,v),
 \label{eq:five-Ag}\\
 \mathcal C_{g,R}^{(1)}(u,v)
 &=\mathcal G(u,v)-C_A\mathcal J(u,v)-C_F\mathcal B(u,v),
 \label{eq:five-Cg}\\
 \mathcal A_{q,R}^{(1)}(u,v)
 &=-\mathcal G(u,v)-\frac{\beta_0}{16u\bar u v\bar v}
   +3(C_A-2C_F)\mathcal K(u,v)-3n_f\mathcal J(u,v)\notag\\*
 &\quad+3C_F\mathcal R(u,v)
   +\frac{n_f}{4u\bar u v\bar v}
    \left[\mathcal L_1(u)+\mathcal L_1(v)-L_{\mu_R}-\frac{11}{12}\right],
 \label{eq:five-Aq}\\
 \mathcal C_{q,R}^{(1)}(u,v)
 &=-\mathcal G(u,v)-\frac{\beta_0}{16u\bar u v\bar v}
   -(C_A-2C_F)\mathcal K(u,v)+n_f\mathcal J(u,v)\notag\\*
 &\quad-C_F\mathcal R(u,v)
   -\frac{n_f}{12u\bar u v\bar v}
    \left[\mathcal L_1(u)+\mathcal L_1(v)-L_{\mu_R}-\frac{11}{12}\right],
 \label{eq:five-Cq}\\
\overline{\mathcal C}_{g,R}^{(1)}(u,v)
&=
-\overline{\mathcal C}_{q,R}^{(1)}(u,v)
=
-\frac{Q^2}{4\Lambda^2}
\left[
\mathcal G(u,v)
+\frac{11C_A}{48u\bar u\,v\bar v}
\right],
\label{eq:Cbar_NLO}
\end{align}
where $\beta_0=(11C_A-2n_f)/3$ and the repeated structures are
\begin{align}
 \mathcal G(u,v)
 ={}&C_A\Biggl\{
 \frac{17\bigl[20(u\bar u+v\bar v)-3\bigr]}{144u\bar u v\bar v}
 +\frac{11(u\bar u+v\bar v)}{24u\bar u v\bar v}
       \bigl[2L_{\mu_R}-\ln(u\bar u v\bar v)\bigr]\notag\\*
 &\hspace{16mm}
 -\frac1{12}\sum_{x\in\{u,v\}}
       \frac{1-2x}{x\bar x}\ln\frac{x}{\bar x}
 \Biggr\}\notag\\*
 &+n_f\frac{u\bar u+v\bar v}{u\bar u v\bar v}
       \left[\frac{\ln(u\bar u v\bar v)}6-\frac59-\frac{L_{\mu_R}}3\right]
 \notag\\*
 &+C_F\Biggl\{
 \frac{3-124(u\bar u+v\bar v)}{36u\bar u v\bar v}
 -\frac{u\bar u+v\bar v}{12u\bar u v\bar v}
      (8L_{\mu_R}+9L_{\mu_F})
 +\frac14\sum_{x\in\{u,v\}}\frac{\mathcal L_2(x)}{x\bar x}
 \notag\\*
 &\hspace{13mm}
 +\frac{1}{24u\bar u v\bar v}
   \sum_{(x,y)\in\{(u,v),(v,u)\}}
   \Bigl[(17x\bar x+4y\bar y)\ln(x\bar x)
          +x\ln x+\bar x\ln\bar x\Bigr]\notag\\*
 &\hspace{13mm}
 +\frac14\left[
     \frac{\mathcal L_1(u)}{u\bar u}\ln(v\bar v)
    +\frac{\mathcal L_1(v)}{v\bar v}\ln(u\bar u)
   \right]\Biggr\},
 \label{eq:five-G}
\end{align}
\begin{align}
 \mathcal B(u,v)
 =\mathcal S\Biggl\{&
 \frac{u\bar u+v\bar v}{(u-v)^3}\Phi(u,v)
 +\frac{(u+v)\ln(uv)+(\bar u+\bar v)\ln(\bar u\bar v)}{(u-v)^2}
 \notag\\*
 &+\frac{uv+\bar u\bar v}{u\bar u v\bar v}
      \left(\frac{17}{18}+\frac23L_{\mu_R}\right)
 \Biggr\},
 \label{eq:five-B}
 \end{align}
 \begin{align}
 \mathcal J(u,v)
 =\mathcal S\Biggl\{&
 -\frac{(u+v-2uv)(u\bar u+v\bar v)}{2(u-v)^5}\Phi(u,v)
 \notag\\*
 &+\frac{(u+v-2uv)\bigl[5(u-v)^2-2(u+v-2uv)\bigr]}
             {24u\bar u v\bar v(u-v)^2}\notag\\*
 &-\sum_{(x,y)\in\{(u,v),(\bar u,\bar v)\}}
   \frac{\ln(xy)}{48\bar x\bar y(x-y)^4}
   \Bigl[3(x+y)^2(\bar x+\bar y)^3\notag\\*
 &\hspace{35mm}
       -2(x+y)^2(\bar x+\bar y)(x-y)^2
       +(2+x+y)(x-y)^4\Bigr]
 \Biggr\},\label{eq:five-J}
\end{align}

\begin{align}
    \mathcal K(u,v)
 =\mathcal S\Biggl\{&
 -\frac{(u+v-2uv)
       \bigl[2u^2v^2-(u+v)(u+v-1)(uv+\bar u\bar v)\bigr]}
       {4u\bar u v\bar v(u-v)^3}\Phi(u,v)\notag\\*
 &-\frac{(uv+\bar u\bar v)^2}
       {4u\bar u v\bar v(u+v-1)}\Phi(u,\bar v)
 +\frac{uv+\bar u\bar v}{4u\bar u v\bar v}
       \ln\frac{u}{\bar u}\ln\frac{v}{\bar v}\notag\\*
 &+\frac{1}{12u\bar u v\bar v(u-v)^2}
   \sum_{(x,y)\in\{(u,v),(\bar u,\bar v)\}}\ln(xy)
   \Bigl\{11(x+y)^2(x+y-1)\notag\\*
 &\hspace{26mm}
    +xy\bigl[32+44xy-20(x+y)+18xy(x+y)-29(x+y)^2\bigr]
   \Bigr\}\notag\\*
 &+\frac{uv+\bar u\bar v}{u\bar u v\bar v}
       \left(\frac{19}{9}+\frac{11}{12}L_{\mu_R}\right)
 \Biggr\},
 \label{eq:five-K}
 \end{align}
 \begin{align}
 \mathcal R(u,v)
 =\frac1{u\bar u v\bar v}\Biggl\{&
 \frac{\mathcal L_2(u)+\mathcal L_2(v)}8
 +\frac{\mathcal L_1(u)\mathcal L_1(v)}4
 -\frac{13}{36}+\frac7{12}L_{\mu_R}-\frac34L_{\mu_F}\notag\\*
 &-\frac1{24}\sum_{x\in\{u,v\}}
       \bigl[(7-4x)\ln x+(3+4x)\ln\bar x\bigr]
 \Biggr\}.
 \label{eq:five-R}
\end{align}
The hard kernel for the pion $F^2$ scalar form factor is given as
\begin{align}
 H_R^{{\rm F^2}(1)}(u,v)
 =\frac1{u\bar u v\bar v}\Biggl\{&
 C_A\Biggl[
 \frac{67}{18}+\frac{11}{6}L_{\mu_R}
 -\frac{11}{12}\ln(u\bar u v\bar v)\notag\\*
 &\hspace{16mm}
 -\frac{(2u-1)(2v-1)}2
       \ln\frac{u}{\bar u}\ln\frac{v}{\bar v}\Biggr]\notag\\*
 &+n_f\left[-\frac59-\frac{L_{\mu_R}}3
                   +\frac{\ln(u\bar u v\bar v)}6\right]\notag\\*
 &+\frac{C_F}{4}\Biggl[
 \mathcal L_2(u)+\mathcal L_2(v)-14-6L_{\mu_F}\notag\\*
 &\hspace{14mm}
 +\sum_{x\in\{u,v\}}\bigl[(3-x)\ln x+(2+x)\ln\bar x\bigr]
 +\ln(u\bar u)\ln(v\bar v)\notag\\*
 &\hspace{14mm}
 +3(2u-1)(2v-1)\ln\frac{u}{\bar u}\ln\frac{v}{\bar v}
 \Biggr]\Biggr\}.
 \label{eq:five-FF}
\end{align}

\subsection{Asymptotic expression at the endpoints}
\label{app:asy}
We also present their asymptotic forms in the relevant endpoint limits, which make the singularity structure of the kernels explicit. In the simultaneous limit $u\to0$ and $v\to0$, we obtain
\begin{align}
\mathcal A_{g,R}^{(1)}(u,v)
 &\sim\frac{C_F(3/2+L_{\mu_R})-C_A/6}{uv},
 \label{eq:five-Ag-endpoint}\\*
 \mathcal C_{g,R}^{(1)}(u,v)
 &\sim-\frac{14C_F+15C_A+12C_FL_{\mu_R}}{36uv},
 \label{eq:five-Cg-endpoint}\\*
\mathcal A_{q,R}^{(1)}(u,v)
 &\sim\frac1{24uv}\Bigl\{
   (33C_A-24C_F-6n_f)L_{\mu_R}
   -18C_F\bigl[\ln(uv)+3\bigr]L_{\mu_F}\notag\\*
 &\hspace{12mm}
   +3C_F\bigl[3\ln^2(uv)+9\ln(uv)+2\pi^2-60\bigr]\notag\\*
 &\hspace{12mm}
   +C_A\bigl[-24\ln(uv)-3\pi^2+79\bigr]
   +n_f\bigl[6\ln(uv)-9\bigr]\Bigr\},
 \label{eq:five-Aq-endpoint}\\*
 \mathcal C_{q,R}^{(1)}(u,v)
 &\sim\frac1{72uv}\Bigl\{
   (-33C_A+24C_F+6n_f)L_{\mu_R}
   +18C_F\bigl[\ln(uv)+3\bigr]L_{\mu_F}\notag\\*
 &\hspace{12mm}
   +C_F\bigl[-9\ln^2(uv)-27\ln(uv)-6\pi^2+172\bigr]\notag\\*
 &\hspace{12mm}
   +C_A\bigl[24\ln(uv)+3\pi^2-67\bigr]
   +n_f\bigl[-6\ln(uv)+13\bigr]\Bigr\},
 \label{eq:five-Cq-endpoint}\\*
 \overline{\mathcal C}_{g,R}^{(1)}(u,v)
&=
-\overline{\mathcal C}_{q,R}^{(1)}(u,v)
\sim
\frac{Q^2(3C_A-2C_F)}{96\Lambda^2 uv}.
\label{eq:Cbar_NLO_endpoint}\\*
 H_R^{F^2(1)}(u,v)
 &\sim\frac1{36uv}\Bigl\{
   (66C_A-12n_f)L_{\mu_R}
   -18C_F\bigl[\ln(uv)+3\bigr]L_{\mu_F}\notag\\*
 &\hspace{12mm}
   +C_A\bigl[134-33\ln(uv)-18\ln u\ln v\bigr]\notag\\*
 &\hspace{12mm}
   +9C_F\bigl[\ln^2(uv)+2\ln u\ln v+3\ln(uv)-14\bigr]
   +n_f\bigl[6\ln(uv)-20\bigr]\Bigr\}.
 \label{eq:five-FF-endpoint}
\end{align}
Owing to the symmetry of the hard-scattering kernel under $u\to\bar{u}$ and $v\to \bar{v}$, the other corner limits can be obtained from the $u\to0$ and $v\to0$ limit by applying the corresponding symmetry transformations.

\subsection{Gegenbauer-projected kernels}
\label{sec:GegenbauerHK}
We denote the NLO Gegenbauer-projected hard-scattering kernels by $\mathcal{H}^{X,(1)}_{mn}$, where $X=A_g,A_q,C_g,C_q,\overline{C}_g,\overline{C}_q,F^2$, and the indices $m$ and $n$ label the Gegenbauer moments associated with the initial- and final-state pion LCDAs, respectively. Below, we list the results required for the numerical analysis, retaining moments up to $m,n=3$ for $\mu_R=\mu_F=\mu$ and $n_f=3$. Here, $L_\mu=\ln(\mu^2/Q^2)$. The projected kernels are
symmetric under $m\leftrightarrow n$, so only entries with
$m\leq n$ are listed.

For the gluonic contribution to the $A$ form factor, the projected kernels are
\begingroup
\footnotesize
\begin{align*}
\mathcal{H}^{A_g,(1)}_{00} &= \frac{175}{3}L_\mu - \frac{288\zeta(3)}{5} + \frac{2578}{15}, &
\mathcal{H}^{A_g,(1)}_{01} &= \frac{123}{2}L_\mu + \frac{7884\zeta(3)}{35} - \frac{12637}{84}, \\[4pt]
\mathcal{H}^{A_g,(1)}_{02} &= \frac{653}{10}L_\mu + \frac{199869}{1400}, &
\mathcal{H}^{A_g,(1)}_{03} &= \frac{7123}{105}L_\mu + \frac{84798409}{529200}, \\[4pt]
\mathcal{H}^{A_g,(1)}_{11} &= 48L_\mu - \frac{46}{105}\bigl(3888\zeta(3)-4819\bigr), &
\mathcal{H}^{A_g,(1)}_{12} &= 48L_\mu + \frac{239760\zeta(3)}{77} - \frac{21258494}{5775}, \\[4pt]
\mathcal{H}^{A_g,(1)}_{13} &= 48L_\mu + \frac{10799}{175}, &
\mathcal{H}^{A_g,(1)}_{22} &= 48L_\mu - \frac{2025000\zeta(3)}{143} + \frac{1025858087}{60060}, \\[4pt]
\mathcal{H}^{A_g,(1)}_{23} &= 48L_\mu + \frac{2450952\zeta(3)}{143} - \frac{161954008451}{7882875}, &
\mathcal{H}^{A_g,(1)}_{33} &= 48L_\mu + \frac{7\bigl(40793269159-33911136000\zeta(3)\bigr)}{3729375}.
\end{align*}
\endgroup
The corresponding quark contribution reads
\begingroup
\footnotesize
\begin{align*}
\mathcal{H}^{A_q,(1)}_{00} &= \frac{379}{6}L_\mu - \frac{36\zeta(3)}{5} + \frac{6157}{30}, &
\mathcal{H}^{A_q,(1)}_{01} &= 135L_\mu + \frac{2}{105}\bigl(37204-8343\zeta(3)\bigr), \\[4pt]
\mathcal{H}^{A_q,(1)}_{02} &= \frac{827}{5}L_\mu - 54\zeta(3) + \frac{1081467}{1400}, &
\mathcal{H}^{A_q,(1)}_{03} &= \frac{19499}{105}L_\mu - 54\zeta(3) + \frac{38767657}{42336}, \\[4pt]
\mathcal{H}^{A_q,(1)}_{11} &= \frac{447}{2}L_\mu + \frac{42228\zeta(3)}{35} - \frac{329593}{840}, &
\mathcal{H}^{A_q,(1)}_{12} &= \frac{2577}{10}L_\mu - \frac{202068\zeta(3)}{77} + \frac{83126261}{18480}, \\[4pt]
\mathcal{H}^{A_q,(1)}_{13} &= \frac{9819}{35}L_\mu - 324\zeta(3) + \frac{228374789}{117600}, &
\mathcal{H}^{A_q,(1)}_{22} &= \frac{2919}{10}L_\mu + \frac{1799280\zeta(3)}{143} - \frac{1010590706}{75075}, \\[4pt]
\mathcal{H}^{A_q,(1)}_{23} &= \frac{11016}{35}L_\mu - \frac{2208438\zeta(3)}{143} + \frac{5160860032033}{252252000}, &
\mathcal{H}^{A_q,(1)}_{33} &=
\begin{aligned}[t]
&\frac{23631}{70}L_\mu + \frac{66416616\zeta(3)}{1105}\\
&-\frac{409973772642593}{5847660000}~.
\end{aligned}
\end{align*}
\endgroup
For the $C$ form factor, the gluon and quark projections are given, respectively, by
\begingroup
\footnotesize
\begin{align*}
\mathcal{H}^{C_g,(1)}_{00} &= \frac{1}{15}\bigl(-85L_\mu + 288\zeta(3) - 561\bigr), &
\mathcal{H}^{C_g,(1)}_{01} &= -\frac{5}{2}L_\mu - \frac{2628\zeta(3)}{35} + \frac{12241}{140}, \\[4pt]
\mathcal{H}^{C_g,(1)}_{02} &= \frac{13}{10}L_\mu + \frac{248321}{12600}, &
\mathcal{H}^{C_g,(1)}_{03} &= \frac{403}{105}L_\mu + \frac{19713809}{529200}, \\[4pt]
\mathcal{H}^{C_g,(1)}_{11} &= -16L_\mu + \frac{19872\zeta(3)}{35} - \frac{77599}{105}, &
\mathcal{H}^{C_g,(1)}_{12} &= -16L_\mu - \frac{79920\zeta(3)}{77} + \frac{13743487}{11550}, \\[4pt]
\mathcal{H}^{C_g,(1)}_{13} &= -16L_\mu - \frac{90674}{1575}, &
\mathcal{H}^{C_g,(1)}_{22} &= -16L_\mu + \frac{675000\zeta(3)}{143} - \frac{68881849}{12012}, \\[4pt]
\mathcal{H}^{C_g,(1)}_{23} &= -16L_\mu - \frac{816984\zeta(3)}{143} + \frac{5961783413}{875875}, &
\mathcal{H}^{C_g,(1)}_{33} &= -16L_\mu + \frac{23444736\zeta(3)}{1105} - \frac{667407982222}{26105625}~.
\end{align*}
\endgroup
The quark projections read
\begingroup
\footnotesize
\begin{align*}
\mathcal{H}^{C_q,(1)}_{00} &= -\frac{209}{6}L_\mu + \frac{12\zeta(3)}{5} - \frac{1153}{10}, &
\mathcal{H}^{C_q,(1)}_{01} &= -63L_\mu + \frac{1854\zeta(3)}{35} - \frac{31553}{105}, \\[4pt]
\mathcal{H}^{C_q,(1)}_{02} &= -\frac{391}{5}L_\mu + 18\zeta(3) - \frac{4432529}{12600}, &
\mathcal{H}^{C_q,(1)}_{03} &= -\frac{9277}{105}L_\mu + 18\zeta(3) - \frac{447600499}{1058400}, \\[4pt]
\mathcal{H}^{C_q,(1)}_{11} &= -\frac{149}{2}L_\mu - \frac{14076\zeta(3)}{35} + \frac{350537}{2520}, &
\mathcal{H}^{C_q,(1)}_{12} &= -\frac{859}{10}L_\mu + \frac{67356\zeta(3)}{77} - \frac{412823489}{277200}, \\[4pt]
\mathcal{H}^{C_q,(1)}_{13} &= -\frac{3273}{35}L_\mu + 108\zeta(3) - \frac{224846341}{352800}, &
\mathcal{H}^{C_q,(1)}_{22} &= -\frac{973}{10}L_\mu - \frac{599760\zeta(3)}{143} + \frac{1013721691}{225225}, \\[4pt]
\mathcal{H}^{C_q,(1)}_{23} &= -\frac{3672}{35}L_\mu + \frac{1472292\zeta(3)}{286} - \frac{1716581680811}{252252000}, &
\mathcal{H}^{C_q,(1)}_{33} &=
\begin{aligned}[t]
&-\frac{7877}{70}L_\mu - \frac{22138872\zeta(3)}{1105}\\
&+\frac{136750451315531}{5847660000}~.
\end{aligned}
\end{align*}
\endgroup
For the $\overline C_a$ form factors, the projected kernels satisfy
$\mathcal{H}^{\overline C_g,(1)}_{mn}
=-\mathcal{H}^{\overline C_q,(1)}_{mn}$.
The independent gluon projections are
\begin{align*}
\mathcal{H}^{\overline C_g,(1)}_{00}
&=
-\frac{Q^2}{\Lambda^2}
\left(
\frac{31}{12}L_\mu+\frac{119}{12}
\right),
&
\mathcal{H}^{\overline C_g,(1)}_{01}
&=
-\frac{Q^2}{\Lambda^2}
\left(
\frac{27}{8}L_\mu+\frac{3163}{240}
\right),
\\[1ex]
\mathcal{H}^{\overline C_g,(1)}_{02}
&=
-\frac{Q^2}{\Lambda^2}
\left(
\frac{173}{40}L_\mu+\frac{78983}{4200}
\right),
&
\mathcal{H}^{\overline C_g,(1)}_{03}
&=
-\frac{Q^2}{\Lambda^2}
\left(
\frac{2083}{420}L_\mu+\frac{49082659}{2116800}
\right),
\\[1ex]
\mathcal{H}^{\overline C_g,(1)}_{11}
&=
\frac{13Q^2}{30\Lambda^2},
&
\mathcal{H}^{\overline C_g,(1)}_{12}
&=
\frac{13007Q^2}{16800\Lambda^2},
\\[1ex]
\mathcal{H}^{\overline C_g,(1)}_{13}
&=
\frac{3151Q^2}{4200\Lambda^2},
&
\mathcal{H}^{\overline C_g,(1)}_{22}
&=
\frac{2489Q^2}{1680\Lambda^2},
\\[1ex]
\mathcal{H}^{\overline C_g,(1)}_{23}
&=
\frac{57469Q^2}{35280\Lambda^2},
&
\mathcal{H}^{\overline C_g,(1)}_{33}
&=
\frac{3481Q^2}{1890\Lambda^2}~.
\end{align*}
Finally, the Gegenbauer projections of the hard-scattering kernel associated with the scalar operator $F^2$ are
\begin{align*}
\mathcal{H}^{F^2,(1)}_{00} &= 162L_\mu + 480, &
\mathcal{H}^{F^2,(1)}_{01} &= 212L_\mu + \frac{3955}{6}, \\[4pt]
\mathcal{H}^{F^2,(1)}_{02} &= \frac{1174}{5}L_\mu + \frac{119089}{150}, &
\mathcal{H}^{F^2,(1)}_{03} &= \frac{3\bigl(816760L_\mu + 2934257\bigr)}{9800}, \\[4pt]
\mathcal{H}^{F^2,(1)}_{11} &= 262L_\mu + \frac{4555}{6}, &
\mathcal{H}^{F^2,(1)}_{12} &= \frac{1424}{5}L_\mu + \frac{130039}{150}, \\[4pt]
\mathcal{H}^{F^2,(1)}_{13} &= \frac{8820840L_\mu + 28058563}{29400}, &
\mathcal{H}^{F^2,(1)}_{22} &= \frac{1538}{5}L_\mu + \frac{144961}{150}, \\[4pt]
\mathcal{H}^{F^2,(1)}_{23} &= \frac{1898232L_\mu + 6167171}{5880}, &
\mathcal{H}^{F^2,(1)}_{33} &= \frac{11832}{35}L_\mu + \frac{11056989}{9800}~.
\end{align*}

\newpage
\bibliographystyle{JHEP}
\bibliography{ref}

@article{Dudek:2012vr,
    author = "Dudek, Jozef and others",
    title = "{Physics Opportunities with the 12 GeV Upgrade at Jefferson Lab}",
    eprint = "1208.1244",
    archivePrefix = "arXiv",
    primaryClass = "hep-ex",
    reportNumber = "JLAB-PHY-12-1599",
    doi = "10.1140/epja/i2012-12187-1",
    journal = "Eur. Phys. J. A",
    volume = "48",
    pages = "187",
    year = "2012"
}

@article{Accardi:2023chb,
    author = "Accardi, A. and others",
    title = "{Strong interaction physics at the luminosity frontier with 22 GeV electrons at Jefferson Lab}",
    eprint = "2306.09360",
    archivePrefix = "arXiv",
    primaryClass = "nucl-ex",
    reportNumber = "JLAB-PHY-23-3840, JLAB-THY-23-3848",
    doi = "10.1140/epja/s10050-024-01282-x",
    journal = "Eur. Phys. J. A",
    volume = "60",
    number = "9",
    pages = "173",
    year = "2024"
}

@article{Boer:2011fh,
    author = "Boer, Daniel and others",
    title = "{Gluons and the quark sea at high energies: Distributions, polarization, tomography}",
    eprint = "1108.1713",
    archivePrefix = "arXiv",
    primaryClass = "nucl-th",
    reportNumber = "SLAC-R-995, INT-PUB-11-034, BNL-96164-2011, JLAB-THY-11-1373",
    month = "8",
    year = "2011"
}

@article{Accardi:2012qut,
    author = "Accardi, A. and others",
    editor = "Deshpande, A. and Meziani, Z. E. and Qiu, J. W.",
    title = "{Electron Ion Collider: The Next QCD Frontier}: {Understanding the glue that binds us all}",
    eprint = "1212.1701",
    archivePrefix = "arXiv",
    primaryClass = "nucl-ex",
    reportNumber = "BNL-98815-2012-JA, JLAB-PHY-12-1652",
    doi = "10.1140/epja/i2016-16268-9",
    journal = "Eur. Phys. J. A",
    volume = "52",
    number = "9",
    pages = "268",
    year = "2016"
}

@article{AbdulKhalek:2021gbh,
    author = "Abdul Khalek, R. and others",
    title = "{Science Requirements and Detector Concepts for the Electron-Ion Collider}: {EIC Yellow Report}",
    eprint = "2103.05419",
    archivePrefix = "arXiv",
    primaryClass = "physics.ins-det",
    reportNumber = "BNL-220990-2021-FORE, JLAB-PHY-21-3198, LA-UR-21-20953",
    doi = "10.1016/j.nuclphysa.2022.122447",
    journal = "Nucl. Phys. A",
    volume = "1026",
    pages = "122447",
    year = "2022"
}

@article{Anderle:2021wcy,
    author = "Anderle, Daniele P. and others",
    title = "{Electron-ion collider in China}",
    eprint = "2102.09222",
    archivePrefix = "arXiv",
    primaryClass = "nucl-ex",
    reportNumber = "Frontiers of Physics, Volume 16 Issue (6):64701, 2021",
    doi = "10.1007/s11467-021-1062-0",
    journal = "Front. Phys. (Beijing)",
    volume = "16",
    number = "6",
    pages = "64701",
    year = "2021"
}

@article{Kobzarev:1962wt,
    author = "Kobzarev, I. Yu. and Okun', L. B.",
    title = "{Gravitational Interaction of Fermions}",
    journal = "Zh. Eksp. Teor. Fiz.",
    volume = "43",
    pages = "1904--1909",
    year = "1962"
}

@article{Pagels:1966zza,
    author = "Pagels, Heinz",
    title = "{Energy-Momentum Structure Form Factors of Particles}",
    doi = "10.1103/PhysRev.144.1250",
    journal = "Phys. Rev.",
    volume = "144",
    pages = "1250--1260",
    year = "1966"
}

@article{Ji:1996ek,
    author = "Ji, Xiang-Dong",
    title = "{Gauge-Invariant Decomposition of Nucleon Spin}",
    eprint = "hep-ph/9603249",
    archivePrefix = "arXiv",
    reportNumber = "MIT-CTP-2517",
    doi = "10.1103/PhysRevLett.78.610",
    journal = "Phys. Rev. Lett.",
    volume = "78",
    pages = "610--613",
    year = "1997"
}

@article{Ji:1994av,
    author = "Ji, Xiang-Dong",
    title = "{A QCD analysis of the mass structure of the nucleon}",
    eprint = "hep-ph/9410274",
    archivePrefix = "arXiv",
    reportNumber = "MIT-CTP-2368",
    doi = "10.1103/PhysRevLett.74.1071",
    journal = "Phys. Rev. Lett.",
    volume = "74",
    pages = "1071--1074",
    year = "1995"
}

@article{Ji:1995sv,
    author = "Ji, Xiang-Dong",
    title = "{Breakup of hadron masses and energy - momentum tensor of QCD}",
    eprint = "hep-ph/9502213",
    archivePrefix = "arXiv",
    reportNumber = "MIT-CTP-2407",
    doi = "10.1103/PhysRevD.52.271",
    journal = "Phys. Rev. D",
    volume = "52",
    pages = "271--281",
    year = "1995"
}

@article{Ji:2021mtz,
    author = "Ji, Xiangdong",
    title = "{Proton mass decomposition: naturalness and interpretations}",
    eprint = "2102.07830",
    archivePrefix = "arXiv",
    primaryClass = "hep-ph",
    doi = "10.1007/s11467-021-1065-x",
    journal = "Front. Phys. (Beijing)",
    volume = "16",
    number = "6",
    pages = "64601",
    year = "2021"
}

@article{Hatta:2018sqd,
    author = "Hatta, Yoshitaka and Rajan, Abha and Tanaka, Kazuhiro",
    title = "{Quark and gluon contributions to the QCD trace anomaly}",
    eprint = "1810.05116",
    archivePrefix = "arXiv",
    primaryClass = "hep-ph",
    reportNumber = "YITP-18-112, J-PARC TH-0136, YITP-18-112,J-PARC TH-0136",
    doi = "10.1007/JHEP12(2018)008",
    journal = "JHEP",
    volume = "12",
    pages = "008",
    year = "2018"
}

@article{Lorce:2017xzd,
    author = "Lorc{\'e}, C{\'e}dric",
    title = "{On the hadron mass decomposition}",
    eprint = "1706.05853",
    archivePrefix = "arXiv",
    primaryClass = "hep-ph",
    doi = "10.1140/epjc/s10052-018-5561-2",
    journal = "Eur. Phys. J. C",
    volume = "78",
    number = "2",
    pages = "120",
    year = "2018"
}

@article{Metz:2020vxd,
    author = "Metz, Andreas and Pasquini, Barbara and Rodini, Simone",
    title = "{Revisiting the proton mass decomposition}",
    eprint = "2006.11171",
    archivePrefix = "arXiv",
    primaryClass = "hep-ph",
    doi = "10.1103/PhysRevD.102.114042",
    journal = "Phys. Rev. D",
    volume = "102",
    pages = "114042",
    year = "2020"
}

@article{Lorce:2021xku,
    author = "Lorc{\'e}, C{\'e}dric and Metz, Andreas and Pasquini, Barbara and Rodini, Simone",
    title = "{Energy-momentum tensor in QCD: nucleon mass decomposition and mechanical equilibrium}",
    eprint = "2109.11785",
    archivePrefix = "arXiv",
    primaryClass = "hep-ph",
    doi = "10.1007/JHEP11(2021)121",
    journal = "JHEP",
    volume = "11",
    pages = "121",
    year = "2021"
}

@article{Yang:2018nqn,
    author = "Yang, Yi-Bo and Liang, Jian and Bi, Yu-Jiang and Chen, Ying and Draper, Terrence and Liu, Keh-Fei and Liu, Zhaofeng",
    title = "{Proton Mass Decomposition from the QCD Energy Momentum Tensor}",
    eprint = "1808.08677",
    archivePrefix = "arXiv",
    primaryClass = "hep-lat",
    doi = "10.1103/PhysRevLett.121.212001",
    journal = "Phys. Rev. Lett.",
    volume = "121",
    number = "21",
    pages = "212001",
    year = "2018"
}

@article{Jaffe:1989jz,
    author = "Jaffe, R. L. and Manohar, Aneesh",
    title = "{The $g_1$ Problem: Fact and Fantasy on the Spin of the Proton}",
    reportNumber = "MIT-CTP-1706-REV, MIT-CTP-1706",
    doi = "10.1016/0550-3213(90)90506-9",
    journal = "Nucl. Phys. B",
    volume = "337",
    pages = "509--546",
    year = "1990"
}

@article{Leader:2013jra,
    author = "Leader, E. and Lorc{\'e}, C.",
    title = "{The angular momentum controversy: What{\textquoteright}s it all about and does it matter?}",
    eprint = "1309.4235",
    archivePrefix = "arXiv",
    primaryClass = "hep-ph",
    doi = "10.1016/j.physrep.2014.02.010",
    journal = "Phys. Rept.",
    volume = "541",
    number = "3",
    pages = "163--248",
    year = "2014"
}

@article{Ji:2020ena,
    author = "Ji, Xiangdong and Yuan, Feng and Zhao, Yong",
    title = "{What we know and what we don{\textquoteright}t know about the proton spin after 30 years}",
    eprint = "2009.01291",
    archivePrefix = "arXiv",
    primaryClass = "hep-ph",
    doi = "10.1038/s42254-020-00248-4",
    journal = "Nature Rev. Phys.",
    volume = "3",
    number = "1",
    pages = "27--38",
    year = "2021"
}

@article{Ji:2020hii,
    author = "Ji, Xiangdong and Yuan, Feng",
    title = "{Transverse spin sum rule of the proton}",
    eprint = "2008.04349",
    archivePrefix = "arXiv",
    primaryClass = "hep-ph",
    doi = "10.1016/j.physletb.2020.135786",
    journal = "Phys. Lett. B",
    volume = "810",
    pages = "135786",
    year = "2020"
}

@article{Polyakov:2002yz,
    author = "Polyakov, M. V.",
    title = "{Generalized parton distributions and strong forces inside nucleons and nuclei}",
    eprint = "hep-ph/0210165",
    archivePrefix = "arXiv",
    reportNumber = "RUB-TP2-14-02",
    doi = "10.1016/S0370-2693(03)00036-4",
    journal = "Phys. Lett. B",
    volume = "555",
    pages = "57--62",
    year = "2003"
}

@article{Polyakov:2018zvc,
    author = "Polyakov, Maxim V. and Schweitzer, Peter",
    title = "{Forces inside hadrons: pressure, surface tension, mechanical radius, and all that}",
    eprint = "1805.06596",
    archivePrefix = "arXiv",
    primaryClass = "hep-ph",
    doi = "10.1142/S0217751X18300259",
    journal = "Int. J. Mod. Phys. A",
    volume = "33",
    number = "26",
    pages = "1830025",
    year = "2018"
}

@article{Shanahan:2018nnv,
    author = "Shanahan, P. E. and Detmold, W.",
    title = "{Pressure Distribution and Shear Forces inside the Proton}",
    eprint = "1810.07589",
    archivePrefix = "arXiv",
    primaryClass = "nucl-th",
    reportNumber = "MIT-CTP/5071",
    doi = "10.1103/PhysRevLett.122.072003",
    journal = "Phys. Rev. Lett.",
    volume = "122",
    number = "7",
    pages = "072003",
    year = "2019"
}

@article{Lorce:2018egm,
    author = "Lorc{\'e}, C{\'e}dric and Moutarde, Herv{\'e} and Trawi{\'n}ski, Arkadiusz P.",
    title = "{Revisiting the mechanical properties of the nucleon}",
    eprint = "1810.09837",
    archivePrefix = "arXiv",
    primaryClass = "hep-ph",
    doi = "10.1140/epjc/s10052-019-6572-3",
    journal = "Eur. Phys. J. C",
    volume = "79",
    number = "1",
    pages = "89",
    year = "2019"
}

@article{Ji:2021mfb,
    author = "Ji, Xiangdong and Liu, Yizhuang",
    title = "{Momentum-Current Gravitational Multipoles of Hadrons}",
    eprint = "2110.14781",
    archivePrefix = "arXiv",
    primaryClass = "hep-ph",
    doi = "10.1103/PhysRevD.106.034028",
    journal = "Phys. Rev. D",
    volume = "106",
    number = "3",
    pages = "034028",
    year = "2022"
}

@article{Ji:2025qax,
    author = "Ji, Xiangdong and Yang, Chen",
    title = "{A journey of seeking pressure and forces in the nucleon}",
    eprint = "2508.16727",
    archivePrefix = "arXiv",
    primaryClass = "hep-ph",
    doi = "10.1016/j.nuclphysb.2026.117342",
    journal = "Nucl. Phys. B",
    volume = "1024",
    pages = "117342",
    year = "2026"
}

@article{Ji:2025gsq,
    author = "Ji, Xiangdong and Yang, Chen",
    title = "{Momentum Flow Mechanisms and Color-Lorentz Forces on Quarks in the Nucleon}",
    eprint = "2503.01991",
    archivePrefix = "arXiv",
    primaryClass = "hep-ph",
    doi = "10.34133/research.1155",
    journal = "Research",
    volume = "9",
    pages = "1155",
    year = "2026"
}

@article{Ji:2026lyj,
    author = "Ji, Xiangdong and Miller, Gerald A. and Yang, Chen",
    title = "{Evidence for Quark Confinement in the Proton}",
    eprint = "2605.00339",
    archivePrefix = "arXiv",
    primaryClass = "hep-ph",
    doi = "10.34133/research.1369",
    journal = "Research",
    volume = "9",
    pages = "1369",
    year = "2026"
}

@article{Ji:1996nm,
    author = "Ji, Xiang-Dong",
    title = "{Deeply virtual Compton scattering}",
    eprint = "hep-ph/9609381",
    archivePrefix = "arXiv",
    reportNumber = "UMD-PP-97-26, MIT-CTP-2568",
    doi = "10.1103/PhysRevD.55.7114",
    journal = "Phys. Rev. D",
    volume = "55",
    pages = "7114--7125",
    year = "1997"
}

@article{Pasquini:2014vua,
    author = "Pasquini, B. and Polyakov, M. V. and Vanderhaeghen, M.",
    title = "{Dispersive evaluation of the D-term form factor in deeply virtual Compton scattering}",
    eprint = "1407.5960",
    archivePrefix = "arXiv",
    primaryClass = "hep-ph",
    doi = "10.1016/j.physletb.2014.10.047",
    journal = "Phys. Lett. B",
    volume = "739",
    pages = "133--138",
    year = "2014"
}

@article{Burkert:2018bqq,
    author = "Burkert, V. D. and Elouadrhiri, L. and Girod, F. X.",
    title = "{The pressure distribution inside the proton}",
    doi = "10.1038/s41586-018-0060-z",
    journal = "Nature",
    volume = "557",
    number = "7705",
    pages = "396--399",
    year = "2018"
}

@article{Dutrieux:2021nlz,
    author = "Dutrieux, H. and Lorc{\'e}, C. and Moutarde, H. and Sznajder, P. and Trawi{\'n}ski, A. and Wagner, J.",
    title = "{Phenomenological assessment of proton mechanical properties from deeply virtual Compton scattering}",
    eprint = "2101.03855",
    archivePrefix = "arXiv",
    primaryClass = "hep-ph",
    doi = "10.1140/epjc/s10052-021-09069-w",
    journal = "Eur. Phys. J. C",
    volume = "81",
    number = "4",
    pages = "300",
    year = "2021"
}

@article{Burkert:2021ith,
    author = "Burkert, V. D. and Elouadrhiri, L. and Girod, F. X.",
    title = "{Determination of shear forces inside the proton}",
    eprint = "2104.02031",
    archivePrefix = "arXiv",
    primaryClass = "nucl-ex",
    month = "4",
    year = "2021"
}

@article{Burkert:2023atx,
    author = "Burkert, V. D. and Elouadrhiri, L. and Girod, F. X.",
    title = "{The mechanical radius of the proton}",
    eprint = "2310.11568",
    archivePrefix = "arXiv",
    primaryClass = "hep-ph",
    month = "10",
    year = "2023"
}

@article{Bhattacharya:2022xxw,
    author = "Bhattacharya, Shohini and Hatta, Yoshitaka and Vogelsang, Werner",
    title = "{Chiral and trace anomalies in deeply virtual Compton scattering}",
    eprint = "2210.13419",
    archivePrefix = "arXiv",
    primaryClass = "hep-ph",
    doi = "10.1103/PhysRevD.107.014026",
    journal = "Phys. Rev. D",
    volume = "107",
    number = "1",
    pages = "014026",
    year = "2023"
}

@article{Bhattacharya:2023wvy,
    author = "Bhattacharya, Shohini and Hatta, Yoshitaka and Vogelsang, Werner",
    title = "{Chiral and trace anomalies in deeply virtual Compton scattering. II. QCD factorization and beyond}",
    eprint = "2305.09431",
    archivePrefix = "arXiv",
    primaryClass = "hep-ph",
    doi = "10.1103/PhysRevD.108.014029",
    journal = "Phys. Rev. D",
    volume = "108",
    number = "1",
    pages = "014029",
    year = "2023"
}

@article{Kharzeev:1995ij,
    author = "Kharzeev, D.",
    editor = "Di Giacomo, A. and Diakonov, Dmitri",
    title = "{Quarkonium interactions in QCD}",
    eprint = "nucl-th/9601029",
    archivePrefix = "arXiv",
    reportNumber = "CERN-TH-95-342, BI-TP-95-41",
    doi = "10.3254/978-1-61499-215-8-105",
    journal = "Proc. Int. Sch. Phys. Fermi",
    volume = "130",
    pages = "105--131",
    year = "1996"
}

@article{Kharzeev:1998bz,
    author = "Kharzeev, D. and Satz, H. and Syamtomov, A. and Zinovjev, G.",
    title = "{J / psi photoproduction and the gluon structure of the nucleon}",
    eprint = "hep-ph/9901375",
    archivePrefix = "arXiv",
    reportNumber = "BI-TP-98-36",
    doi = "10.1007/s100529900047",
    journal = "Eur. Phys. J. C",
    volume = "9",
    pages = "459--462",
    year = "1999"
}

@article{Hatta:2018ina,
    author = "Hatta, Yoshitaka and Yang, Di-Lun",
    title = "{Holographic $J/\psi$ production near threshold and the proton mass problem}",
    eprint = "1808.02163",
    archivePrefix = "arXiv",
    primaryClass = "hep-ph",
    reportNumber = "YITP-18-79",
    doi = "10.1103/PhysRevD.98.074003",
    journal = "Phys. Rev. D",
    volume = "98",
    number = "7",
    pages = "074003",
    year = "2018"
}

@article{Hatta:2019lxo,
    author = "Hatta, Yoshitaka and Rajan, Abha and Yang, Di-Lun",
    title = "{Near threshold J/{\ensuremath{\psi}} and {\ensuremath{\Upsilon}} photoproduction at JLab and RHIC}",
    eprint = "1906.00894",
    archivePrefix = "arXiv",
    primaryClass = "hep-ph",
    reportNumber = "YITP-19-46",
    doi = "10.1103/PhysRevD.100.014032",
    journal = "Phys. Rev. D",
    volume = "100",
    number = "1",
    pages = "014032",
    year = "2019"
}

@article{Kharzeev:2021qkd,
    author = "Kharzeev, Dmitri E.",
    title = "{Mass radius of the proton}",
    eprint = "2102.00110",
    archivePrefix = "arXiv",
    primaryClass = "hep-ph",
    doi = "10.1103/PhysRevD.104.054015",
    journal = "Phys. Rev. D",
    volume = "104",
    number = "5",
    pages = "054015",
    year = "2021"
}

@article{Boussarie:2020vmu,
    author = "Boussarie, Renaud and Hatta, Yoshitaka",
    title = "{QCD analysis of near-threshold quarkonium leptoproduction at large photon virtualities}",
    eprint = "2004.12715",
    archivePrefix = "arXiv",
    primaryClass = "hep-ph",
    doi = "10.1103/PhysRevD.101.114004",
    journal = "Phys. Rev. D",
    volume = "101",
    number = "11",
    pages = "114004",
    year = "2020"
}

@article{Hatta:2021can,
    author = "Hatta, Yoshitaka and Strikman, Mark",
    title = "{$\phi$-meson lepto-production near threshold and the strangeness $D$-term}",
    eprint = "2102.12631",
    archivePrefix = "arXiv",
    primaryClass = "hep-ph",
    doi = "10.1016/j.physletb.2021.136295",
    journal = "Phys. Lett. B",
    volume = "817",
    pages = "136295",
    year = "2021"
}

@article{Guo:2021ibg,
    author = "Guo, Yuxun and Ji, Xiangdong and Liu, Yizhuang",
    title = "{QCD Analysis of Near-Threshold Photon-Proton Production of Heavy Quarkonium}",
    eprint = "2103.11506",
    archivePrefix = "arXiv",
    primaryClass = "hep-ph",
    doi = "10.1103/PhysRevD.103.096010",
    journal = "Phys. Rev. D",
    volume = "103",
    number = "9",
    pages = "096010",
    year = "2021"
}

@article{Guo:2023qgu,
    author = "Guo, Yuxun and Ji, Xiangdong and Yuan, Feng",
    title = "{Proton{\textquoteright}s gluon GPDs at large skewness and gravitational form factors from near threshold heavy quarkonium photoproduction}",
    eprint = "2308.13006",
    archivePrefix = "arXiv",
    primaryClass = "hep-ph",
    doi = "10.1103/PhysRevD.109.014014",
    journal = "Phys. Rev. D",
    volume = "109",
    number = "1",
    pages = "014014",
    year = "2024"
}

@article{Guo:2023pqw,
    author = "Guo, Yuxun and Ji, Xiangdong and Liu, Yizhuang and Yang, Jinghong",
    title = "{Updated analysis of near-threshold heavy quarkonium production for probe of proton{\textquoteright}s gluonic gravitational form factors}",
    eprint = "2305.06992",
    archivePrefix = "arXiv",
    primaryClass = "hep-ph",
    doi = "10.1103/PhysRevD.108.034003",
    journal = "Phys. Rev. D",
    volume = "108",
    number = "3",
    pages = "034003",
    year = "2023"
}

@article{Sun:2021gmi,
    author = "Sun, Peng and Tong, Xuan-Bo and Yuan, Feng",
    title = "{Perturbative QCD analysis of near threshold heavy quarkonium photoproduction at large momentum transfer}",
    eprint = "2103.12047",
    archivePrefix = "arXiv",
    primaryClass = "hep-ph",
    doi = "10.1016/j.physletb.2021.136655",
    journal = "Phys. Lett. B",
    volume = "822",
    pages = "136655",
    year = "2021"
}

@article{Sun:2021pyw,
    author = "Sun, Peng and Tong, Xuan-Bo and Yuan, Feng",
    title = "{Near threshold heavy quarkonium photoproduction at large momentum transfer}",
    eprint = "2111.07034",
    archivePrefix = "arXiv",
    primaryClass = "hep-ph",
    doi = "10.1103/PhysRevD.105.054032",
    journal = "Phys. Rev. D",
    volume = "105",
    number = "5",
    pages = "054032",
    year = "2022"
}

@article{Mamo:2021tzd,
    author = "Mamo, Kiminad A. and Zahed, Ismail",
    title = "{Electroproduction of heavy vector mesons using holographic QCD: From near threshold to high energy regimes}",
    eprint = "2106.00722",
    archivePrefix = "arXiv",
    primaryClass = "hep-ph",
    doi = "10.1103/PhysRevD.104.066023",
    journal = "Phys. Rev. D",
    volume = "104",
    number = "6",
    pages = "066023",
    year = "2021"
}

@article{Mamo:2022eui,
    author = "Mamo, Kiminad A. and Zahed, Ismail",
    title = "{J/{\ensuremath{\psi}} near threshold in holographic QCD: A and D gravitational form factors}",
    eprint = "2204.08857",
    archivePrefix = "arXiv",
    primaryClass = "hep-ph",
    doi = "10.1103/PhysRevD.106.086004",
    journal = "Phys. Rev. D",
    volume = "106",
    number = "8",
    pages = "086004",
    year = "2022"
}

@article{Wang:2021dis,
    author = "Wang, Rong and Kou, Wei and Xie, Ya-Ping and Chen, Xurong",
    title = "{Extraction of the proton mass radius from the vector meson photoproductions near thresholds}",
    eprint = "2102.01610",
    archivePrefix = "arXiv",
    primaryClass = "hep-ph",
    doi = "10.1103/PhysRevD.103.L091501",
    journal = "Phys. Rev. D",
    volume = "103",
    number = "9",
    pages = "L091501",
    year = "2021"
}

@article{Duran:2022xag,
    author = "Duran, B. and others",
    title = "{Determining the gluonic gravitational form factors of the proton}",
    eprint = "2207.05212",
    archivePrefix = "arXiv",
    primaryClass = "nucl-ex",
    doi = "10.1038/s41586-023-05730-4",
    journal = "Nature",
    volume = "615",
    number = "7954",
    pages = "813--816",
    year = "2023"
}

@article{Burkert:2023wzr,
    author = "Burkert, V. D. and Elouadrhiri, L. and Girod, F. X. and Lorc{\'e}, C. and Schweitzer, P. and Shanahan, P. E.",
    title = "{Colloquium: Gravitational form factors of the proton}",
    eprint = "2303.08347",
    archivePrefix = "arXiv",
    primaryClass = "hep-ph",
    reportNumber = "JLAB-PHY-23-3774",
    doi = "10.1103/RevModPhys.95.041002",
    journal = "Rev. Mod. Phys.",
    volume = "95",
    number = "4",
    pages = "041002",
    year = "2023"
}

@article{Hagler:2003jd,
    author = "Hagler, Philipp and Negele, John W. and Renner, Dru Bryant and Schroers, W. and Lippert, T. and Schilling, K.",
    collaboration = "LHPC, SESAM",
    title = "{Moments of nucleon generalized parton distributions in lattice QCD}",
    eprint = "hep-lat/0304018",
    archivePrefix = "arXiv",
    reportNumber = "MIT-CPT-3353",
    doi = "10.1103/PhysRevD.68.034505",
    journal = "Phys. Rev. D",
    volume = "68",
    pages = "034505",
    year = "2003"
}

@article{Gockeler:2003jfa,
    author = "Gockeler, M. and Horsley, R. and Pleiter, D. and Rakow, Paul E. L. and Schafer, A. and Schierholz, G. and Schroers, W.",
    collaboration = "QCDSF",
    title = "{Generalized parton distributions from lattice QCD}",
    eprint = "hep-ph/0304249",
    archivePrefix = "arXiv",
    reportNumber = "DESY-03-051, LU-ITP-2003-006, EDINBURGH-2003-04, LTH-573, MIT-CTP-3365",
    doi = "10.1103/PhysRevLett.92.042002",
    journal = "Phys. Rev. Lett.",
    volume = "92",
    pages = "042002",
    year = "2004"
}

@article{Hagler:2007xi,
    author = "Hagler, Ph. and others",
    collaboration = "LHPC",
    title = "{Nucleon Generalized Parton Distributions from Full Lattice QCD}",
    eprint = "0705.4295",
    archivePrefix = "arXiv",
    primaryClass = "hep-lat",
    reportNumber = "DESY-07-077, JLAB-THY-07-651, TUM-T39-07-09",
    doi = "10.1103/PhysRevD.77.094502",
    journal = "Phys. Rev. D",
    volume = "77",
    pages = "094502",
    year = "2008"
}

@article{Deka:2013zha,
    author = "Deka, M. and others",
    title = "{Lattice study of quark and glue momenta and angular momenta in the nucleon}",
    eprint = "1312.4816",
    archivePrefix = "arXiv",
    primaryClass = "hep-lat",
    reportNumber = "RIKEN-QHP-109",
    doi = "10.1103/PhysRevD.91.014505",
    journal = "Phys. Rev. D",
    volume = "91",
    number = "1",
    pages = "014505",
    year = "2015"
}

@article{Shanahan:2018pib,
    author = "Shanahan, P. E. and Detmold, W.",
    title = "{Gluon gravitational form factors of the nucleon and the pion from lattice QCD}",
    eprint = "1810.04626",
    archivePrefix = "arXiv",
    primaryClass = "hep-lat",
    reportNumber = "MIT-CTP/5069",
    doi = "10.1103/PhysRevD.99.014511",
    journal = "Phys. Rev. D",
    volume = "99",
    number = "1",
    pages = "014511",
    year = "2019"
}

@article{Hackett:2023nkr,
    author = "Hackett, Daniel C. and Oare, Patrick R. and Pefkou, Dimitra A. and Shanahan, Phiala E.",
    title = "{Gravitational form factors of the pion from lattice QCD}",
    eprint = "2307.11707",
    archivePrefix = "arXiv",
    primaryClass = "hep-lat",
    reportNumber = "MIT-CTP/5585",
    doi = "10.1103/PhysRevD.108.114504",
    journal = "Phys. Rev. D",
    volume = "108",
    number = "11",
    pages = "114504",
    year = "2023"
}

@article{Hackett:2023rif,
    author = "Hackett, Daniel C. and Pefkou, Dimitra A. and Shanahan, Phiala E.",
    title = "{Gravitational Form Factors of the Proton from Lattice QCD}",
    eprint = "2310.08484",
    archivePrefix = "arXiv",
    primaryClass = "hep-lat",
    reportNumber = "MIT-CTP/5630, FERMILAB-PUB-23-592-T",
    doi = "10.1103/PhysRevLett.132.251904",
    journal = "Phys. Rev. Lett.",
    volume = "132",
    number = "25",
    pages = "251904",
    year = "2024"
}

@article{Tong:2021ctu,
    author = "Tong, Xuan-Bo and Ma, Jian-Ping and Yuan, Feng",
    title = "{Gluon gravitational form factors at large momentum transfer}",
    eprint = "2101.02395",
    archivePrefix = "arXiv",
    primaryClass = "hep-ph",
    doi = "10.1016/j.physletb.2021.136751",
    journal = "Phys. Lett. B",
    volume = "823",
    pages = "136751",
    year = "2021"
}

@article{Tong:2022zax,
    author = "Tong, Xuan-Bo and Ma, Jian-Ping and Yuan, Feng",
    title = "{Perturbative calculations of gravitational form factors at large momentum transfer}",
    eprint = "2203.13493",
    archivePrefix = "arXiv",
    primaryClass = "hep-ph",
    doi = "10.1007/JHEP10(2022)046",
    journal = "JHEP",
    volume = "10",
    pages = "046",
    year = "2022"
}

@article{Coriano:2026gyl,
    author = "Corian{\`o}, Claudio and Li, Hsiang-nan and Melle, Dario",
    title = "{The pion gravitational form factors and the trace anomaly in QCD factorization}",
    eprint = "2606.03055",
    archivePrefix = "arXiv",
    primaryClass = "hep-ph",
    doi = "10.1016/j.physletb.2026.140749",
    journal = "Phys. Lett. B",
    volume = "880",
    pages = "140749",
    year = "2026"
}

@article{Coriano:2026ssa,
    author = "Corian{\`o}, Claudio and Li, Hsiang-nan and Melle, Dario and Torcellini, Leonardo",
    title = "{Gravitational form factor of the pion in perturbative QCD with a dilaton interaction}",
    eprint = "2606.10993",
    archivePrefix = "arXiv",
    primaryClass = "hep-ph",
    doi = "10.1103/y1qs-vbld",
    journal = "Phys. Rev. D",
    volume = "114",
    number = "7",
    pages = "074045",
    year = "2026"
}

@article{Hagiwara:2024wqz,
    author = "Hagiwara, Yoshikazu and Tong, Xuan-Bo and Xiao, Bo-Wen",
    title = {{Understanding gravitational form factors with the Weizs{\"a}cker-Williams method}},
    eprint = "2401.12840",
    archivePrefix = "arXiv",
    primaryClass = "hep-ph",
    doi = "10.1103/PhysRevD.111.L051503",
    journal = "Phys. Rev. D",
    volume = "111",
    number = "5",
    pages = "L051503",
    year = "2025"
}

@article{Uji:2026zvn,
    author = "Uji, Tomoya",
    title = "{Energy-momentum tensor at small $x$ from gravitational transverse-momentum distributions}",
    eprint = "2610.08456",
    archivePrefix = "arXiv",
    primaryClass = "hep-ph",
    month = "10",
    year = "2026"
}

@article{Donoghue:1991qv,
    author = "Donoghue, John F. and Leutwyler, H.",
    title = "{Energy and momentum in chiral theories}",
    reportNumber = "UMHEP-321, NSF-ITP-91-119I",
    doi = "10.1007/BF01560453",
    journal = "Z. Phys. C",
    volume = "52",
    pages = "343--351",
    year = "1991"
}

@article{Belitsky:2002jp,
    author = "Belitsky, Andrei V. and Ji, X.",
    title = "{Chiral structure of nucleon gravitational form-factors}",
    eprint = "hep-ph/0203276",
    archivePrefix = "arXiv",
    reportNumber = "DOE-ER-40762-252, MD-PP-02-043",
    doi = "10.1016/S0370-2693(02)02025-7",
    journal = "Phys. Lett. B",
    volume = "538",
    pages = "289--297",
    year = "2002"
}

@article{Diehl:2006ya,
    author = "Diehl, M. and Manashov, A. and Schafer, A.",
    title = "{Chiral perturbation theory for nucleon generalized parton distributions}",
    eprint = "hep-ph/0608113",
    archivePrefix = "arXiv",
    reportNumber = "DESY-06-123",
    doi = "10.1140/epja/i2006-10098-4",
    journal = "Eur. Phys. J. A",
    volume = "29",
    pages = "315--326",
    year = "2006",
    note = "[Erratum: Eur.Phys.J.A 56, 220 (2020)]"
}

@article{Alharazin:2020yjv,
    author = "Alharazin, H. and Djukanovic, D. and Gegelia, J. and Polyakov, M. V.",
    title = "{Chiral theory of nucleons and pions in the presence of an external gravitational field}",
    eprint = "2006.05890",
    archivePrefix = "arXiv",
    primaryClass = "hep-ph",
    doi = "10.1103/PhysRevD.102.076023",
    journal = "Phys. Rev. D",
    volume = "102",
    number = "7",
    pages = "076023",
    year = "2020"
}

@article{Gegelia:2021wnj,
    author = "Gegelia, Jambul and Polyakov, Maxim V.",
    title = "{A bound on the nucleon Druck-term from chiral EFT in curved space-time and mechanical stability conditions}",
    eprint = "2104.13954",
    archivePrefix = "arXiv",
    primaryClass = "hep-ph",
    doi = "10.1016/j.physletb.2021.136572",
    journal = "Phys. Lett. B",
    volume = "820",
    pages = "136572",
    year = "2021"
}

@article{Alharazin:2023uhr,
    author = "Alharazin, H.",
    title = "{Gravitational form factors of the nucleon and one pion graviproduction in chiral EFT}",
    eprint = "2312.09675",
    archivePrefix = "arXiv",
    primaryClass = "hep-ph",
    doi = "10.1103/PhysRevD.109.016009",
    journal = "Phys. Rev. D",
    volume = "109",
    number = "1",
    pages = "016009",
    year = "2024"
}

@article{Tanaka:2025pny,
    author = "Tanaka, Mitsuru and Fujii, Daisuke and Kawaguchi, Mamiya",
    title = "{Gravitational form factors of the nucleon in the Skyrme model based on scale-invariant chiral perturbation theory}",
    eprint = "2507.21220",
    archivePrefix = "arXiv",
    primaryClass = "hep-ph",
    doi = "10.1103/gzj5-7bln",
    journal = "Phys. Rev. D",
    volume = "112",
    number = "5",
    pages = "054048",
    year = "2025"
}

@article{Anikin:2019kwi,
    author = "Anikin, I. V.",
    title = "{Gravitational form factors within light-cone sum rules at leading order}",
    eprint = "1902.00094",
    archivePrefix = "arXiv",
    primaryClass = "hep-ph",
    doi = "10.1103/PhysRevD.99.094026",
    journal = "Phys. Rev. D",
    volume = "99",
    number = "9",
    pages = "094026",
    year = "2019"
}

@article{Aliev:2020aih,
    author = "Aliev, T. M. and Barakat, T. and {\c{S}}im{\c{s}}ek, K.",
    title = "{Gravitational formfactors of the $\rho$, $\pi$, and $K$ mesons in the light-cone QCD sum rules}",
    eprint = "2009.07926",
    archivePrefix = "arXiv",
    primaryClass = "hep-ph",
    doi = "10.1103/PhysRevD.103.054001",
    journal = "Phys. Rev. D",
    volume = "103",
    pages = "054001",
    year = "2021"
}

@article{Donoghue:1990xh,
    author = "Donoghue, John F. and Gasser, J. and Leutwyler, H.",
    title = "{The Decay of a Light Higgs Boson}",
    reportNumber = "CERN-TH-5644/90, BUTP-89-06",
    doi = "10.1016/0550-3213(90)90474-R",
    journal = "Nucl. Phys. B",
    volume = "343",
    pages = "341--368",
    year = "1990"
}

@article{Cao:2024zlf,
    author = "Cao, Xiong-Hui and Guo, Feng-Kun and Li, Qu-Zhi and Yao, De-Liang",
    title = "{Dispersive determination of nucleon gravitational form factors}",
    eprint = "2411.13398",
    archivePrefix = "arXiv",
    primaryClass = "hep-ph",
    doi = "10.1038/s41467-025-62278-9",
    journal = "Nature Commun.",
    volume = "16",
    pages = "6979",
    year = "2025"
}

@article{Cao:2025dkv,
    author = "Cao, Xiong-Hui and Guo, Feng-Kun and Li, Qu-Zhi and Wu, Bo-Wen and Yao, De-Liang",
    title = "{Gravitational form factors of pions, kaons and nucleons from dispersion relations}",
    eprint = "2507.05375",
    archivePrefix = "arXiv",
    primaryClass = "hep-ph",
    doi = "10.1140/epjs/s11734-025-02025-9",
    journal = "Eur. Phys. J. ST",
    volume = "235",
    number = "9",
    pages = "2121--2154",
    year = "2026"
}

@article{Abidin:2008hn,
    author = "Abidin, Zainul and Carlson, Carl E.",
    title = "{Gravitational Form Factors in the Axial Sector from an AdS/QCD Model}",
    eprint = "0804.0214",
    archivePrefix = "arXiv",
    primaryClass = "hep-ph",
    doi = "10.1103/PhysRevD.77.115021",
    journal = "Phys. Rev. D",
    volume = "77",
    pages = "115021",
    year = "2008"
}

@article{Abidin:2009hr,
    author = "Abidin, Zainul and Carlson, Carl E.",
    title = "{Nucleon electromagnetic and gravitational form factors from holography}",
    eprint = "0903.4818",
    archivePrefix = "arXiv",
    primaryClass = "hep-ph",
    doi = "10.1103/PhysRevD.79.115003",
    journal = "Phys. Rev. D",
    volume = "79",
    pages = "115003",
    year = "2009"
}

@article{Mamo:2021krl,
    author = "Mamo, Kiminad A. and Zahed, Ismail",
    title = "{Nucleon mass radii and distribution: Holographic QCD, Lattice QCD and GlueX data}",
    eprint = "2103.03186",
    archivePrefix = "arXiv",
    primaryClass = "hep-ph",
    doi = "10.1103/PhysRevD.103.094010",
    journal = "Phys. Rev. D",
    volume = "103",
    number = "9",
    pages = "094010",
    year = "2021"
}

@article{Fujita:2022jus,
    author = "Fujita, Mitsutoshi and Hatta, Yoshitaka and Sugimoto, Shigeki and Ueda, Takahiro",
    title = "{Nucleon D-term in holographic quantum chromodynamics}",
    eprint = "2206.06578",
    archivePrefix = "arXiv",
    primaryClass = "hep-th",
    reportNumber = "YITP-22-08, KUNS-2931",
    doi = "10.1093/ptep/ptac110",
    journal = "PTEP",
    volume = "2022",
    number = "9",
    pages = "093B06",
    year = "2022"
}

@article{Fujii:2024rqd,
    author = "Fujii, Daisuke and Iwanaka, Akihiro and Tanaka, Mitsuru",
    title = "{Gravitational form factors of pion from top-down holographic QCD}",
    eprint = "2407.21113",
    archivePrefix = "arXiv",
    primaryClass = "hep-ph",
    doi = "10.1103/PhysRevD.110.L091501",
    journal = "Phys. Rev. D",
    volume = "110",
    number = "9",
    pages = "L091501",
    year = "2024"
}

@article{Sain:2025kup,
    author = "Sain, Amrita and Choudhary, Poonam and Gurjar, Bheemsehan and Mondal, Chandan and Chakrabarti, Dipankar and Mukherjee, Asmita",
    title = "{Gluon gravitational form factors of the proton in a light-front spectator model}",
    eprint = "2503.12574",
    archivePrefix = "arXiv",
    primaryClass = "hep-ph",
    doi = "10.1103/PhysRevD.111.094011",
    journal = "Phys. Rev. D",
    volume = "111",
    number = "9",
    pages = "094011",
    year = "2025"
}

@article{Deng:2026ydm,
    author = "Deng, Jiali and Wang, Xiaolong and Zhou, Yang and Hou, Defu",
    title = "{Gravitational form factors of the pion in light-front holographic QCD}",
    eprint = "2607.07305",
    archivePrefix = "arXiv",
    primaryClass = "hep-ph",
    month = "7",
    year = "2026"
}

@article{Guo:2025jiz,
    author = "Guo, Yuxun and Yuan, Feng and Zhao, Wenbin",
    title = "{Bayesian Inferring Nucleon Gravitational Form Factors via Near-Threshold J/{\ensuremath{\psi}} Photoproduction}",
    eprint = "2501.10532",
    archivePrefix = "arXiv",
    primaryClass = "hep-ph",
    doi = "10.1103/3x7r-ythq",
    journal = "Phys. Rev. Lett.",
    volume = "135",
    number = "11",
    pages = "111902",
    year = "2025"
}

@article{Freedman:1974gs,
    author = "Freedman, Daniel Z. and Muzinich, Ivan J. and Weinberg, Erick J.",
    title = "{On the Energy-Momentum Tensor in Gauge Field Theories}",
    reportNumber = "BNL-18740, COO-2220-18",
    doi = "10.1016/0003-4916(74)90448-5",
    journal = "Annals Phys.",
    volume = "87",
    pages = "95",
    year = "1974"
}

@article{Nielsen:1975ph,
    author = "Nielsen, N. K.",
    title = "{Gauge Invariance and Broken Conformal Symmetry}",
    reportNumber = "PRINT-75-0532 (AARHUS)",
    doi = "10.1016/0550-3213(75)90378-8",
    journal = "Nucl. Phys. B",
    volume = "97",
    pages = "527--540",
    year = "1975"
}

@article{Nielsen:1977sy,
    author = "Nielsen, N. K.",
    title = "{The Energy Momentum Tensor in a Nonabelian Quark Gluon Theory}",
    doi = "10.1016/0550-3213(77)90040-2",
    journal = "Nucl. Phys. B",
    volume = "120",
    pages = "212--220",
    year = "1977"
}

@article{Collins:1976yq,
    author = "Collins, John C. and Duncan, Anthony and Joglekar, Satish D.",
    title = "{Trace and Dilatation Anomalies in Gauge Theories}",
    reportNumber = "COO-2220-88",
    doi = "10.1103/PhysRevD.16.438",
    journal = "Phys. Rev. D",
    volume = "16",
    pages = "438--449",
    year = "1977"
}

@article{Tarrach:1981bi,
    author = "Tarrach, R.",
    title = "{The renormalization of FF}",
    reportNumber = "DESY-81-052",
    doi = "10.1016/0550-3213(82)90301-7",
    journal = "Nucl. Phys. B",
    volume = "196",
    pages = "45--61",
    year = "1982"
}

@article{Ji:2021qgo,
    author = {Ji, Xiangdong and Liu, Yizhuang and Sch{\"a}fer, Andreas},
    title = "{Scale symmetry breaking, quantum anomalous energy and proton mass decomposition}",
    eprint = "2105.03974",
    archivePrefix = "arXiv",
    primaryClass = "hep-ph",
    doi = "10.1016/j.nuclphysb.2021.115537",
    journal = "Nucl. Phys. B",
    volume = "971",
    pages = "115537",
    year = "2021"
}

@article{Kumano:2017lhr,
    author = "Kumano, S. and Song, Qin-Tao and Teryaev, O. V.",
    title = "{Hadron tomography by generalized distribution amplitudes in pion-pair production process $\gamma^* \gamma \rightarrow \pi^0 \pi^0 $ and gravitational form factors for pion}",
    eprint = "1711.08088",
    archivePrefix = "arXiv",
    primaryClass = "hep-ph",
    reportNumber = "J-PARC-TH-0086, KEK-TH-1959, J-PARC-TH-0086 (Erratum: KEK-TH-2096, J-PARC-TH-0155)",
    doi = "10.1103/PhysRevD.97.014020",
    journal = "Phys. Rev. D",
    volume = "97",
    number = "1",
    pages = "014020",
    year = "2018"
}

@article{Lorce:2022tiq,
    author = "Lorc{\'e}, C{\'e}dric and Pire, Bernard and Song, Qin-Tao",
    title = "{Kinematical higher-twist corrections in $\gamma^* \gamma \to M \bar M $}",
    eprint = "2209.11140",
    archivePrefix = "arXiv",
    primaryClass = "hep-ph",
    reportNumber = "CPHT-RR055.092022",
    doi = "10.1103/PhysRevD.106.094030",
    journal = "Phys. Rev. D",
    volume = "106",
    number = "9",
    pages = "094030",
    year = "2022"
}

@article{Hatta:2025ryj,
    author = "Hatta, Yoshitaka and Schoenleber, Jakob",
    title = "{Sullivan Process near Threshold and the Pion Gravitational Form Factors}",
    eprint = "2502.12061",
    archivePrefix = "arXiv",
    primaryClass = "hep-ph",
    doi = "10.1103/y9fq-y84c",
    journal = "Phys. Rev. Lett.",
    volume = "134",
    number = "25",
    pages = "251901",
    year = "2025"
}

@article{Avdeenko:1981twg,
    author = "Avdeenko, V. A. and Korenblit, S. E. and Chernyak, V. L.",
    title = "{Asymptotic behavior of electromagnetic and weak nucleon formfactors in QCD}",
    reportNumber = "Irkutsk-79-23",
    journal = "Sov. J. Nucl. Phys.",
    volume = "33",
    number = "2",
    pages = "252--256",
    year = "1981"
}

@article{Lepage:1979za,
    author = "Lepage, G. Peter and Brodsky, Stanley J.",
    title = "{Exclusive Processes in Quantum Chromodynamics: The Form Factors of Baryons at Large Momentum Transfer}",
    reportNumber = "SLAC-PUB-2348",
    doi = "10.1103/PhysRevLett.43.545",
    journal = "Phys. Rev. Lett.",
    volume = "43",
    number = "8",
    pages = "545--549",
    year = "1979",
    note = "[Erratum: Phys.Rev.Lett. 43, 1625--1626 (1979)]"
}

@article{Brodsky:1981kj,
    author = "Brodsky, Stanley J. and Lepage, G. Peter",
    title = "{Helicity Selection Rules and Tests of Gluon Spin in Exclusive QCD Processes}",
    reportNumber = "SLAC-PUB-2746",
    doi = "10.1103/PhysRevD.24.2848",
    journal = "Phys. Rev. D",
    volume = "24",
    pages = "2848",
    year = "1981"
}

@article{Efremov:1979qk,
    author = "Efremov, A. V. and Radyushkin, A. V.",
    title = "{Factorization and Asymptotical Behavior of Pion Form-Factor in QCD}",
    reportNumber = "JINR-P2-12900",
    doi = "10.1016/0370-2693(80)90869-2",
    journal = "Phys. Lett. B",
    volume = "94",
    pages = "245--250",
    year = "1980"
}

@article{Chernyak:1977as,
    author = "Chernyak, V. L. and Zhitnitsky, A. R.",
    title = "{Asymptotic Behavior of Hadron Form-Factors in Quark Model. (In Russian)}",
    journal = "JETP Lett.",
    volume = "25",
    pages = "510",
    year = "1977"
}

@article{Chernyak:1980dj,
    author = "Chernyak, V. L. and Zhitnitsky, A. R.",
    title = "{Asymptotics of Hadronic Form-Factors in the Quantum Chromodynamics. (In Russian)}",
    journal = "Sov. J. Nucl. Phys.",
    volume = "31",
    pages = "544--552",
    year = "1980"
}

@article{Lepage:1980fj,
    author = "Lepage, G. Peter and Brodsky, Stanley J.",
    title = "{Exclusive Processes in Perturbative Quantum Chromodynamics}",
    reportNumber = "SLAC-PUB-2478",
    doi = "10.1103/PhysRevD.22.2157",
    journal = "Phys. Rev. D",
    volume = "22",
    pages = "2157",
    year = "1980"
}

@article{Chernyak:1983ej,
    author = "Chernyak, V. L. and Zhitnitsky, A. R.",
    title = "{Asymptotic Behavior of Exclusive Processes in QCD}",
    reportNumber = "IYF-83-103, IYF-83-104, IYF-83-105, IYF-83-106, IYF-83-107, IYF-83-108",
    doi = "10.1016/0370-1573(84)90126-1",
    journal = "Phys. Rept.",
    volume = "112",
    pages = "173",
    year = "1984"
}

@article{Chernyak:1984bm,
    author = "Chernyak, V. L. and Zhitnitsky, I. R.",
    title = "{Nucleon Wave Function and Nucleon Form-Factors in QCD}",
    doi = "10.1016/0550-3213(84)90114-7",
    journal = "Nucl. Phys. B",
    volume = "246",
    pages = "52--74",
    year = "1984"
}

@article{Belitsky:2002kj,
    author = "Belitsky, Andrei V. and Ji, Xiang-dong and Yuan, Feng",
    title = "{A Perturbative QCD analysis of the nucleon's Pauli form-factor F(2)(Q**2)}",
    eprint = "hep-ph/0212351",
    archivePrefix = "arXiv",
    reportNumber = "DOE-ER-400762-271, UM-PP-03-031",
    doi = "10.1103/PhysRevLett.91.092003",
    journal = "Phys. Rev. Lett.",
    volume = "91",
    pages = "092003",
    year = "2003"
}

@article{Chen:2024fhj,
    author = "Chen, Long-Bin and Chen, Wen and Feng, Feng and Hu, Siwei and Jia, Yu",
    title = "{Next-to-Leading-Order QCD Corrections to Nucleon Dirac Form Factors}",
    eprint = "2406.19994",
    archivePrefix = "arXiv",
    primaryClass = "hep-ph",
    doi = "10.1103/r25y-rsz5",
    journal = "Phys. Rev. Lett.",
    volume = "135",
    number = "13",
    pages = "131903",
    year = "2025"
}

@article{Huang:2024ugd,
    author = "Huang, Yong-Kang and Shi, Bo-Xuan and Wang, Yu-Ming and Zhao, Xue-Chen",
    title = "{Next-to-Leading-Order QCD Predictions for the Nucleon Form Factors}",
    eprint = "2407.18724",
    archivePrefix = "arXiv",
    primaryClass = "hep-ph",
    doi = "10.1103/xz46-8ptk",
    journal = "Phys. Rev. Lett.",
    volume = "135",
    number = "6",
    pages = "061901",
    year = "2025"
}

@article{Field:1981wx,
    author = "Field, R. D. and Gupta, Rajan and Otto, Steve and Chang, Lee",
    title = "{Beyond Leading Order {QCD} Perturbative Corrections to the Pion Form-factor}",
    reportNumber = "UFTP-80-21",
    doi = "10.1016/0550-3213(81)90022-5",
    journal = "Nucl. Phys. B",
    volume = "186",
    pages = "429--474",
    year = "1981"
}

@article{Chen:2023byr,
    author = "Chen, Long-Bin and Chen, Wen and Feng, Feng and Jia, Yu",
    title = "{Next-to-Next-to-Leading-Order QCD Corrections to Pion Electromagnetic Form Factors}",
    eprint = "2312.17228",
    archivePrefix = "arXiv",
    primaryClass = "hep-ph",
    doi = "10.1103/PhysRevLett.132.201901",
    journal = "Phys. Rev. Lett.",
    volume = "132",
    number = "20",
    pages = "201901",
    year = "2024",
    note = "[Erratum: Phys.Rev.Lett. 134, 229901 (2025)]"
}

@article{Ji:2024iak,
    author = "Ji, Yao and Shi, Bo-Xuan and Wang, Jian and Wang, Ye-Fan and Wang, Yu-Ming and Yu, Hui-Xin",
    title = "{Next-to-Next-to-Leading-Order QCD Prediction for the Pion Form Factor}",
    eprint = "2411.03658",
    archivePrefix = "arXiv",
    primaryClass = "hep-ph",
    reportNumber = "TUM-HEP-1533/24",
    doi = "10.1103/yx51-lc3q",
    journal = "Phys. Rev. Lett.",
    volume = "134",
    number = "22",
    pages = "221901",
    year = "2025"
}

@article{Joglekar:1975nu,
    author = "Joglekar, Satish D. and Lee, Benjamin W.",
    title = "{General Theory of Renormalization of Gauge Invariant Operators}",
    reportNumber = "FERMILAB-PUB-75-050-THY, FERMILAB-PUB-75-050-T",
    doi = "10.1016/0003-4916(76)90225-6",
    journal = "Annals Phys.",
    volume = "97",
    pages = "160",
    year = "1976"
}

@article{BELINFANTE1939887,
    author = "Belinfante, F. J.",
    title = "{On the spin angular momentum of mesons}",
    doi = "10.1016/s0031-8914(39)90090-x",
    journal = "Physica",
    volume = "6",
    number = "7-12",
    pages = "887--898",
    year = "1939"
}

@Article{belinfante1940current,
  author    = {Belinfante, Frederik J},
  journal   = {Physica},
  title     = {On the current and the density of the electric charge, the energy, the linear momentum and the angular momentum of arbitrary fields},
  year      = {1940},
  number    = {5},
  pages     = {449--474},
  volume    = {7},
  publisher = {Elsevier},
  doi = {10.1016/S0031-8914(40)90091-X},
}

@article{Freese:2021czn,
    author = "Freese, Adam and Miller, Gerald A.",
    title = "{Forces within hadrons on the light front}",
    eprint = "2102.01683",
    archivePrefix = "arXiv",
    primaryClass = "hep-ph",
    reportNumber = "NT@UW-21-02",
    doi = "10.1103/PhysRevD.103.094023",
    journal = "Phys. Rev. D",
    volume = "103",
    pages = "094023",
    year = "2021"
}

@article{Tanaka:2018nae,
    author = "Tanaka, Kazuhiro",
    title = "{Three-loop formula for quark and gluon contributions to the QCD trace anomaly}",
    eprint = "1811.07879",
    archivePrefix = "arXiv",
    primaryClass = "hep-ph",
    reportNumber = "J-PARC TH-0140",
    doi = "10.1007/JHEP01(2019)120",
    journal = "JHEP",
    volume = "01",
    pages = "120",
    year = "2019"
}

@article{Tanaka:2018wea,
    author = "Tanaka, Kazuhiro",
    title = "{Operator relations for gravitational form factors of a spin-0 hadron}",
    eprint = "1806.10591",
    archivePrefix = "arXiv",
    primaryClass = "hep-ph",
    reportNumber = "J-PARC-TH-0120",
    doi = "10.1103/PhysRevD.98.034009",
    journal = "Phys. Rev. D",
    volume = "98",
    number = "3",
    pages = "034009",
    year = "2018"
}

@article{Tanaka:2022wzy,
    author = "Tanaka, Kazuhiro",
    title = "{Twist-four gravitational form factor at NNLO QCD from trace anomaly constraints}",
    eprint = "2212.09417",
    archivePrefix = "arXiv",
    primaryClass = "hep-ph",
    doi = "10.1007/JHEP03(2023)013",
    journal = "JHEP",
    volume = "03",
    pages = "013",
    year = "2023"
}

@article{Radyushkin:1977gp,
    author = "Radyushkin, A. V.",
    title = "{Deep Elastic Processes of Composite Particles in Field Theory and Asymptotic Freedom}",
    eprint = "hep-ph/0410276",
    archivePrefix = "arXiv",
    reportNumber = "JINR-P2-10717, JLAB-THY-04-35",
    month = "6",
    year = "1977"
}

@article{Efremov:1978rn,
    author = "Efremov, A. V. and Radyushkin, A. V.",
    title = "{Asymptotical Behavior of Pion Electromagnetic Form-Factor in QCD}",
    reportNumber = "JINR-E2-11983",
    doi = "10.1007/BF01032111",
    journal = "Theor. Math. Phys.",
    volume = "42",
    pages = "97--110",
    year = "1980"
}

@article{Lepage:1979zb,
    author = "Lepage, G. Peter and Brodsky, Stanley J.",
    title = "{Exclusive Processes in Quantum Chromodynamics: Evolution Equations for Hadronic Wave Functions and the Form-Factors of Mesons}",
    reportNumber = "SLAC-PUB-2343",
    doi = "10.1016/0370-2693(79)90554-9",
    journal = "Phys. Lett. B",
    volume = "87",
    pages = "359--365",
    year = "1979"
}

@article{Dittes:1983dy,
    author = "Dittes, F. M. and Radyushkin, A. V.",
    title = "{TWO LOOP CONTRIBUTION TO THE EVOLUTION OF THE PION WAVE FUNCTION}",
    doi = "10.1016/0370-2693(84)90016-9",
    journal = "Phys. Lett. B",
    volume = "134",
    pages = "359--362",
    year = "1984"
}

@article{Mikhailov:1984ii,
    author = "Mikhailov, S. V. and Radyushkin, A. V.",
    title = "{Evolution Kernels in {QCD}: Two Loop Calculation in Feynman Gauge}",
    reportNumber = "JINR-P2-84-534",
    doi = "10.1016/0550-3213(85)90213-5",
    journal = "Nucl. Phys. B",
    volume = "254",
    pages = "89--126",
    year = "1985"
}

@article{Mueller:1994cn,
    author = "Mueller, Dieter",
    title = "{The Evolution of the pion distribution amplitude in next-to-leading-order}",
    eprint = "hep-ph/9411338",
    archivePrefix = "arXiv",
    reportNumber = "SLAC-PUB-6622",
    doi = "10.1103/PhysRevD.51.3855",
    journal = "Phys. Rev. D",
    volume = "51",
    pages = "3855--3864",
    year = "1995"
}

@article{Feng:2021kha,
    author = "Feng, Feng and Xie, Yi-Fan and Zhou, Qiu-Chen and Tang, Shan-Rong",
    title = "{HepLib: A C++ library for high energy physics}",
    eprint = "2103.08507",
    archivePrefix = "arXiv",
    primaryClass = "hep-ph",
    doi = "10.1016/j.cpc.2021.107982",
    journal = "Comput. Phys. Commun.",
    volume = "265",
    pages = "107982",
    year = "2021"
}

@article{Hahn:2000kx,
    author = "Hahn, Thomas",
    title = "{Generating Feynman diagrams and amplitudes with FeynArts 3}",
    eprint = "hep-ph/0012260",
    archivePrefix = "arXiv",
    reportNumber = "KA-TP-23-2000",
    doi = "10.1016/S0010-4655(01)00290-9",
    journal = "Comput. Phys. Commun.",
    volume = "140",
    pages = "418--431",
    year = "2001"
}

@article{Duhr:2019tlz,
    author = "Duhr, Claude and Dulat, Falko",
    title = "{PolyLogTools {\textemdash} polylogs for the masses}",
    eprint = "1904.07279",
    archivePrefix = "arXiv",
    primaryClass = "hep-th",
    reportNumber = "CP3-19-17, CERN-TH-2019-045, SLAC-PUB-17423",
    doi = "10.1007/JHEP08(2019)135",
    journal = "JHEP",
    volume = "08",
    pages = "135",
    year = "2019"
}

@article{Mikhailov:2016klg,
    author = "Mikhailov, S. V. and Pimikov, A. V. and Stefanis, N. G.",
    title = "{Systematic estimation of theoretical uncertainties in the calculation of the pion-photon transition form factor using light-cone sum rules}",
    eprint = "1604.06391",
    archivePrefix = "arXiv",
    primaryClass = "hep-ph",
    reportNumber = "RUB-TPII-01-2016",
    doi = "10.1103/PhysRevD.93.114018",
    journal = "Phys. Rev. D",
    volume = "93",
    number = "11",
    pages = "114018",
    year = "2016"
}

@article{Bakulev:2001pa,
    author = "Bakulev, A. P. and Mikhailov, S. V. and Stefanis, N. G.",
    title = "{QCD based pion distribution amplitudes confronting experimental data}",
    eprint = "hep-ph/0103119",
    archivePrefix = "arXiv",
    reportNumber = "RUB-TPII-03-01",
    doi = "10.1016/S0370-2693(01)00517-2",
    journal = "Phys. Lett. B",
    volume = "508",
    pages = "279--289",
    year = "2001",
    note = "[Erratum: Phys.Lett.B 590, 309--310 (2004)]"
}

@article{Ball:2006wn,
    author = "Ball, Patricia and Braun, V. M. and Lenz, A.",
    title = "{Higher-twist distribution amplitudes of the K meson in QCD}",
    eprint = "hep-ph/0603063",
    archivePrefix = "arXiv",
    reportNumber = "IPPP-06-01, DCPT-06-02, CERN-PH-TH-2006-026",
    doi = "10.1088/1126-6708/2006/05/004",
    journal = "JHEP",
    volume = "05",
    pages = "004",
    year = "2006"
}

@article{Stefanis:2020rnd,
    author = "Stefanis, N. G.",
    title = "{Pion-photon transition form factor in light cone sum rules and tests of asymptotics}",
    eprint = "2006.10576",
    archivePrefix = "arXiv",
    primaryClass = "hep-ph",
    reportNumber = "RUB-TPII-01/2020",
    doi = "10.1103/PhysRevD.102.034022",
    journal = "Phys. Rev. D",
    volume = "102",
    number = "3",
    pages = "034022",
    year = "2020"
}

@article{Cheng:2020vwr,
    author = "Cheng, Shan and Khodjamirian, Alexander and Rusov, Aleksey V.",
    title = "{Pion light-cone distribution amplitude from the pion electromagnetic form factor}",
    eprint = "2007.05550",
    archivePrefix = "arXiv",
    primaryClass = "hep-ph",
    reportNumber = "SI-HEP-2020-15, P3H-20-032, IPPP/20/27",
    doi = "10.1103/PhysRevD.102.074022",
    journal = "Phys. Rev. D",
    volume = "102",
    number = "7",
    pages = "074022",
    year = "2020"
}

@article{Stefanis:2014nla,
    author = "Stefanis, N. G.",
    title = "{What binds quarks together at different momentum scales? A conceptual scenario}",
    eprint = "1405.0959",
    archivePrefix = "arXiv",
    primaryClass = "hep-ph",
    reportNumber = "RUB-TPII-03-2014",
    doi = "10.1016/j.physletb.2014.10.018",
    journal = "Phys. Lett. B",
    volume = "738",
    pages = "483--487",
    year = "2014"
}

@article{Chang:2013pq,
    author = "Chang, Lei and Cloet, I. C. and Cobos-Martinez, J. J. and Roberts, C. D. and Schmidt, S. M. and Tandy, P. C.",
    title = "{Imaging dynamical chiral symmetry breaking: pion wave function on the light front}",
    eprint = "1301.0324",
    archivePrefix = "arXiv",
    primaryClass = "nucl-th",
    doi = "10.1103/PhysRevLett.110.132001",
    journal = "Phys. Rev. Lett.",
    volume = "110",
    number = "13",
    pages = "132001",
    year = "2013"
}

@article{Raya:2015gva,
    author = "Raya, K. and Chang, L. and Bashir, A. and Cobos-Martinez, J. J. and Guti{\'e}rrez-Guerrero, L. X. and Roberts, C. D. and Tandy, P. C.",
    title = "{Structure of the neutral pion and its electromagnetic transition form factor}",
    eprint = "1510.02799",
    archivePrefix = "arXiv",
    primaryClass = "nucl-th",
    doi = "10.1103/PhysRevD.93.074017",
    journal = "Phys. Rev. D",
    volume = "93",
    number = "7",
    pages = "074017",
    year = "2016"
}

@article{Lan:2019rba,
    author = "Lan, Jiangshan and Mondal, Chandan and Jia, Shaoyang and Zhao, Xingbo and Vary, James P.",
    title = "{Pion and kaon parton distribution functions from basis light front quantization and QCD evolution}",
    eprint = "1907.01509",
    archivePrefix = "arXiv",
    primaryClass = "nucl-th",
    doi = "10.1103/PhysRevD.101.034024",
    journal = "Phys. Rev. D",
    volume = "101",
    number = "3",
    pages = "034024",
    year = "2020"
}

@article{Choi:2014ifm,
    author = "Choi, Ho-Meoyng and Ji, Chueng-Ryong",
    title = "{Consistency of the light-front quark model with chiral symmetry in the pseudoscalar meson analysis}",
    eprint = "1412.2507",
    archivePrefix = "arXiv",
    primaryClass = "hep-ph",
    doi = "10.1103/PhysRevD.91.014018",
    journal = "Phys. Rev. D",
    volume = "91",
    number = "1",
    pages = "014018",
    year = "2015"
}

@article{Chang:2016ouf,
    author = "Chang, Qin and Brodsky, Stanley J. and Li, Xin-Qiang",
    title = "{Light-front holographic distribution amplitudes of pseudoscalar mesons and their application to $B$-meson decays}",
    eprint = "1612.05298",
    archivePrefix = "arXiv",
    primaryClass = "hep-ph",
    doi = "10.1103/PhysRevD.95.094025",
    journal = "Phys. Rev. D",
    volume = "95",
    number = "9",
    pages = "094025",
    year = "2017"
}

@article{RQCD:2019osh,
    author = {Bali, Gunnar S. and Braun, Vladimir M. and B{\"u}rger, Simon and G{\"o}ckeler, Meinulf and Gruber, Michael and Hutzler, Fabian and Korcyl, Piotr and Sch{\"a}fer, Andreas and Sternbeck, Andr{\'e} and Wein, Philipp},
    collaboration = "RQCD",
    title = "{Light-cone distribution amplitudes of pseudoscalar mesons from lattice QCD}",
    eprint = "1903.08038",
    archivePrefix = "arXiv",
    primaryClass = "hep-lat",
    doi = "10.1007/JHEP08(2019)065",
    journal = "JHEP",
    volume = "08",
    pages = "065",
    year = "2019",
    note = "[Addendum: JHEP 11, 037 (2020)]"
}

@article{LatticeParton:2022zqc,
    author = "Hua, Jun and others",
    collaboration = "Lattice Parton",
    title = "{Pion and Kaon Distribution Amplitudes from Lattice QCD}",
    eprint = "2201.09173",
    archivePrefix = "arXiv",
    primaryClass = "hep-lat",
    doi = "10.1103/PhysRevLett.129.132001",
    journal = "Phys. Rev. Lett.",
    volume = "129",
    number = "13",
    pages = "132001",
    year = "2022"
}

@article{LPCnew,
    author = "Wang {\it et al.} [LPC Collaboration], J. H.  ",
    title = "{to appear}"}

@article{Meli__1999,
    author = "Melic, B. and Nizic, B. and Passek, K.",
    title = "{Complete next-to-leading order perturbative QCD prediction for the pion form-factor}",
    eprint = "hep-ph/9802204",
    archivePrefix = "arXiv",
    reportNumber = "IRB-TH-4-97",
    doi = "10.1103/PhysRevD.60.074004",
    journal = "Phys. Rev. D",
    volume = "60",
    pages = "074004",
    year = "1999"
}

@article{Bakulev_2013,
    author = "Bakulev, Alexander P. and Khandramai, Vyacheslav L.",
    title = "{FAPT: a Mathematica package for calculations in QCD Fractional Analytic Perturbation Theory}",
    eprint = "1204.2679",
    archivePrefix = "arXiv",
    primaryClass = "hep-ph",
    doi = "10.1016/j.cpc.2012.08.014",
    journal = "Comput. Phys. Commun.",
    volume = "184",
    number = "1",
    pages = "183--193",
    year = "2013"
}

@article{Xu:2023izo,
    author = "Xu, Yin-Zhen and Ding, Minghui and Raya, Kh{\'e}pani and Roberts, Craig D. and Rodr{\'\i}guez-Quintero, Jos{\'e} and Schmidt, Sebastian M.",
    title = "{Pion and kaon electromagnetic and gravitational form factors}",
    eprint = "2311.14832",
    archivePrefix = "arXiv",
    primaryClass = "hep-ph",
    reportNumber = "NJU-INP 081/23",
    doi = "10.1140/epjc/s10052-024-12518-x",
    journal = "Eur. Phys. J. C",
    volume = "84",
    number = "2",
    pages = "191",
    year = "2024"
}

@article{Wang:2024lrm,
    author = "Wang, Bigeng and He, Fangcheng and Wang, Gen and Draper, Terrence and Liang, Jian and Liu, Keh-Fei and Yang, Yi-Bo",
    collaboration = "{\ensuremath{\chi}}QCD",
    title = "{Trace anomaly form factors from lattice QCD}",
    eprint = "2401.05496",
    archivePrefix = "arXiv",
    primaryClass = "hep-lat",
    doi = "10.1103/PhysRevD.109.094504",
    journal = "Phys. Rev. D",
    volume = "109",
    number = "9",
    pages = "094504",
    year = "2024"
}
\end{document}